\documentclass[12pt,preprint]{aastex}
\usepackage{epsfig}
\usepackage{graphicx}
\usepackage{rotating}
\usepackage{multirow}
\usepackage{color}
\usepackage{lscape}
\usepackage{mathrsfs,amssymb}
\usepackage{amsmath}
\usepackage{morefloats}
\usepackage{comment}
\newcommand	\beq	{\begin{equation}}	%{\begin{displaymath}}
\newcommand	\eeq	{\end{equation}}	%{\end{displaymath}}
\newcommand       \Angstrom     {\,{\rm \AA}}

\newcommand       \K            {\,{\rm K}}
\newcommand       \pc           {\,{\rm pc}}

\newcommand       \simlt        {\lesssim}
\newcommand       \simgt        {\gtrsim}
\newcommand       \gtsim        {\gtrsim}

\newcommand       \um           {\mu{\rm m}}
\newcommand       \mum          {\,{\rm \mu m}}
\newcommand       \ppm          {\,{\rm ppm}}

\newcommand       \simali       {\sim\,}
\newcommand       \magni        {\,{\rm mag}}

\def    \obs		{{\rm obs}}
\def    \mod	{{\rm mod}}
\def    \Nobs	{N_{\rm obs}}
\def    \Nmod	{N_{\rm mod}}
\def    \dof	        {{\rm dof}}

\newcommand{\spitzerirs}{{\em Spitzer}/IRS\ }

\newcommand{\etal}{\textrm{et al.\ }}

\newcommand{\eg}{\textrm{e.g., }}

\newcommand \tausil      {\tau_{9.7}}

\renewcommand\i   {\hbox{$i$}}

\newcommand       \NC         {N_{\rm C}}

\newcommand{\noprint}[1]{}

\def\lax{{$\mathrel{\hbox{\rlap{\hbox{\lower4pt\hbox{$\sim$}}}\hbox{$<$}}}$}}
\def\gax{{$\mathrel{\hbox{\rlap{\hbox{\lower4pt\hbox{$\sim$}}}\hbox{$>$}}}$}}

\countdef\decade=200
\advance\decade by \year
\countdef\hours=201
\advance\hours by \time
\divide\hours by 60
\countdef\mins=202
\advance\mins by \hours
\multiply\mins by 60
\multiply\hours by 100
\countdef\miltime=203
\advance\miltime by \hours
\advance\miltime by \time
\advance\miltime by -\mins

\begin{document}
\title{A New Technique for Measuring Polycyclic Aromatic Hydrocarbon Emission in Different Environments}
%\begin{comment}
\author{
Yanxia~Xie\altaffilmark{1},  
Luis~C. Ho\altaffilmark{1,2}, 
Aigen~Li\altaffilmark{3}, and 
Jinyi~Shangguan\altaffilmark{2} 
}
\altaffiltext{1}{Kavli Institute for Astronomy and Astrophysics, 
                       Peking University, Beijing 100871, China;  
                       {\sf yanxia.xie@pku.edu.cn, lho.pku@gmail.com}
                       }
\altaffiltext{2}{Department of Astronomy, School of Physics, 
                      Peking University, Beijing 100871, China, 
                       {\sf shangguan@pku.edu.cn}
                       }                       
\altaffiltext{3}{Department of Physics and Astronomy, 
                       University of Missouri, 
                       Columbia, MO 65211, USA
                       {\sf lia@missouri.edu}
                       }
%\end{comment}                       

\begin{abstract}
We present a new method to decompose the emission features of polycyclic 
aromatic hydrocarbons (PAHs) from mid-infrared spectra using theoretical PAH 
templates in conjunction with modified blackbody components for the dust 
continuum and an extinction term.  The primary goal is to obtain robust 
measurements of the PAH features, which are sensitive to the star formation 
rate, in a variety of extragalactic environments.  We demonstrate the 
effectiveness of our technique, starting with the simplest Galactic 
high-latitude clouds to extragalactic systems of ever-increasing complexity, 
from normal star-forming galaxies to low-luminosity active galaxies, quasars, 
and heavily obscured infrared-luminous galaxies.  In addition to providing 
accurate measurements of the PAH emission, including the upper limits thereof, our 
fits can reproduce reasonably well the overall continuum shape and 
constrain the line-of-sight extinction.  Our new PAH line flux measurements 
differ systematically and significantly from those of previous methods 
by $\simali$15\% to as much as a factor of $\simali$6. 
The decomposed PAH spectra show remarkable similarity among different systems, 
suggesting a uniform set of conditions responsible for their excitation.
\end{abstract}

\keywords{dust, extinction --- galaxies: active --- galaxies: ISM --- galaxies: nuclei --- galaxies: Seyfert --- infrared: ISM}

\clearpage

\section{Introduction\label{sec:intro}}
 
Quantifying star formation in galaxies is essential to  understanding their evolution. 
To this end, different observables have been used  to calibrate the 
star formation rate in galaxies (Kennicutt \& Evans 2012).  The 
emission features from polycyclic aromatic hydrocarbons (PAHs) have long 
been proposed as a useful tracer of star formation on galactic scales.  
Peeters \etal (2004) showed that local star-forming regions and nearby 
starburst galaxies obey a tight empirical correlation 
between the 6.2$\mum$ PAH emission feature
and the far-infrared (FIR) luminosity, 
implying that the PAH 6.2$\mum$ feature
is an effective tracer of the star formation rate.  
The advent of the {\it Infrared Space Observatory}
and especially the Infrared Spectrograph (IRS) 
onboard the {\it Spitzer Space Telescope} 
(Houck \etal 2004; Werner et al. 2004) has enabled 
the unambiguous detection of a set of 
distinct spectral features at 3.3, 6.2, 7.7, 8.6, and 11.3$\mum$, 
which are ubiquitously detected in the mid-infrared (MIR) spectra 
of a variety of Galactic and extragalactic sources.  For instance, these 
features are seen in protoplanetary disks around intermediate-mass, 
pre-main-sequence Herbig Ae/Be stars (\eg Acke \etal 2010),
and their low-mass analog T Tauri stars (\eg Geers et al.\ 2006), 
as well as debris disks around main-sequence stars
(\eg Seok \& Li 2015).  Moreover, they are also 
prevalent in external star-forming and starburst galaxies, accounting for more 
than 10\% of their total IR luminosity (Smith et al.\ 2007; Tielens 2008).  
The relatively broad wavelength coverage and prominence of these bands 
facilitate their detection in star-forming regions both in local and distant 
(\eg $z\simgt1$) galaxies (\eg Huang et al. 2009; Magdis \etal 2011).  The 
emission feature at 11.3$\mum$ can be robustly detected even in some active 
galactic nuclei (AGNs; Alonso-Herrero \etal 2014, 2016).  The exact 
nature of the carriers of these emission bands remains unknown, with PAHs 
being the leading candidate (\eg L\'eger \& Puget 1984; Allamandola et al.  
1985). The PAH molecules are transiently heated by ultraviolet (UV) photons from 
newly formed stars, and they are predominantly associated with 
photodissociation regions. The observed close coupling between PAH molecules 
and star-forming regions supports this hypothesis.

Rigorous application or interpretation of the PAH features requires that they 
be quantitatively measured.  This is challenging, however, for several reasons. 
First, the PAH features lie on top of a complex continuum, whose intrinsic 
underlying shape is quite uncertain because it arises, depending on the 
physical nature of the source, from the superposition of dust emission from 
different components with different temperatures.  The AGNs complicate matters 
further by introducing very hot dust from a central torus.  At the shortest IR 
wavelengths, even stellar emission may be nonnegligible.  Second, the PAH 
lines have intrinsically broad, extended profiles, which, in concert with their
close wavelength separations, render them heavily blended, especially given 
the rather low spectral resolution of instruments such as the IRS.  In the 
spectral region $\simali$6--9$\mum$, 
for example, there are so many overlapping PAH features 
that it is almost impossible to define a line-free continuum region.
Moreover, some PAH features are also blended with ionic fine-structure 
lines (\eg the 12.7$\mum$ PAH feature with the 12.8$\mum$ [Ne~{\footnotesize II}] line).
Lastly, all of the above components, line and continuum, may be 
affected further by the broad silicate emission or absorption features 
at $\simali$10 and $\simali$18$\mum$.  
Dust absorption is especially severe in some of 
the most luminous, high-redshift systems.

PAHFIT is one of the most popular tools used to decompose MIR spectra of galaxies 
(Smith \etal 2007). This technique fits the underlying continuum with a 
combination of stellar continuum, represented by blackbody emission with 
a temperature fixed at 5000\,K,  plus eight modified blackbodies with 
preassigned temperatures (from 35 to 300 K) to reproduce the dust 
continuum.  The PAH features are modeled simultaneously with Drude functions of 
fixed full width at half maximum (FWHM) and wavelength positions, and single 
Gaussian functions are added to fit the molecular hydrogen and ionic 
fine-structure lines.  

Another frequently used method performs a spline fit to continuum anchor 
points apparently free of line emission, and the PAH flux is derived by 
integrating the line emission above the local continuum (\eg Peeters \etal 
2004; Brandl \etal 2006).  In this approach, the choice of continuum anchor 
points strongly depends on the subjective, visual inspection of the spectrum,
which varies greatly from galaxy to galaxy.  Moreover, as mentioned previously,
the broad, overlapping PAH features have such extended wings that there are 
hardly any truly line-free continuum regions available.  Comparison between 
PAH strengths measured using PAHFIT and spline fit shows large scatter and 
systematic discrepancies.  In general, the spline fit method underestimates 
the PAH line fluxes by factors of $\simali$2--6, depending on the feature (Smith 
\etal 2007).

In their spectral energy distribution (SED)  fitting software
``Continuum And Feature Extraction'' (CAFE),
Marshall \etal (2007) constructed a semiempirical PAH template by applying
PAHFIT to an average spectrum of starburst galaxies (Brandl \etal 2006).  
After subtracting the continuum, they extracted individual profiles 
for a large number of PAH features between $\simali$3 and 19$\mum$.   CAFE 
fits the SED in two steps.  In the first step, a model is fit to the continuum 
after the PAH template is subtracted from the overall spectrum.  Then, the
residual, continuum-subtracted PAH spectrum is fit by adjusting the strengths 
of individual PAH features.

This paper presents a new technique to measure the PAH features 
in the MIR spectra of galaxies by making use of 
a theoretically calculated PAH template 
in combination with three dust components,
allowing all the parameters 
to vary simultaneously and self-consistently. 
We illustrate the effectiveness of our method by applying it first to one of the simplest 
environments in which PAH emission is found, the Galactic high-latitude clouds 
(HLCs; Blitz et al. 1984; Magnani et al. 1985).  We then extend it to a range 
of extragalactic systems of increasing levels of complexity, from nearby 
normal star-forming galaxies, to AGNs with a range of activity spanning 
low-ionization nuclear emission-line regions (LINERs; Heckman 1980) to 
Seyferts and quasars, and finally to highly obscured ultraluminous IR 
galaxies (ULIRGs).  Our technique can measure the absolute strength 
of the PAH emission features and derive upper limits on them 
when undetected.  At the same time, we can describe 
the underlying dust continuum with a minimum number of 
dust components (Y. Xie et al. 2018, in preparation) and simultaneously 
solve for the line-of-sight extinction.

We organize the paper as follows.  Section 2 introduces the MIR spectra 
used in the paper.  Section 3 explains our new method and describes the 
different components that make up our model.  The application of the new 
method to objects of different levels of complexity is presented in \S 4.  
We discuss the implications of our results in \S 5 and summarize in \S 6.

\section{MIR Emission Features and {\it Spitzer}/IRS Spectroscopic Data 
\label{sec:gas_line}}

\begin{figure} 
\begin{center}
\resizebox{\hsize}{!}
{\includegraphics{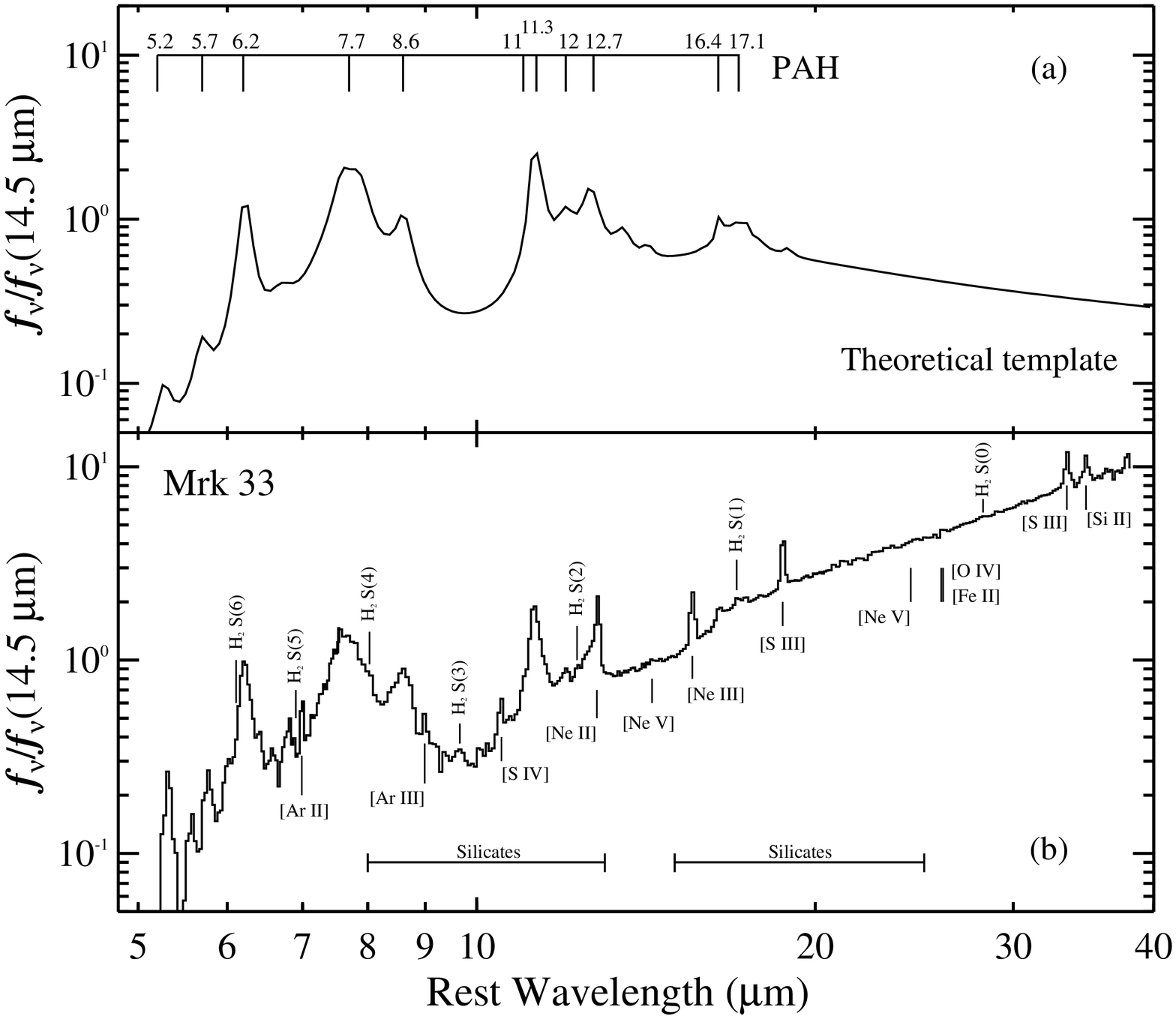}}
\caption{\footnotesize
({\it a}) Sample theoretical PAH spectrum from Draine \& Li (2007), calculated 
with starlight intensity scale factor $U\,=\,1$, illustrating the location of 
the prominent PAH emission features in the MIR.  ({\it b}) Sample IRS spectrum 
of the star-forming galaxy Mrk\,33, highlighting the locations of the most 
prominent ionic and molecular hydrogen emission lines and broad silicate 
absorption features at $\simali$10 and 18$\mum$.
\label{fig:gas_line_cont} }
\end{center}
\end{figure}

The MIR spectrum of galaxies contains rich information on gas and dust.  A 
wealth of ionic emission lines of different ionization potentials and levels 
(e.g., [S~{\footnotesize IV}] 10.51$\mum$, [Ne~{\footnotesize II}] 12.81$\mum$, 
[Ne~{\footnotesize III}] 15.56$\mum$, [Ne~{\footnotesize V}] 14.32, 24.32$\mum$, [S~{\footnotesize III}] 18.71, 33.48$\mum$, [O~{\footnotesize IV}] 25.8$\mum$, 
[Si~{\footnotesize II}] 34.82$\mum$) provides powerful tools to diagnose the excitation, 
density, and metallicity of the ionized gas (\eg Genzel \etal 1998; Lutz 
\etal 1998; Dale \etal 2006; Hao \etal 2009; Veilleux \etal 2009; Sargsyan 
\etal 2011).  The MIR spectral range also covers a suite of pure rotational 
lines of molecular hydrogen: $\rm H_{2}$ $S$(6) 6.11$\mum$, $\rm H_{2}$ 
$S$(5) 6.91$\mum$, $\rm H_{2}$ $S$(4) 8.03$\mum$, $\rm H_{2}$ $S$(3) 9.67 
$\mum$, $\rm H_{2}$ $S$(2)12.27$\mum$, $\rm H_{2}$ $S$(1) 17.03$\mum$ , and 
$\rm H_{2}$ $S$(0) 28.22$\mum$, among which $\rm H_{2}$ $S$(5) 6.91$\mum$ 
is blended with [Ar~{\footnotesize II}] 6.99$\mum$. Dust emission features are key 
components in the MIR.  Apart from PAH emission at 6.2, 7.7, 8.6, 11.3, 12.7, 
16.4, and 17.1$\mum$, silicate dust exhibits two broad bumps/troughs at 
$\simali$10 and $\simali$18$\mum$. Here [Ne~{\footnotesize II}] 12.8$\mum$ is blended 
with PAH 12.7$\mum$, and $\rm H_{2}$ $S$(1) 17.03$\mum$ 
is blended with PAH features at 16.4 and 17.1$\mum$.  
To complicate matters further, the line emission sits 
on top of continuum emission from dust spanning a range of temperatures and 
grain sizes. In particular, the MIR region contains continuum emission from 
very small grains that are heated via single-photon absorption (Y. Xie et 
al. 2018, in preparation).  The main dust and gas features are illustrated in 
Figure~\ref{fig:gas_line_cont}. 
 
All of the MIR spectra data used in the current paper were obtained with the 
IRS instrument onboard {\it Spitzer}. 
The IRS has both a low-resolution and a high-resolution mode. The 
low-resolution data cover the 5.2--38 $\um$ wavelength range with a resolving 
power of 
$R \equiv \lambda/\Delta \lambda = 57-128$; the high-resolution mode covers 
9.9--37.2$\mum$ with $R \approx 600$.  The low-resolution spectra are split 
into the short low (SL; $\lambda < 14\mum$) and the long low (LL; $\lambda 
\simgt14.5\mum$) modes.  In particular, the main PAH features populate the SL 
spectra, which, unfortunately, have a lower signal-to-noise ratio because the 
slit size of the SL mode is $\simali$3 times smaller than that of the LL mode.

For the data utilized in this paper, we give priority to the spectra archived 
in the Cornell AtlaS of {\it Spitzer}/IRS Sources (CASSIS), which were 
extracted according to the extent of the sources. CASSIS includes 
$\simali$13,000 low-resolution spectra of $>$\,11,000 distinct sources observed in 
standard staring mode and provides publishable-quality spectra  (Lebouteiller 
\etal 2011).  Spectra for galaxies not included in CASSIS were collected from 
the literature.  For the Galactic HLCs, Ingalls \etal (2011) published 
spectra of 15 clouds that have strong molecular hydrogen-line emission in 
the MIR.   Our analysis includes the four HLCs that have complete IRS 
low-resolution spectra from 5 to 38$\mum$.  The reduced spectra for nearby 
star-forming galaxies and low-luminosity AGNs are taken from the NASA/IPAC 
InfraRed Science Archive 
(IRSA).\footnote{\tt http://irsa.ipac.caltech.edu/data/SPITZER/SINGS/}

\section{ Methodology 
             \label{sec:model}}

We develop a new technique to decompose MIR IRS spectra into their dominant 
emission components from PAH features and the dust continuum, simultaneously 
accounting for dust extinction.  Our technique greatly reduces the number of 
free parameters (to eight in total) and is easy to apply to astronomical 
sources with various levels of complexity.  We adopt a linear combination of a 
theoretical template for PAH emission, three modified blackbodies for the 
continuum, plus an extinction component.  After shifting the spectra to the 
rest frame, we mask strong, unblended ionic and molecular hydrogen emission 
lines, assuming that the lines are unresolved at the minimum instrumental 
resolution of FWHM = 0.34$\mum$.   We do not mask or attempt to fit [Ne~{\footnotesize II}] 
12.8$\mum$ and $\rm H_{2}$ $S$(1) 17.03$\mum$, both of which are blended 
with PAH features, because their intensity is negligible compared to that of 
PAH emission and will not affect the fit significantly.

In light of the complexity of the spectral region in question, we perform the 
analysis in a stepwise fashion, starting with fits to the relatively simple 
Galactic HLCs and then proceeding to extragalactic environments of 
ever-increasing complexity, from star-forming galaxies, to low-luminosity 
active galaxies, to more powerful quasars, and finally to highly obscured 
AGNs and nuclear starbursts in ULIRGs. The 
luminous sources encompass a wide range of PAH strengths, from objects with
unambiguously strong features to those with barely any signs of PAH emission,
for which we are able to place meaningful upper limits.

\subsection{The Model}

\begin{figure} 
\begin{center}
\resizebox{0.8\hsize}{!}
{\includegraphics{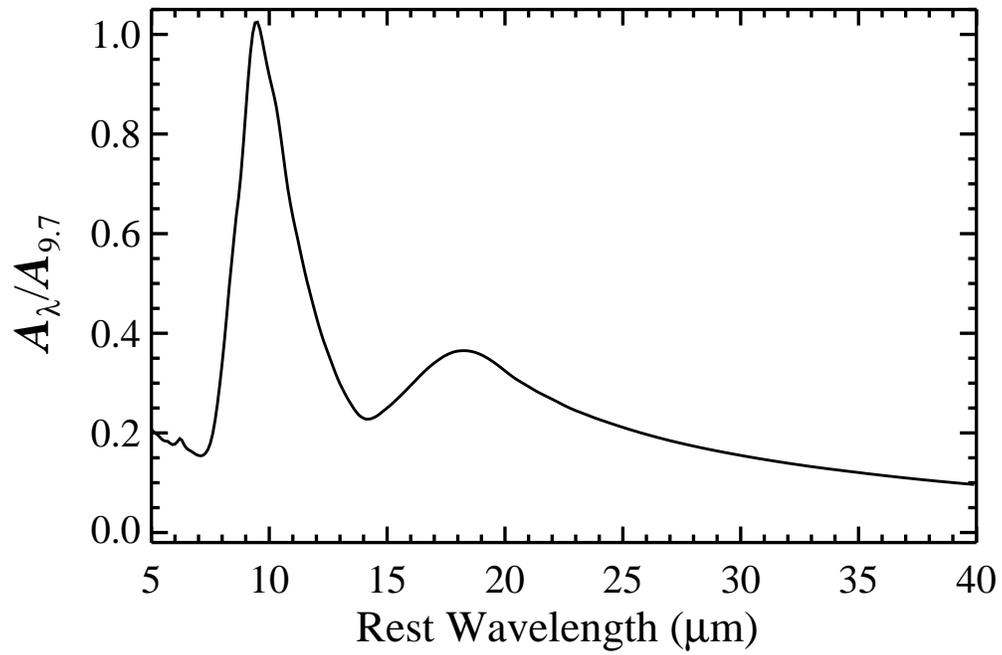}}
\caption{\footnotesize
The MIR extinction curve normalized at 9.7$\mum$ adopted in the current work, from Wang \etal 
(2015). 
\label{fig:ext} }
\end{center}
\end{figure}

The spectral model used to fit the MIR spectra contains five components: three 
modified blackbodies of different temperatures (hot, warm, and cold) to 
represent the continuum emission, a theoretical PAH template calculated from 
Draine \& Li (2007), and an extinction term that affects the continuum and 
line emission.  The model, expressed as

\begin{equation}
\label{eq:num_1}
F_\nu ({\rm mod}) = 
%F_\nu =
\left [A^{\rm PAH} J^{\rm PAH}_{\nu} + 
    \it A^{h}B_{\nu}(T^{h})/{\lambda}^{\beta} + %
        A^{w}B_{\nu}(T^{w})/{\lambda}^{\beta} + %
        A^{c}B_{\nu}(T^{c})/{\lambda}^{\beta} \right]%
  \rm exp({-\it\tau_{\lambda}}), %_{0}*\tau(\lambda)]
\end{equation}
contains eight free parameters: the scale factor $A^{\rm PAH}$ of the 
PAH template $J^{\rm PAH}_{\nu}$; the scale factors $A^{h}$, $A^{w}$, and 
$A^{c}$ for the hot, warm, and cold dust components represented as modified 
blackbodies $B_{\nu}(T)$ of temperature $T^{h}$, $T^{w}$, and $T^{c}$; the 
dust emissivity index $\beta$, assumed to be the same for all three modified 
blackbodies and fixed\footnote{Assuming that all dust particles are 
spherical and are composed of electrons and ions, they can be treated as a 
dipole of a classical Lorentz harmonic oscillator, oscillating under the 
force of an electromagnetic field. For submicron-sized grains and very small grains 
that satisfy the Rayleigh limit in the IR ($2\,\pi\,a/\lambda \ll 
1$), the absorption by dust of incident radiation can be approximated as 
$\simali$$\lambda^{-2}$ (i.e., $\beta=2$), in accordance with the solution 
for the motion of a harmonic oscillator, 
where $\lambda$ is the wavelength of the incident 
electromagnetic wave. In practical situations, 
$\beta$ may vary from $\simali$1.5 to $\simali$2,
which indicates that the vibration of the dust grain deviates from that of 
ideal harmonic oscillation (see Draine 2003 and Li 2009 for more details). 
The determination of the exact value of $\beta$ requires full information on 
the IR SED. In this paper, we fix $\beta$ to 2.} 
to $\beta = 2$;
$\tau_{\lambda}$, is the total optical depth 
applied to all components, 
with $\tau_\lambda=\tau_{9.7}\times\left(A_\lambda/A_{9.7}\right)$, 
where $A_\lambda/A_{9.7}$ is the interstellar extinction curve of 
Wang \etal (2015) normalized at 9.7$\mum$ (Figure~\ref{fig:ext})
and $\tau_{9.7}$ is the optical depth at 9.7$\mum$,
the only adjustable parameter for the extinction term.

In principle, the observed MIR continuum arises from a mixture of dust 
components spanning a wide range of temperatures.  However, as we have very 
limited constraints on the exact dust temperature distribution, we simply 
adopt a minimum number (three) of modified blackbodies to represent the dust 
continuum emission. The temperatures derived from the current model should 
only be regarded as representative of the average values of certain dust 
components.

\subsection{Theoretical PAH Template \label{sec:pah}}

The PAH model attributes the observed 3.3, 6.2, 7.7, 8.6, 11.3, 
and 12.7$\mum$ features to the vibrational modes of PAHs, 
with the 3.3$\mum$ feature 
assigned to C--H stretching modes, the 6.2 and 7.7$\mum$ features to
C--C stretching modes, the 8.6$\mum$ feature to C--H in-plane bending modes,
and the 11.3 and 12.7$\mum$ features to C--H out-of-plane bending modes 
(L\'eger \& Puget 1984; Allamandola et al.\ 1985).  The relative strengths of 
these bands depend on the charge, size, and molecular structure of the PAH 
molecule (see Li 2004).  While the emission from ionized PAHs dominates at 6.2, 7.7, and 8.6 
$\mum$, neutral PAHs emit strongly at 3.3 and 11.3$\mum$ (see Hudgins \& 
Allamandola 2004 and references therein).  On the other hand, with a larger 
heat capacity-which is proportional to $\NC$, the number of C atoms-a larger 
PAH molecule will be heated to a lower peak temperature (i.e., $T_{\rm peak}
\propto \NC^{-1/4}$) upon absorption of a single stellar photon.  Therefore, 
the bulk emission of larger PAHs occurs at longer wavelengths than that of smaller 
PAHs (Li \& Mann 2012).  Finally, compared to catacondensed PAHs, which have a 
more open structure, compact, pericondensed PAHs (e.g., coronene, ovalene, and 
circumcoronene) have a lower hydrogen-to-carbon ratio (H/C).  Thus, for PAHs 
of the same size (i.e., $\NC$), pericondensed PAHs are expected to have a 
lower 11.3-7.7$\mum$ band ratio (e.g., Hony et al.\ 2001;
Vermeij et al.\ 2002; Shannon et al.\ 2016). 

The PAH model readily explains the general PAH band patterns observed in 
various regions in terms of a mixture of neutral and charged PAHs of different 
sizes (e.g., see Allamandola et al.\ 1999; Li \& Draine 2001; Peeters et al.\ 
2004; Draine \& Li 2001; Bauschlicher et al.\ 2010; Cami et al.\ 2011; 
Rosenberg et al.\ 2011; Boersma et al.\ 2013, 2014; Andrews et al.\ 2015).
Draine \& Li (2007) calculated the IR emission spectra of PAHs\footnote{The 
PAHs are assumed to be compact and pericondensed, as catacondensed PAHs are less 
stable in hostile interstellar environments (see Eq.\,4 of Li \& Draine 2001).}
for the Galactic diffuse interstellar medium (ISM), 
with the PAH charging\footnote{The PAH charging is measured by the PAH ionization fraction
$\phi_{\rm ion}$, the probability of finding a PAH molecule in a nonzero 
charge state.  It depends on the quantity $U\sqrt{T_{\rm gas}}/n_e$ (Bakes \& 
Tielens 1994; Weingartner \& Draine 2001b), where $U$ is the starlight density 
in units of the solar neighborhood interstellar radiation field (ISRF; Mathis 
et al.\ 1983, hereafter MMP83), $T_{\rm gas}$ is the gas temperature, and 
$n_e$ is the electron number density.} determined from the balance between 
photoionization and electronic recombination (Bakes \& Tielens 1994; 
Weingartner \& Draine 2001b), and with the PAH size distribution constrained 
by the broadband photometric data of {\it COBE}/DIRBE at 3.5, 4.9, 12 and 25 
$\mum$, together with the spectroscopic data at 2.8--12$\mum$ obtained by the 
{\it Infrared Telescope in Space} (Onaka et al.\ 1996; Tanaka et al.\ 1996)
and 5--40$\mum$ data from {\it Spitzer}/IRS 
(Smith \etal 2007).\footnote{The Galactic UV extinction curve 
cannot constrain the size distribution of PAHs because, 
even in the far-UV, PAHs are in the Rayleigh regime, 
and the extinction cross sections of PAHs on a per unit volume basis 
are size-independent; therefore, the UV extinction curve only constrains
the quantity of PAHs, not the detailed size distribution (Wang et al. 2015).}

Li \& Draine (2001) unified PAHs and graphite into a single population of dust 
grains: ``carbonaceous'' grains.  The so-called carbonaceous grain population 
extends from grains with graphitic properties at radii $a\gtsim50\Angstrom$ 
down to particles with PAH-like properties at very small sizes.\footnote{
Following Li \& Draine (2001), the ``radius'' $a$ of a PAH containing $\NC$ C 
atoms is defined to be the radius of a sphere with the carbon density of 
graphite containing the same number of C atoms (i.e., 
$a\approx1.286 N_{\rm C}^{1/3}\Angstrom$).} Using the size-distribution 
functional form of Weingartner \& Draine (2001a), Draine \& Li (2007) 
described the size distribution of the carbonaceous grain population with two 
lognormal components and a component similar to an exponentially cut off power
law.  The lognormal components are for PAHs and are characterized by the peak 
size $a_{0,j}$ (where $j=1,2$), the width $\sigma_{j}$, and 
${\rm \left(C/H\right)}_{j}$, the amount of C (relative to H) locked up in 
each component.  Draine \& Li (2007) derived the best-fit parameters to be 
$a_{0,1}\approx4\Angstrom$, $\sigma_{1}\approx0.4$, and 
${\rm \left(C/H\right)}_{1}\approx45\ppm$ (where ppm refers to parts per 
million) for the smaller component and $a_{0,2}\approx20\Angstrom$, 
$\sigma_{2}\approx0.55$, and ${\rm \left(C/H\right)}_{2}\approx15\ppm$ for the 
larger component.  For these lognormal components, the mass distributions 
peak at $a_{0,j}\exp\left(3\sigma_{j}^2\right) \approx 6.5\Angstrom$ for $j=1$
and $\simali$50$\Angstrom$ for $j=2$ (see Figure~11 of Draine \& Li 2007).

\begin{figure} 
\begin{center}
\resizebox{0.8\hsize}{!}
{\includegraphics{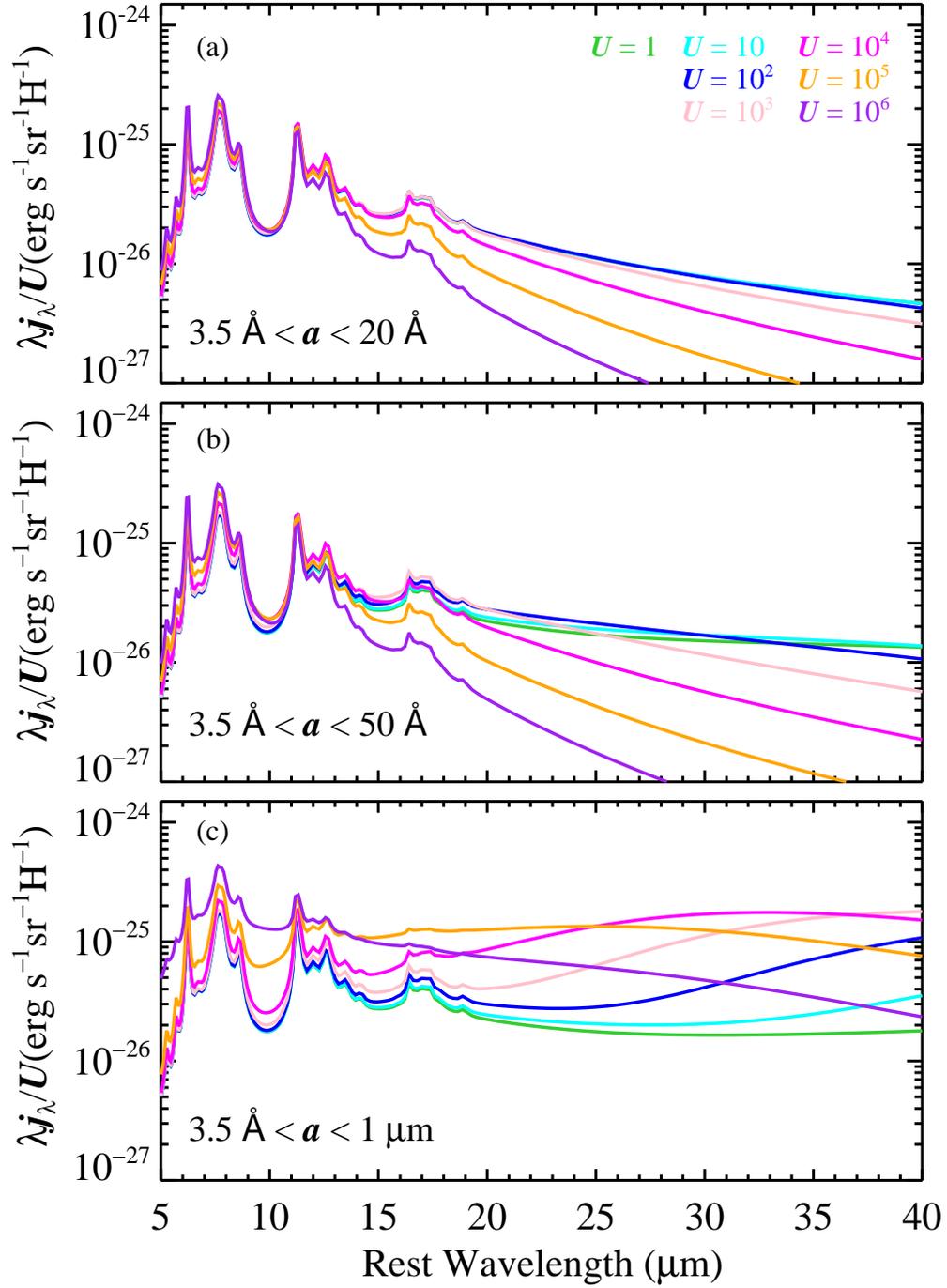}}
\caption{\footnotesize
Theoretical emission spectra of carbonaceous dust for selected 
starlight intensity scale factors $U$ ranging from 1 to $10^6$, for 
({\it a}) grain sizes $< 20\,\Angstrom$, 
({\it b}) grain sizes $< 50\,\Angstrom$, and
({\it c}) the full  grain size distribution. 
\label{fig:template} }
\end{center}
\end{figure}

In Figure~\ref{fig:template}, we show the IR spectra in the 5--40$\mum$ 
wavelength range emitted by the carbonaceous grain population for which the 
size distribution was constrained by Draine \& Li (2007) using the functional 
form of Weingartner \& Draine (2001a).  The grains are illuminated by an 
MMP83-type ISRF but with the starlight intensity enhanced by a factor of 
$U=1, 10^2, ..., 10^6$.  To explore the role of PAH size in contributing to 
the major PAH bands at 6.2, 7.7, 8.6, 11.3, and 12.7$\mum$, we consider three 
size ranges: (i) $3.5\Angstrom < a < 20\Angstrom$, (ii) $3.5\Angstrom < a < 
50\Angstrom$, and (ii) $3.5\Angstrom < a < 1\mum$.\footnote{The lower size 
cutoff is set at $a=3.5\Angstrom$ (corresponding to $\NC = 20$), since PAHs 
smaller than $\NC=20$ are not stable in the Galactic ISM (see Li \& Draine 
2001 and references therein).}  While the first two size ranges are mostly for 
PAHs, the third size range contains both PAHs and large graphitic grains. 

Figure~\ref{fig:template} demonstrates that the major PAH bands at 6.2, 7.7, 
8.6, 11.3, and 12.7$\mum$ are predominantly emitted by PAHs of 
$a<20\Angstrom$.  For $1\simlt U\simlt10^6$, the emission spectra of PAHs of 
$a<50\Angstrom$ are not appreciably different from those of $a<20\Angstrom$.
Also, for $1\simlt U\simlt10^4$, the emission spectra of the entire 
carbonaceous grain population of $3.5\Angstrom< a<1\mum$ are not appreciably 
different from those of PAHs of $a<20\Angstrom$.  Only with $U\simgt10^5$, the 
emission spectra of the entire carbonaceous grain population of $3.5\Angstrom< 
a<1\mum$ show considerable deviations from those of PAHs of $a<20\Angstrom$.
This is because, exposed to an ISRF of $U\simgt10^5$, grains of several tens 
of nanometers attain an equilibrium temperature high enough to emit in the MIR 
and raise the continuum at $\lambda>5\mum$.  

Figure~\ref{fig:template} also shows that the emission spectra of PAHs at 
$\lambda<15\mum$ of both $a<20\Angstrom$ and $a<50\Angstrom$, after being 
scaled by $U$, are essentially independent of $U$.  This is expected from the 
single-photon heating nature of PAHs.\footnote{Single-photon heating implies 
that the shape of the high-$T$ end of the temperature ($T$) probability 
distribution function for a PAH molecule is the same for different levels of 
starlight intensity, and, therefore, the emission spectra (after being scaled by $U$)
remain the same (see Li \& Mann 2012).  What really matters is the mean photon 
energy, which determines to what peak temperature a PAH molecule will reach 
upon absorption of such a photon.} This is also true for the entire 
carbonaceous grain population of $3.5\Angstrom< a<1\mum$, provided 
$U\simlt10^4$.  For $U\simgt10^5$, the PAH emission spectra at $\lambda<15\mum$
differ appreciably from those of $U\simlt10^4$ due to the continuum emission
from grains of several tens of nanometers, which are heated sufficiently to 
emit at $\lambda>5\mum$.

Taking all these together, in the following for the PAH model template, we will 
adopt the PAH emission spectra calculated from PAHs of $a<20\Angstrom$ 
illuminated by $U=1$.  As discussed earlier, the PAH emission spectra at 
$\lambda<15\mum$ (after being scaled by $U$) are essentially independent of $U$ and 
are predominantly emitted by PAHs of $a<20\Angstrom$, the only exception being 
that, in regions with an extremely intense ISRF, the model spectra 
could have an unusually strong underlying continuum 
emitted by grains of several tens of nanometers.   
%
%\textcolor{red}{
As will be demonstrated in \S\ref{sec:application}, the selection of this 
particular PAH template is justified by the fact 
that it works very well for a wide range of environments.%}

We should note that the Draine \& Li (2007) PAH model 
is actually for ``astronomical'' PAHs 
in the sense that, in modeling 
the observed PAH emission spectra, 
an empirical approach was taken 
to construct ``astro-PAH'' absorption properties 
that are consistent with both laboratory measurements 
and quantum-chemical computations
(e.g., see Figure~2 of Draine \& Li 2007) 
and spectroscopic observations of PAH emission 
in various astrophysical environments (Smith et al.\ 2007). 
Astro-PAHs do not represent any specific PAH molecules  
but approximate the actual absorption properties 
of the PAH mixture in astrophysical regions.
Furthermore, neither laboratory-measured nor 
quantum-chemically computed spectra of individual PAH 
molecules accurately resemble the band positions and 
widths of observed PAH emission spectra.  For example,
the PAH bands of laboratory-measured spectra,
which are commonly detected in absorption, 
are too narrow to be compatible with 
astronomical spectra,
which are commonly seen in emission.
On the other hand, the quantum-chemically computed 
spectra suffer from the fact that the band positions and strength are 
often not accurate and need to be scaled
(Yang et al.\ 2017a, 2017b).
Finally, the laboratory measurements are so far 
limited to PAHs smaller than $\simali$50 C atoms
(Hudgins \& Allamandola 2005),
while interstellar PAHs can be much larger
(see Figure~7 of Draine \& Li 2007). 
Li \& Draine (2001) derived a mean PAH size 
of $N_{\rm C} \approx 100$
in the diffuse ISM. 
Hence, it is desirable to synthesize 
a set of vibrational bands for ``astro-PAHs''
by adopting empirical PAH band positions, widths,
and cross sections to
mimic astronomical observations
(e.g., Smith et al.\ 2007).%}

\subsection{Algorithm and Error Estimation}

We obtain the best fit for each galaxy using the IDL code {\tt MPFIT}, a 
$\chi^{2}$-minimization routine based on the Levenberg-Marquardt algorithm 
(Markwardt 2009). The goodness-of-fit is measured by the reduced $\chi^2$,

\begin{equation}
\label{eq:num_2}
\chi^2_{\rm global}/\dof = \sum_{j=1}^{\Nobs}
\left\{\frac{F_\nu(\mod)-F_\nu(\obs)}
{\sigma_\nu(\obs)}\right\}^2/\left\{\Nobs-\Nmod\right\} ~~,
\end{equation}

\noindent
where $F_\nu(\mod)$ is the calculated model flux density, $F_\nu(\obs)$ is 
the observed flux density, $\sigma_\nu(\obs)$ is the observed  $1\,\sigma$ 
uncertainty at each flux density point, $\Nobs$ is the number of data points, 
and $\Nmod$ is the total number of free parameters in the model.\footnote{ 
The signal-to-noise ratio of an IRS spectrum in the $\simali$5--14.5$\mum$ 
interval is lower than that in the $\simali$14.5--38$\mum$ interval. This is 
caused by the different observation modules: the 5--14.5$\mum$ SL module has a 
slit width of 3\farcs6, and the 14.5--38$\mum$ LL module has a slit width of 
11\farcs2.  To better fit the 5--8$\mum$ continuum emission, PAHs, and the 
9.7$\mum$ silicate feature, we arbitrarily increase the weights for the data 
points at 5--8$\mum$ by a factor of 4 and those at 8--14.5$\mum$ by a factor of 
3 relative to those at 14.5--38$\mum$.}

\begin{figure} 
\begin{center}
\resizebox{\hsize}{!}
{\includegraphics{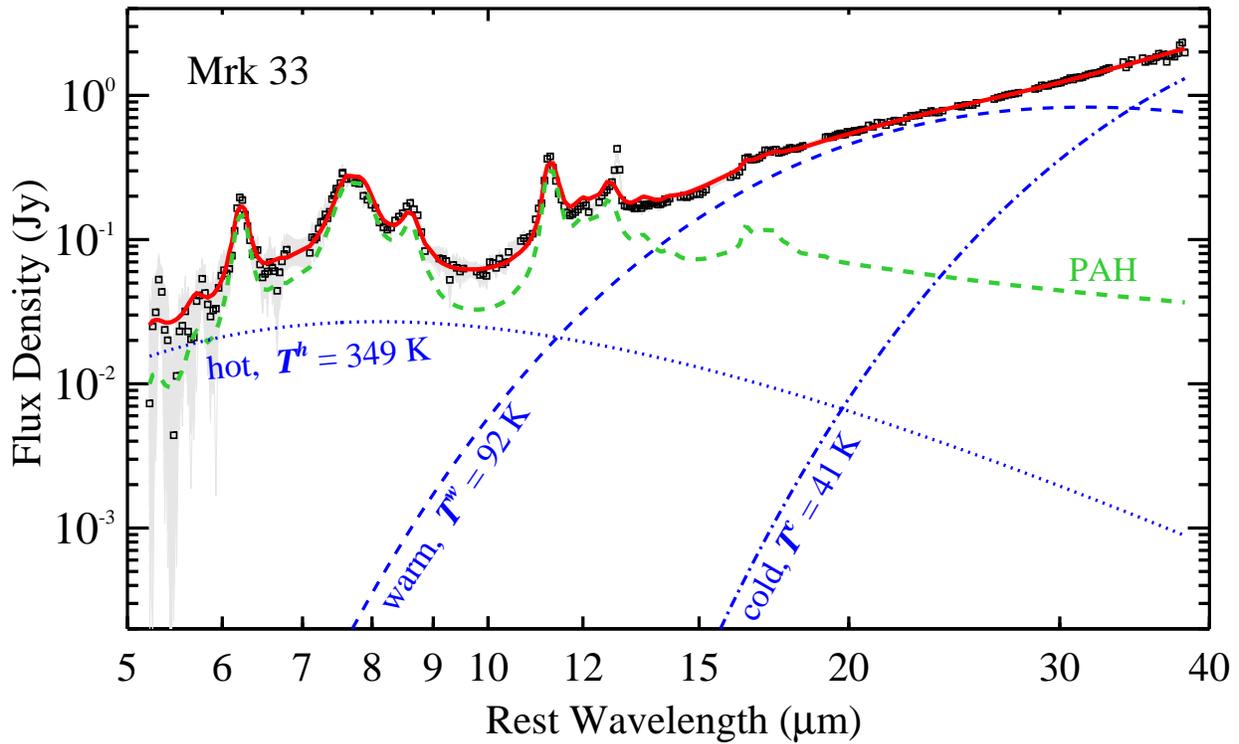}}
\caption{\footnotesize
Illustration of the method to determine the errors of the best-fit parameters 
for Mrk\,33.  The observed IRS spectrum is plotted as black squares.  The 
best-fit model (solid red line) comprises the theoretical PAH template 
(green dashed line) and three modified blackbody components (blue lines) 
for the hot (dotted), warm (dashed), and cold (dot-dashed) dust continuum.  
The gray shaded region denotes simulated spectra generated from 500 
Monte Carlo realizations. 
\label{fig:show_mc} }
\end{center}
\end{figure}

We estimate the uncertainties for the eight model parameters by performing 
Monte Carlo simulations.  For the IRS spectrum of each source, we assume that 
each flux density point statistically follows a normal distribution.  The 
dispersion is characterized by the observed 1 $\sigma$ error, which comprises 
statistical and systematic errors. The latter arises from the flux differences 
between the two nods of the IRS spectra, sky background contamination, and IRS 
pointing and flux calibration errors (Lebouteiller \etal 2011).  A new random 
spectrum is generated, and then the same model is fit to derive the best-fit 
parameter set.  We perform 500 simulations for each galaxy, and the final 
fitted spectrum is calculated with the median value of each parameter from 
their simulated distributions. The uncertainty is derived from the standard 
deviation of the distribution. Figure~\ref{fig:show_mc} illustrates our 
technique for Mrk\,33.

\section{Applications \label{sec:application}}

\subsection{Galactic HLCs \label{sec:sample_A}}
                   
To test our new technique, we first apply it to four Galactic HLCs, whose MIR 
dust spectrum is relatively pure and simple.  HLCs, initially reported as FIR
``cirrus'' (Low \etal 1984), were discovered by the {\it Infrared Astronomical
Satellite (IRAS)}\ at Galactic latitudes $|b| \simgt 25$ deg and 100\pc\ away 
from the Galactic center. They are mostly translucent clouds that are 
cospatial with atomic H\,{\footnotesize{I}} gas and molecular CO gas (Blitz et al. 1984; 
Magnani et al. 1985).  As HLCs generally have very small optical extinctions 
($A_{V}$ \lax\ 1 mag; A. Mishra \etal 2018, in preparation),
their MIR extinction, $\tau_{9.7\mum} \approx A_V/18$ (Roche \& Aitken 1984), 
is expected to be negligible.  Thus, contamination by silicate absorption 
features (at $\simali$10 and 18$\mum$) is minimal.  Moreover, as HLCs reside far 
from the Galactic plane and contain little embedded ongoing star formation 
activity, there will be minimal emission from silicate dust heated by internal 
ionizing photons.  Consequently, the overall 5--15$\mum$ emission of HLCs 
is expected to arise simply from PAH molecules and grains of different 
temperatures. The basic parameters of the HLCs analyzed in this paper are 
tabulated in Table~\ref{tab:basic_hgl}. 

\begin{figure*}[ht] 
\begin{center}
\resizebox{0.8\vsize}{!}
{\includegraphics{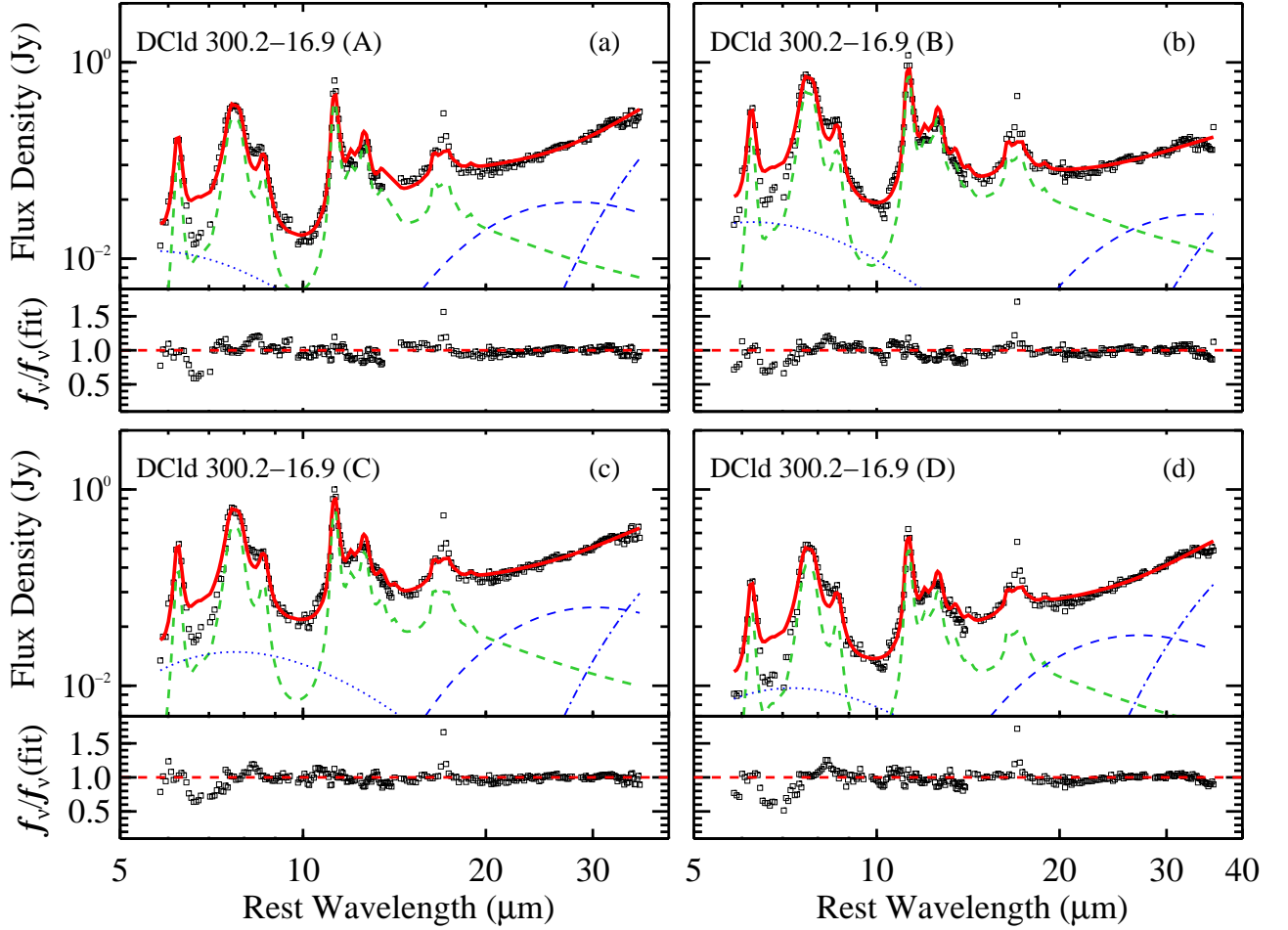}} 
\caption{\footnotesize 
Decomposition of the IRS spectra of Galactic HLCs ({\it a}) 
DCld\,300.2--16.9\,(A), ({\it b}) DCld\,300.2--16.9\,(B), ({\it c}) 
DCld\,300.2--16.9\,(C), and ({\it d}) DCld\,300.2--16.9\,(D).  For each 
source, the top panel shows  the observed spectrum (black squares) and the 
best-fit model (solid red line), comprising a theoretical PAH template (green 
dashed line) and three modified blackbody components (blue lines) for the hot 
(dotted), warm (dashed), and cold (dot-dashed) dust continuum.  The bottom 
panel shows the ratio of the observed spectrum with the best-fit model.  
\label{fig:hgl}
             }      
\end{center}
\end{figure*}

The MIR spectra of the four HLCs display prominent PAH emission at 6.2, 7.7, 
8.6, 11.3, and 12.7$\mum$ (Figure~\ref{fig:hgl}). Their complete IRS data 
also show strong molecular hydrogen lines of $\rm H_{2}$ $S$(2) 12.3, 
$\rm H_{2}$ $S$(1) 17.0, and $\rm H_{2}$ $S$(0) 28.2$\mum$. Since
$\rm H_{2}$ $S$(2) 12.3 and $\rm H_{2}$ $S$(1) 17.0$\mum$ are blended with PAH 
features and their relative strength is significantly lower compared to 
adjacent PAH features, we retain these two lines in the fit, but we omit the 
strongest line, $\rm H_{2}$ $S$(0) 28.2$\mum$.  It is apparent that the PAH 
template, in conjunction with three modified blackbodies, can fairly well 
describe the PAH emission and continuum emission of HLCs.  The flux density 
ratio between the data and the fit (bottom of each subpanel) also 
indicates that the global difference is less than 10\%.  The largest 
discrepancy occurs around 17$\mum$, which coincides with $\rm H_{2}$ $S$(1).
The model slightly overpredicts the emission between PAH 6.2 and 7.7$\mum$, 
and the wing of 8.6$\mum$ is underestimated.  

We note that the dust temperature (and standard deviation\footnote{The 
uncertainties on the dust temperatures, particularly for the cold component, 
are generally very small and likely underestimated because of the small errors 
of the flux densities at $\lambda\gtsim20\mum$. The true uncertainties are 
dominated by systematic effects such as the omission of FIR data and the 
choice of PAH template, which we discuss in Section 5.1.}) required to produce 
the continuum emission in the 
four HLCs is $460\pm59$, $102\pm10$, and $45\pm1$ K on average for the hot, 
warm, and cold components, respectively.  The temperature for the hot 
component is especially high for HLCs, which have a deficit of UV radiation.  
Additionally, it seems that the temperature variation from cloud to cloud is 
rather small for the hot and warm components, despite the rather different 
spatial distributions of the HLCs.  In the next section, we will show that 
this fact remains true in extragalactic H~{\footnotesize II} regions. 

The best-fit MIR extinction (Table~\ref{tab:extinction}) is negligible, 
consistent with the expectation from the low values of $A_V$ toward HLCs.  We 
place an upper limit on $\tau_{9.7}$ by iteratively adjusting its value while 
keeping the other model parameters fixed.  The goodness-of-fit is judged by 
$\mathrm {\chi^2(sil)}$ defined over 9--11$\mum$ to closely bracket the silicate 
absorption feature at 10$\mum$.  We estimate an upper limit on $\tau_{9.7}$ 
when $\rm \chi^2(sil)$ changes by 1\%.

\subsection{Star-forming Galaxies \label{sec:sample_B}}
 
\begin{figure*}[ht] 
\begin{center}
\resizebox{0.7\hsize}{!}
{\includegraphics{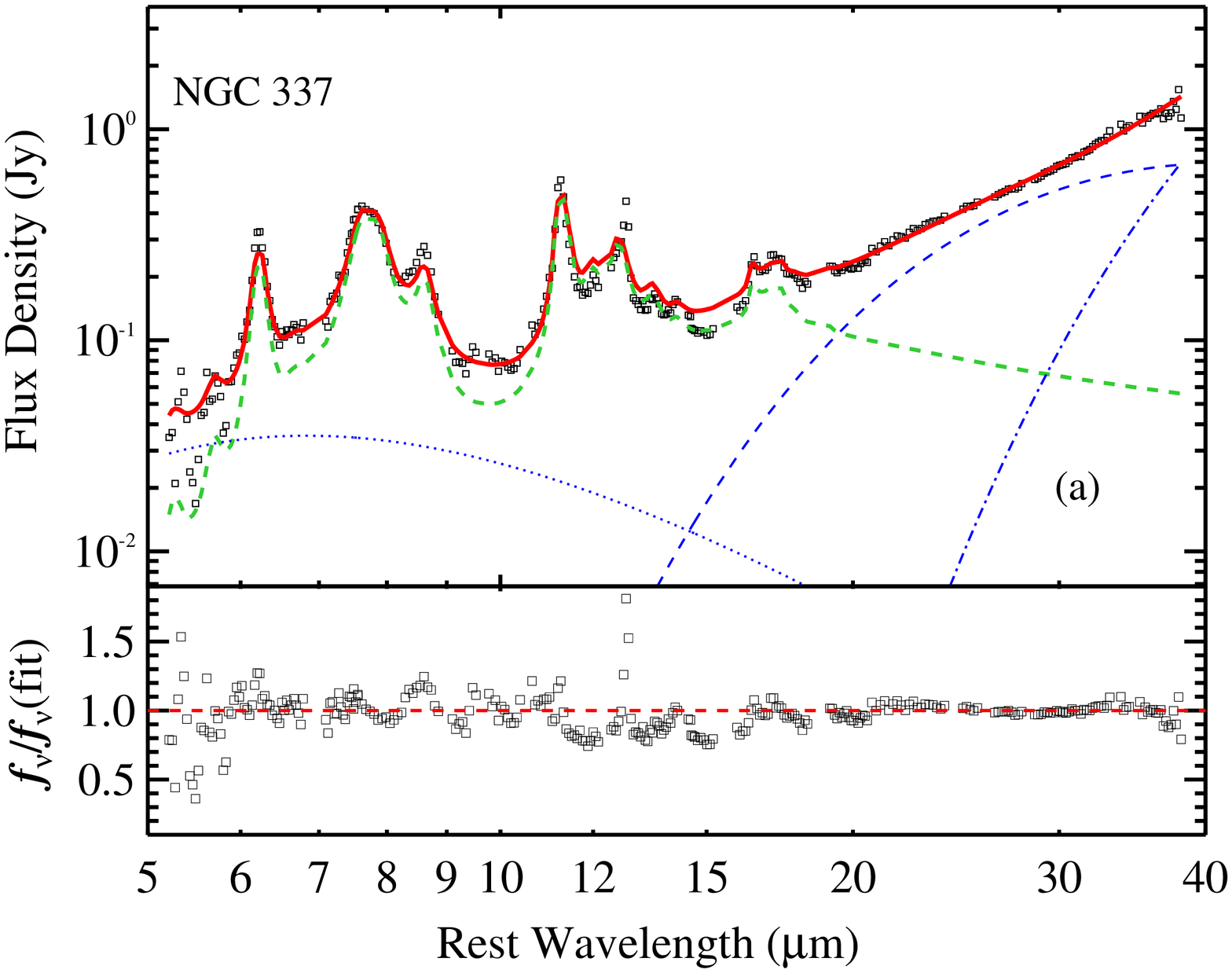}} \\ 
\resizebox{0.7\hsize}{!}
{\includegraphics{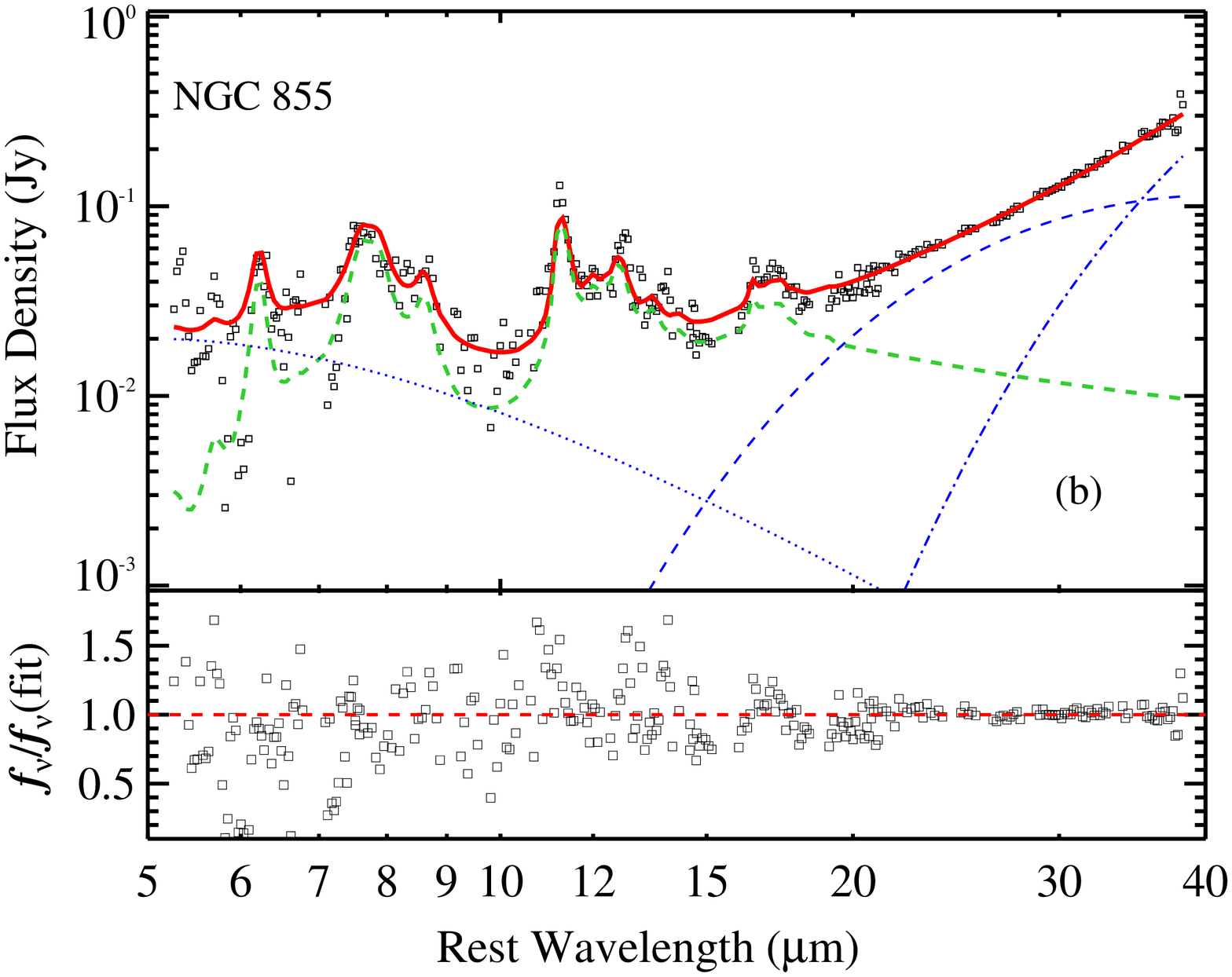}} \\
\caption{\footnotesize 
Decomposition of the IRS spectra of the SINGS H~{\scriptsize{II}} galaxies ({\it a}) NGC\,337 
and ({\it b}) NGC\,855.  For each source, the top panel shows the observed 
spectrum (black squares) and the best-fit model (solid red line), comprising a 
theoretical PAH template (green dashed line) and three modified blackbody 
components (blue lines) for the hot (dotted), warm (dashed), and cold 
(dot-dashed) dust continuum.  The bottom panel shows the ratio of the 
observed spectrum with the best-fit model.  The results for all SINGS H~{\scriptsize{II}}
galaxies analyzed in this study are available in the online version of the 
paper.
\label{fig:hii_irs_fit} }      
 \end{center}
 \end{figure*}

Proceeding to the next level of complexity, we apply our technique to nearby 
galaxies whose dominant or only known source of excitation is star formation.  
We utilize data from the Spitzer Infrared Nearby Galaxy Survey (SINGS; Kennicutt \etal 
2003; Smith \etal 2007), which comprises 75 nearby galaxies with distances 
$\simlt \rm 30\,Mpc$, including both star-forming (H~{\footnotesize II}) nuclei, LINERs, and 
Seyfert galaxies.
SINGS is designed to span a wide range of galaxy morphology, luminosity, star formation rate,
and extinction.  The sample has comprehensive multiwavelength data from 
X-rays to radio, allowing derivation of a variety of physical properties, 
including those of dust and gas.  Among the 75 SINGS galaxies,  we first 
choose the 24 classified as H~{\footnotesize II} nuclei (Table~\ref{tab:basic_sings}). The 
best-fit results for two cases are illustrated in Figure~\ref{fig:hii_irs_fit}; 
the fits for the rest are available in the electronic figure set. %\input{fig6set.tex} 

Our five-component model can successfully recover the PAH and continuum 
emission.   In some galaxies, there is a dip around 15$\mum$ that the model 
model fails to fit. Such a feature is prominent in NGC\,337, 628, 925, 2403, 
2976, 4254, 4625, and 7793. We note that this feature is absent from HLCs. 
The median temperature of the cold, warm, and hot dust 
components are $34\pm8$, $70\pm8$, and $386\pm78$ K, respectively.  The 
hot dust component has a temperature distribution of $\simali$300--400$\K$, similar 
to that of HLCs.  This strikingly high dust temperature for extragalactic H~{\footnotesize II} 
regions and Galactic HLCs cannot originate from dust in temperature 
equilibrium, as the IR SEDs of the SINGS 
star-forming galaxies and HLCs peak 
around 100$\mum$ (Dale \etal 2012; A. Mishra \etal 2018, in preparation).  Nor 
can the observed high temperatures be explained by an intense UV radiation 
field because HLCs have no embedded star formation and are located far from 
the Galactic disk ($70\pm15$ pc). For H~{\footnotesize II} regions, if such a high temperature 
is generated from hot dust heated by intense UV photons, we would expect to 
see silicate emission around 10 and 18$\mum$, which can be produced in an 
environment with high photon density and optically thin MIR conditions.  Yet 
the H~{\footnotesize II} nuclei show no emission or only mild silicate absorption. The fairly 
uniform high temperatures detected in H~{\footnotesize II} nuclei and HLCs indicate some 
common physical process at work.  Y. Xie \etal (2018, in preparation) propose that this hot 
component is produced by nanometer-sized dust grains transiently heated by 
single-photon absorption.  

Our best-fit results also yield estimates of the MIR extinction.  The majority 
(19/24) of the H~{\footnotesize II} nuclei have no detectable extinction, with upper limits of 
$\tau_{9.7}$ \lax\ 0.01.

\subsection{Low-luminosity AGNs \label{sec:sample_C}}
                   
\begin{figure*}[ht] 
\begin{center}
\resizebox{0.7\hsize}{!}
{\includegraphics{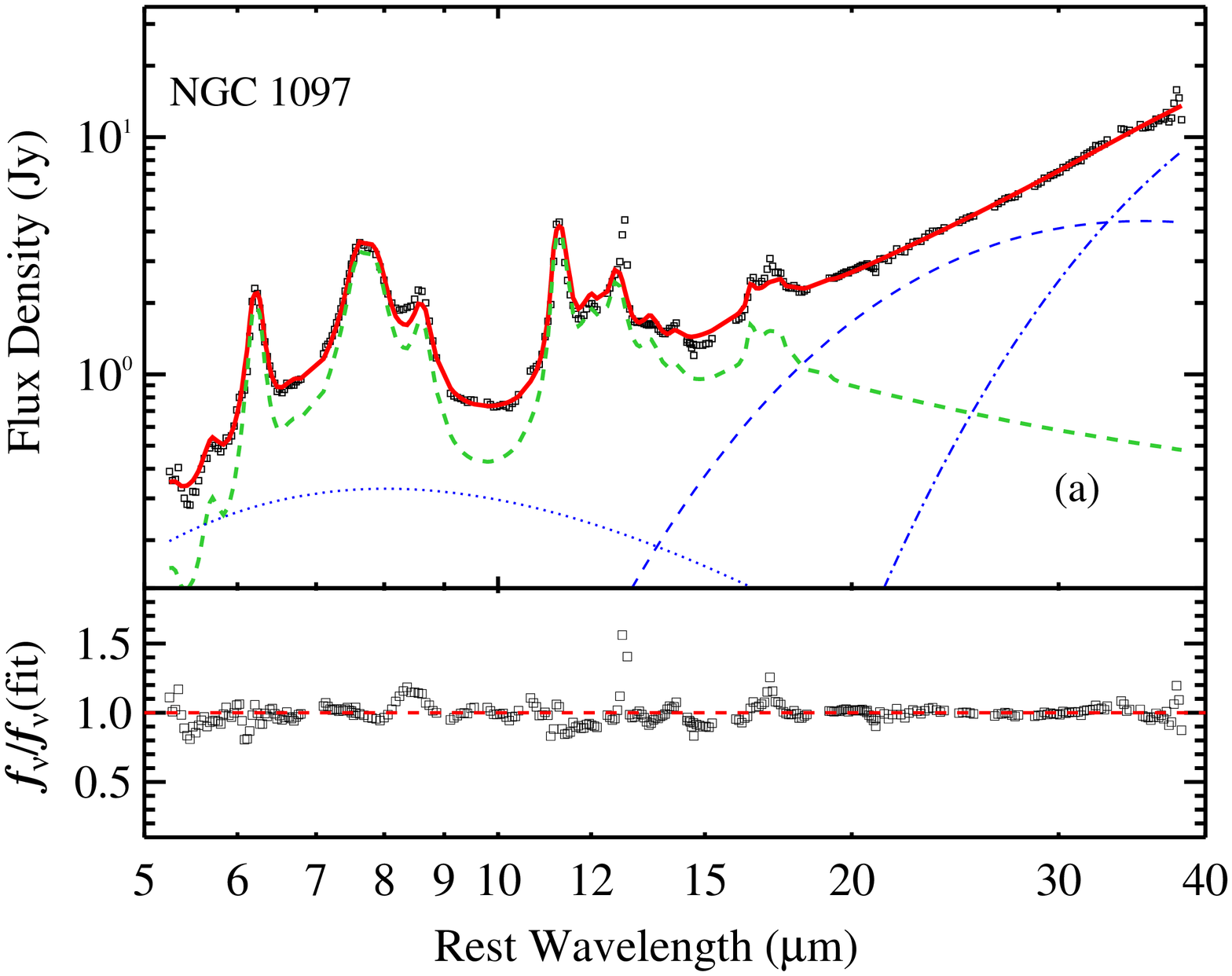}} \\ 
\resizebox{0.7\hsize}{!}
{\includegraphics{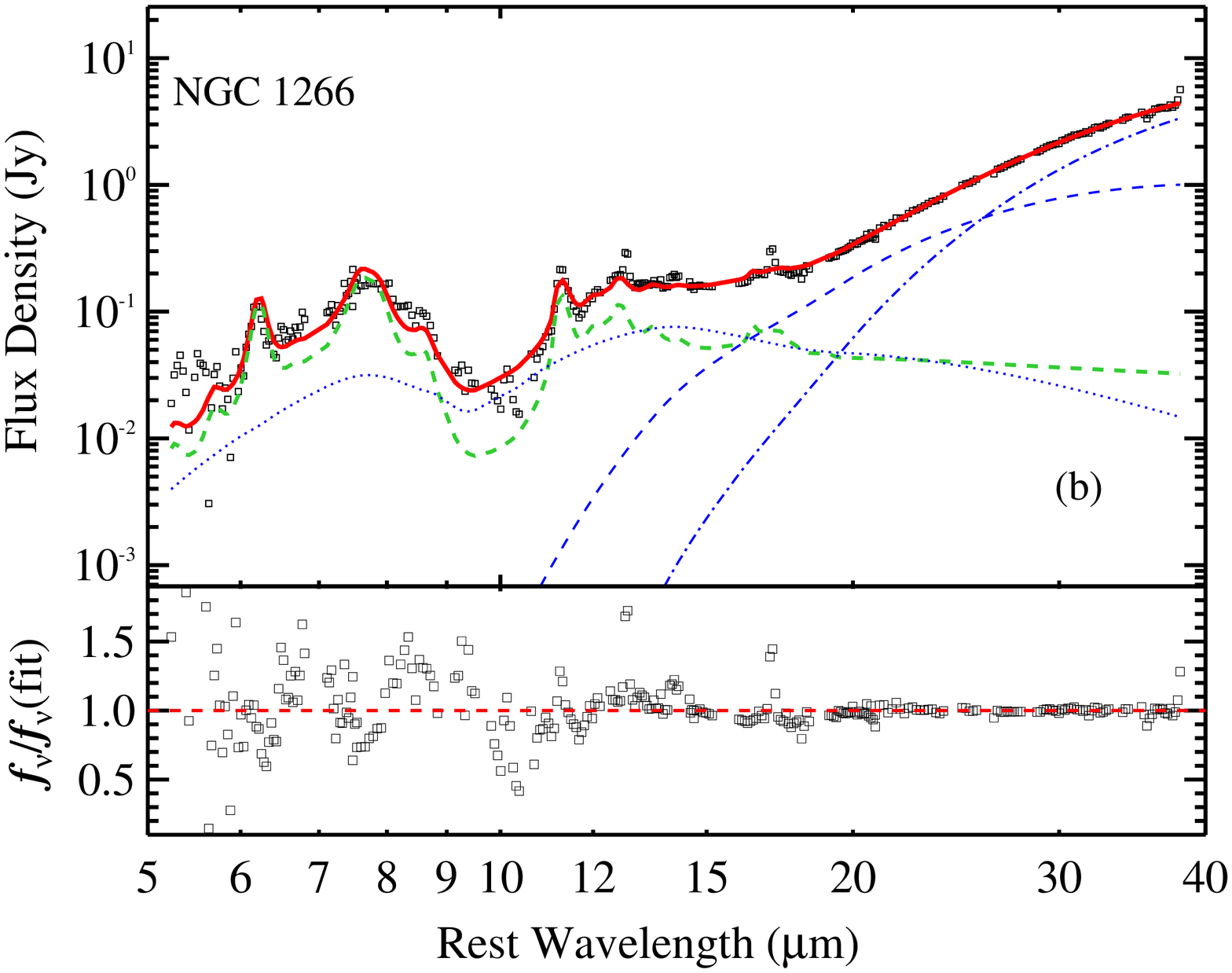}} \\
\caption{\footnotesize 
Decomposition of the IRS spectra of the SINGS AGNs ({\it a}) NGC\,1097
and ({\it b}) NGC\,1266.  For each source, the top panel shows the observed 
spectrum (black squares) and the best-fit model (solid red line), comprising a 
theoretical PAH template (green dashed line) and three modified blackbody 
components (blue lines) for the hot (dotted), warm (dashed), and cold 
(dot-dashed) dust continuum.  The bottom panel shows the ratio of the 
observed spectrum with the best-fit model.  The results for all SINGS AGN 
galaxies analyzed in this study are available in the online version of the 
paper.
\label{fig:agn_irs_fit} }      
\end{center}
\end{figure*}

We continue with a subset of 22 nearby, well-studied, low-luminosity AGNs from 
SINGS with high-quality IRS spectra (Table~\ref{tab:basic_sings}).   In these 
systems, the MIR light comes from a mixture of emission from the AGN torus and 
star-forming regions from the central kpc-scale region of the galaxy.  All the 
sources display prominent PAH features.  The best-fit results for two cases 
are illustrated in Figure~\ref{fig:agn_irs_fit}; the fits for the rest are 
available in the electronic figure set. %\input{fig7set.tex}

The global MIR emission for the AGN sample is also well fit with a combination
of a PAH template and three modified blackbodies. As in the H~{\footnotesize II} nuclei, seven 
objects (NGC\,3521, 3938, 4579, 4669, 4736, 5033 and 5055) show a prominent dip 
feature around 15$\mum$ that is not well captured by the model.  The entire 
PAH spectrum is generally very well recovered.  A notable exception, also 
observed in HLCs and H~{\footnotesize II} nuclei, is that the left wing of PAH 8.6$\mum$ is 
slightly underestimated in 10 out of 22 objects.  The signal-to-noise ratio of 
the IRS spectrum of NGC 2915 and NGC 4579 is very low for $\lambda \simlt 
14.5\mum$, and their PAH 6.2$\mum$ and 8.6$\mum$ measurements are reported as 
nondetections.

As for the continuum, the median temperatures of the cold ($45\pm8$ K) and 
hot ($391\pm266$ K) dust components are comparable to those of H~{\footnotesize II} nuclei, 
while the temperature of the warm component ($101\pm56$ K) is more elevated 
and has a larger dispersion.

\subsection{High-luminosity AGNs \label{sec:sample_D}}

The SINGS sample contains only relatively low-luminosity AGNs.  
It is known that PAH emission is present in some high-luminosity quasars (\eg Shi \etal 2014; 
Xie et al. 2017), wherein the radiation from the AGN dominates the UV and 
optical emission of the entire host galaxy and a hot dusty torus complicates 
the MIR band.  We apply our technique to a few representative Palomar-Green (PG; 
Schmidt \& Green 1983) quasars.  To demonstrate the potential of our 
decomposition method, we select as examples three PG quasars that have 
distinctive MIR characteristics (Table~\ref{tab:basic_unique}).  We show that 
our methodology can be extended to more complicated galaxy systems that 
contain bright AGNs. Both PAH features and the overall continuum are well 
reproduced in the fits, and upper limits on the PAH strength can be placed 
for systems with undetected PAH emission.

\subsubsection{PG\,1351+236} 

\begin{figure*}[ht] 
\begin{center}
\resizebox{0.8\hsize}{!}
{\includegraphics{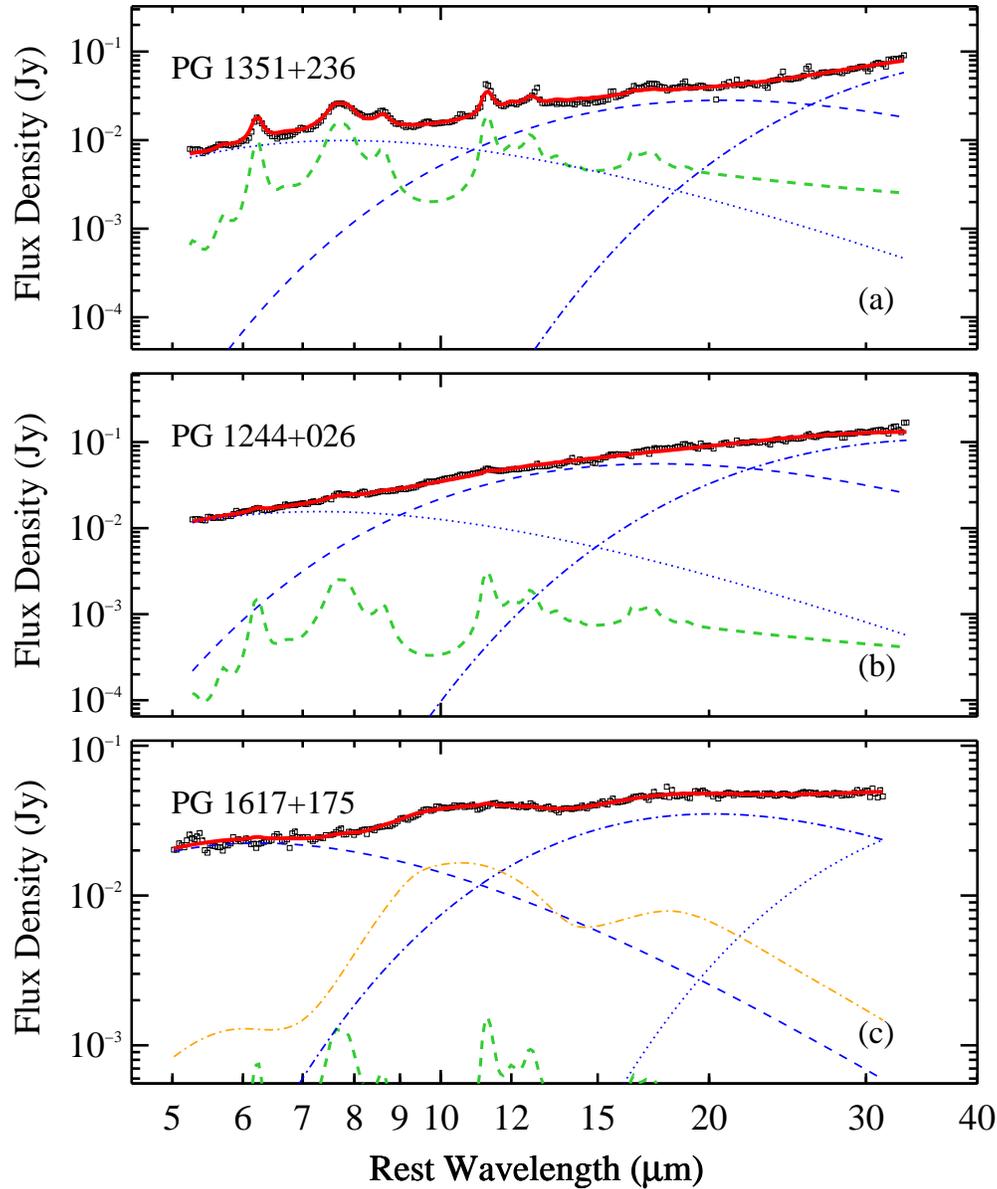}}
\caption{\footnotesize 
Decomposition of the IRS spectra of the quasars ({\it a}) PG\,1351+236, which 
shows strong PAH emission; ({\it b}) PG\,1244+026, which shows very weak PAH 
emission and a flat featureless continuum; and ({\it c}) PG\,1617+175, which 
shows no PAH features but strong silicate emission at 10 and 18$\mum$. 
For each source, we show the observed spectrum (black squares) and the 
best-fit model (solid red line), comprising a theoretical PAH template (green 
dashed line) and three modified blackbody components (blue lines) for the hot 
(dotted), warm (dashed), and cold (dot-dashed) dust continuum.  The 
silicate emission component in PG\,1617+175 is shown as an orange 
dot-dashed line. 
\label{fig:agn_high_fit} }      
\end{center}
\end{figure*}

The MIR spectrum of PG\,1351+236 is characterized by moderately strong PAH 
emission features at 6.2, 7.7, 8.6, and 11.3$\mum$ sitting on top of a flat 
continuum (Figure~\ref{fig:agn_high_fit} (a)).  There are also several ionic 
lines of high ionization potential, including [S~{\footnotesize IV}] 10.5$\mum$, [Ne~{\footnotesize V}] 14.3 $\mum$, [Ne~{\footnotesize V}] 24.3$\mum$, and [O~{\footnotesize IV}] 25.89$\mum$. 
Our generic five-component 
model well captures the PAH spectrum, while successfully depicting the MIR 
continuum emission. The derived temperatures are $375\pm3$, $141\pm1$, and 
$59\pm1$ K for the hot, warm, and cold components, respectively. The 
extinction toward this source is consistent with zero.

\subsubsection{PG\,1244+026}

By contrast, PG\,1244+026 exhibits only a flat, featureless continuum with no 
detectable PAH or ionic line emission. The best-fit result of this galaxy is 
show in Figure~\ref{fig:agn_high_fit} (b). The fit of the overall 
continuum is fairly well reproduced by the superposition of three continuum 
components with temperatures $394\pm5$ K, $165\pm1$ K, and $77\pm1$ K.
We place an upper limit on the strength of the PAH component, which accounts 
for only 1.3\% of the total MIR luminosity.

\subsubsection{PG\,1617+175}

PG\,1617+175 represents another example of a quasar with little to no PAH 
or ionic line emission, but in this case, the continuum is not featureless
(Figure~\ref{fig:agn_high_fit} (c)).  Two prominent emission features 
from amorphous silicate dust appear at 10 and 18$\mum$.  Apart from the 
standard three blackbodies for the continuum, we add an additional warm 
silicate component, which is assumed to be optically thin in the MIR. Equation~2 is 
supplemented with an extra component, $\kappa_\nu B_{\nu}(T)$, 
where $\kappa_\nu$ represents the mass absorption coefficient of silicate 
dust as a function of frequency,\footnote{The function $\kappa_\nu$ 
depends on the grain size and chemical composition of a specific dust. Here we 
adopt parameters for silicate dust that can represent the majority of PG 
quasars, with grain size $a\,=\,1.5\mum$ and ``astronomical silicate'' 
(Draine \& Lee 1984). The mass absorption coefficient is calculated from Mie 
theory; see Xie \etal (2017) and references therein for more details.} and
$B_{\nu}(T)$ is the Planck function for silicate emission of 
temperature $T$.  This relatively simple model can successfully describe the 
MIR emission of PG\,1617+175: the dust temperatures for the hot, warm, and cold 
blackbodies are $464\pm5$, $144\pm1$, and $62\pm1$ K, respectively, and the 
silicate emission component has a temperature of $410\pm17$ K. The PAH emission 
contributes less than 1\% of the total MIR power. 

A cautionary note: the silicate profile of AGNs that exhibit strong MIR m
silicate emission can vary significantly from source to source.  The silicate 
dust parameters that we adopt for PG\,1617+175 may not be applicable to other
sources.

\subsection{Heavily Obscured Galaxies \label{sec:sample_E}}
                   
Thus far, we have applied our technique to galaxies that suffer from little or 
no MIR extinction (\S\ref{sec:sample_A}--\S\ref{sec:sample_C}) or that are 
optically thin (\S\ref{sec:sample_D}). However, the nuclear regions of some 
galaxies, especially merging systems such as ULIRGs, are heavily dust-obscured.
The MIR emission of these galaxies is strongly extincted, and observationally 
it shows deep silicate absorption at 10 and 18$\mum$ (\eg Spoon \etal 2006), 
which substantially further complicates the decomposition of the PAHs.  This 
phenomenon is also expected to be common in the high-redshift universe, where 
galaxies are generically more gas- (and hence dust-) rich.  In this section, we 
extend our technique to more complex galaxy systems whose central source (AGN 
or nuclear starburst) is extremely enshrouded by dust and gas. We use as 
illustration two sources, one with PAH detected and another without 
(Table~\ref{tab:basic_unique}).

\subsubsection{ Arp\,220}

The optically heavily obscured late-stage merger Arp\,220 is the nearest (78 
Mpc) ULIRG.  Its radio morphology suggests that the galaxy comprises two 
nuclei separated by 370\,pc (Norris 1988; Barcos-Mu\~noz \etal 2015).  While 
the presence of X-ray emission and radio variability reveals the presence of a 
weak AGN (Iwasawa \etal 2001; Batejat \etal 2012), most of the IR luminosity 
in Arp\,220 is powered by a nuclear starburst (Armus \etal 2007).  

\begin{figure*}[ht] 
\begin{center}
\resizebox{0.8\hsize}{!}
{\includegraphics{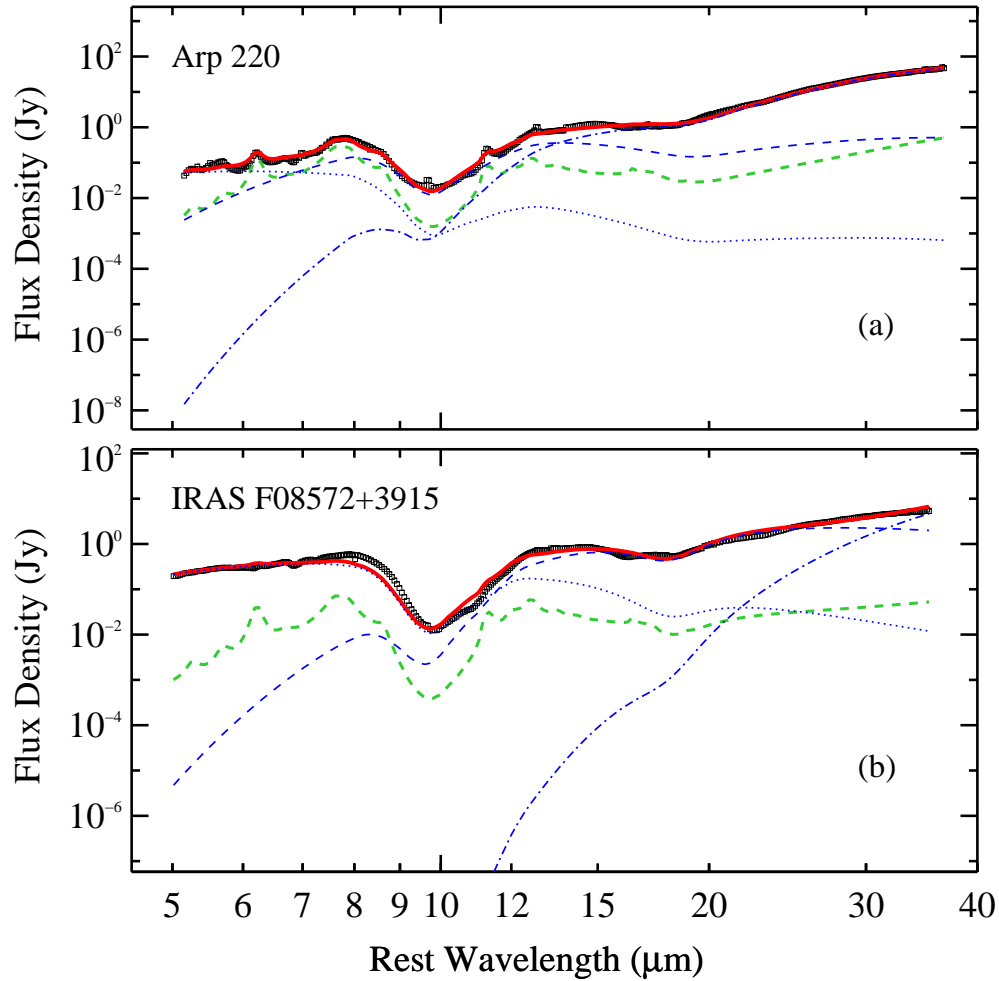}}
\caption{\footnotesize 
Decomposition of the IRS spectra of the highly obscured, IR-luminous galaxies
({\it a}) Arp\,220, which shows PAH emission and silicate absorption, and 
({\it b}) IRAS\,F08572+3915, which does not have detectable PAH emission but has 
strong silicate absorption at 10 and 18$\mum$.  For each source, we show the 
observed spectrum (black squares) and the best-fit model (solid red line), 
comprising a theoretical PAH template (green dashed line) and three modified 
blackbody components (blue lines) for the hot (dotted), warm (dashed), and 
cold (dot-dashed) dust continuum.  
\label{fig:obscured_irs_fit} }      
\end{center}
\end{figure*}

The MIR spectrum of Arp\,220 is characterized by deep silicate absorption 
troughs around 9.7 and 18$\mum$, but PAH emission at 6.2, 7.7, and 11.3$\mum$
can be clearly seen.  Our fitting scheme successfully accommodates both 
(Figure~\ref{fig:obscured_irs_fit} (a)).  The temperatures of the three 
continuum components are 1500, $192\pm1$, and $82\pm0.1$ K. 
The temperature reached by the hot component (1500 K) is the maximum value 
allowed and corresponds to the sublimation limit of graphite dust. 
The higher temperature detected here presumably reflects significant 
heating from the embedded AGN. 

Interestingly, the observed silicate absorption profile in Arp\,220 differs 
from the silicate profile of the Galactic extinction curve of Wang \etal 
(2015), which peaks at a shorter wavelength ($\simali$9.5$\mum$) and has a broader 
width. The extinction curve of Wang \etal (2015) is constrained via modeling 
observed Galactic extinctions from optical to MIR toward multiple lines of 
sight with theoretical ``astronomical silicate'' and graphite dust from Draine 
\& Lee (1984). The ``astronomical silicate'' profile was calibrated using the
observed 9.7$\mum$ silicate emission profile from the inner region of the 
Orion H~{\footnotesize II} (Trapezium) region, which is found to be very similar to absorption 
features in dense molecular clouds in terms of having a shorter wavelength 
peak and broader width (Gillett \etal 1975). The poor match between the 9.7 
$\mum$ silicate absorption profile of Arp\,220 and that of Wang et al.'s 
extinction curve might be caused by variations in ISM
properties in different environments.  We therefore adopt the extinction curve 
derived from the diffuse ISM, whose 9.7$\mum$ silicate feature generally has a 
longer peak wavelength and narrower width, considering extinction curves based 
on observations in the solar circle (Chiar \& Tielens 2006) and toward the 
Galactic center (Smith \etal 2007)\footnote{The extinction curve of Smith et 
al. (2007) is composed of a power law and silicate extinction peaking at 9.7 
and 18$\mum$. The silicate component consists of the observed absorption 
profile that covers 8--13$\mum$ (Kemper \etal 2004) and a Drude profile 
with a peak at 18$\mum$ and a FWHM of 4.45$\mum$ to form the extinction 
above 13$\mum$. The curve is extended to $\lambda < 8\mum$ by assuming 
$\tau_{\lambda}=\tau_{8}\rm\times exp[2.03(\lambda-8)]$, 
with $\tau_{8}$ being the observed silicate absorption at 8$\mum$. 
A power law with an index of 1.7 is adopted to 
represent the carbon dust extinction and is joined to the silicate profile to form 
the complete extinction curve. See Smith \etal (2007) for more details.}.  We 
find that both extinction curves can match the silicate absorption profile 
equally well at 9.7 and 18$\mum$ for Arp 220; 
Figure~\ref{fig:obscured_irs_fit} (a) illustrates the best-fit result 
obtained using the curve of Chiar \& Tielens (2006). Chiar \& Tielens (2006) 
found that the silicate profile and relative feature strength (9.7/18) of the 
local ISM extinction curve can be better explained using porous pyroxene 
silicate combined with a small fraction of amorphous olivine, while for the 
diffuse ISM in the Galactic center an amorphous olivine mass fraction of more 
than $80\%$ is required to explain the silicate profile at 9.7$\mum$ (Kemper 
\etal 2004).  Nevertheless, the silicate absorption profiles in Arp\,220 
definitely share closer characteristics with the diffuse ISM.

\subsubsection{ IRAS\,F08572+3915}

IRAS\,F08572+3915 is another late-stage merger ULIRG, one that possesses double
 nuclei detected in the X-rays (Iwasawa \etal 2011). Optical emission-line 
ratios indicate the presence of an obscured AGN in the northwestern core 
(Yuan \etal 2010). There is no PAH emission detected in the MIR, and the whole 
spectrum is dominated entirely by two deep absorption troughs at 10 and 18 
$\mum$, which arise from amorphous silicate dust. Besides amorphous silicate, 
IRAS\,F08572+3915 also shows unambiguous crystalline silicate absorption at 
11, 16, and 18$\mum$ (Spoon \etal 2006).  

Figure~\ref{fig:obscured_irs_fit} (b) shows that our technique can 
recover the observed silicate absorption features, together with the continuum 
emission, reasonably well. The dust continuum is characterized by temperatures
of $690\pm90$, $250\pm42$, and $69\pm12$ K for the hot, warm, and cold 
components, respectively. The PAH component contributes $< 1\%$ to the total 
MIR luminosity, and we are able to only place an upper limit on the PAH 
strength for this galaxy.  We note, in passing, that the popular PAHFIT code 
performs especially poorly for galaxies such as IRAS\,F08572+3915 that have a
highly complex continuum.  In this particular example, the PAH flux derived 
from PAHFIT is significantly overestimated.

Similar to Arp\,220, IRAS\,F08572+3915 shows a redder peak and broader width 
for the 9.7$\mum$ silicate absorption profile with respect to the extinction 
curve of Wang \etal (2015), but its 18$\mum$ absorption profile appears normal.
The extinction curve of Chiar \& Tielens (2006) can fit the observed 9.7$\mum$ 
silicate profile but not that at 18$\mum$, which shows a bluer peak in 
the observations.  The best overall fit is achieved by the extinction curve of 
Smith \etal (2007), which is based on observations of the diffuse ISM toward 
the Galactic center.

\section{Discussion \label{sec:discussion}}

We demonstrated that a simple technique comprising a theoretical PAH template, 
three modified blackbodies, and an extinction term can successfully decompose 
the PAH spectrum and the main dust components.  
We apply this technique to a variety of systems, from
the simplest HLCs to galaxies with increasingly complex properties.  We 
successfully recover the strength of the PAH emission (or upper limits thereof),
characteristic temperatures of the dust continuum, and the level of dust 
extinction. 
Below, we discuss the robustness of our technique (\S\ref{sec:robustness}), and
we perform a detailed comparison 
between the PAH strengths measured by our technique 
with those derived from other methods (\S\ref{sec:compare}).

\subsection{Robustness Demonstration\label{sec:robustness}}

\subsubsection{The Effect of the FIR\label{sec:fir_sed}}

Our methodology fits the MIR spectrum with a simple model comprising a PAH 
template and three modified blackbodies to represent the continuum.  We do not 
consider longer-wavelength data in the FIR, and it is legitimate to wonder to 
what extent our results might be affected by this choice.  Here we test this 
effect by explicitly fitting two objects with and without including FIR data 
from {\it Herschel}, which allows us to extend the SED out to 500 $\mu$m.

Prior to combining the IRS spectra with the {\it Herschel}\ photometry, we need
to consider a possible aperture mismatch between the two data sets.  The slit 
width of the SL module ($\simali$5.1--14.5$\mum$) is 3\farcs6, and for the LL 
module ($\simali$14.5--38$\mum$) it is 10\farcs5, whereas the coarsest 
point-spread function of SPIRE on {\it Herschel}\ has FWHM = 36\farcs6 at 500 
$\mum$.  In this paper, the flux density of the SL module is scaled to that 
of the LL module to compensate for possible flux loss due to the aperture 
difference.  We use {\it IRAS}\ 25$\mum$ images to verify that the vastly 
different resolutions between the MIR and FIR sample the same physical scale.  
We require that the galaxies included in our analysis are fully enclosed 
within the LL slit width.

\begin{figure}
\begin{center}
\resizebox{\hsize}{!}
{\includegraphics{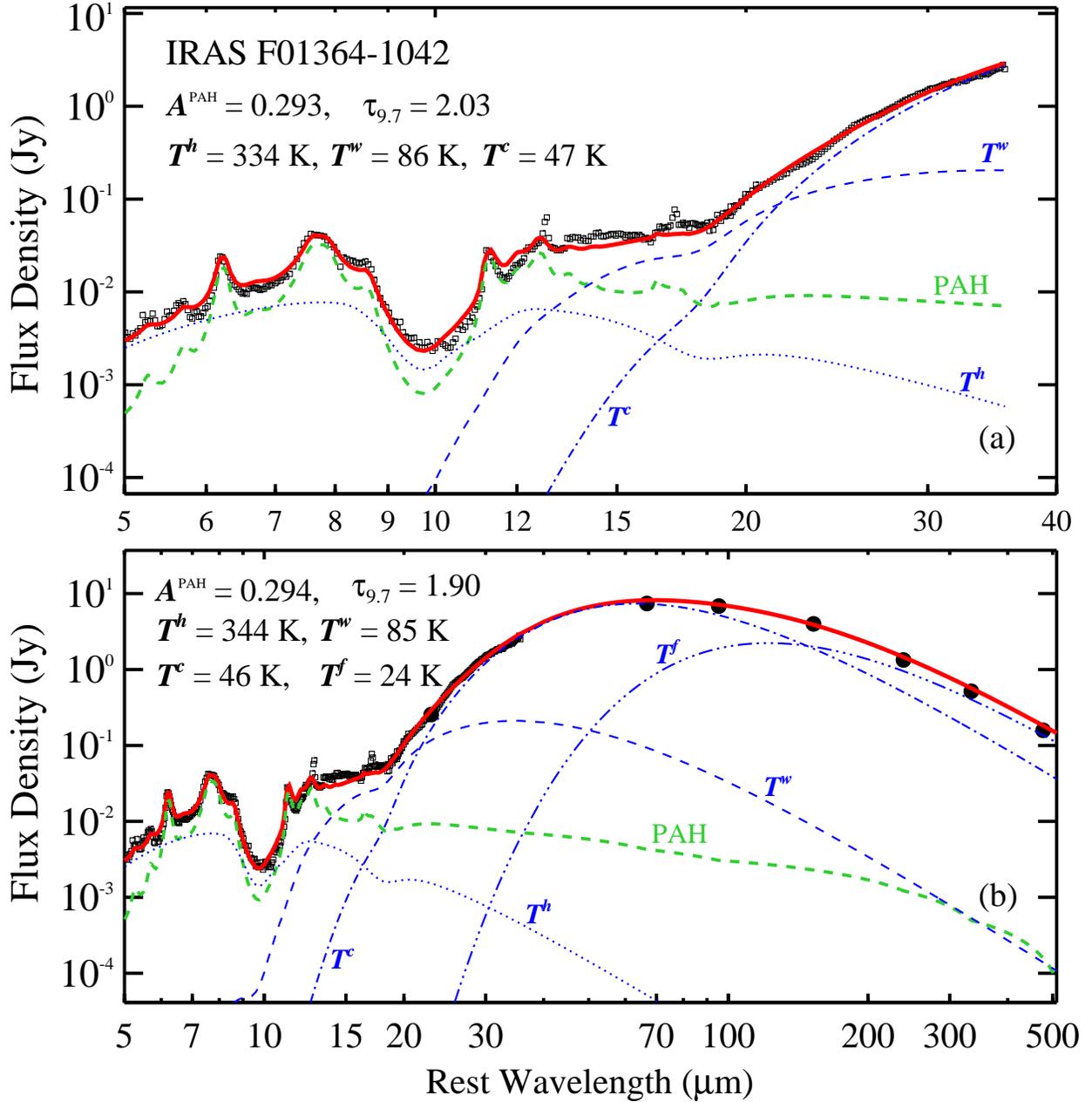}}\\
\caption{\footnotesize
({\it a}) Decomposition of the IRS spectrum of IRAS F01364$-$1042 (black 
squares). The best-fit model (solid red line) comprises a theoretical PAH 
template (green dashed line) and three modified blackbody components (blue 
lines) for the hot (dotted), warm (dashed), and cold (dot-dashed) dust 
continuum.  ({\it b}) Decomposition of the IRS spectrum of IRAS F01364$-$1042 
with the SED extended to the FIR using photometric measurements from 
{\it Herschel}\ at 70, 100, 160, 250, 350, and 500$\mum$.  An additional 
cold dust component with $T^f = 24$ K is required to fit the long-wavelength 
data, but the best-fit parameters of the original fit to the IRS are not
significantly affected.
\label{fig:fir1}             }
\end{center}
\end{figure}

\begin{figure} 
\begin{center}
\resizebox{\hsize}{!}
{\includegraphics{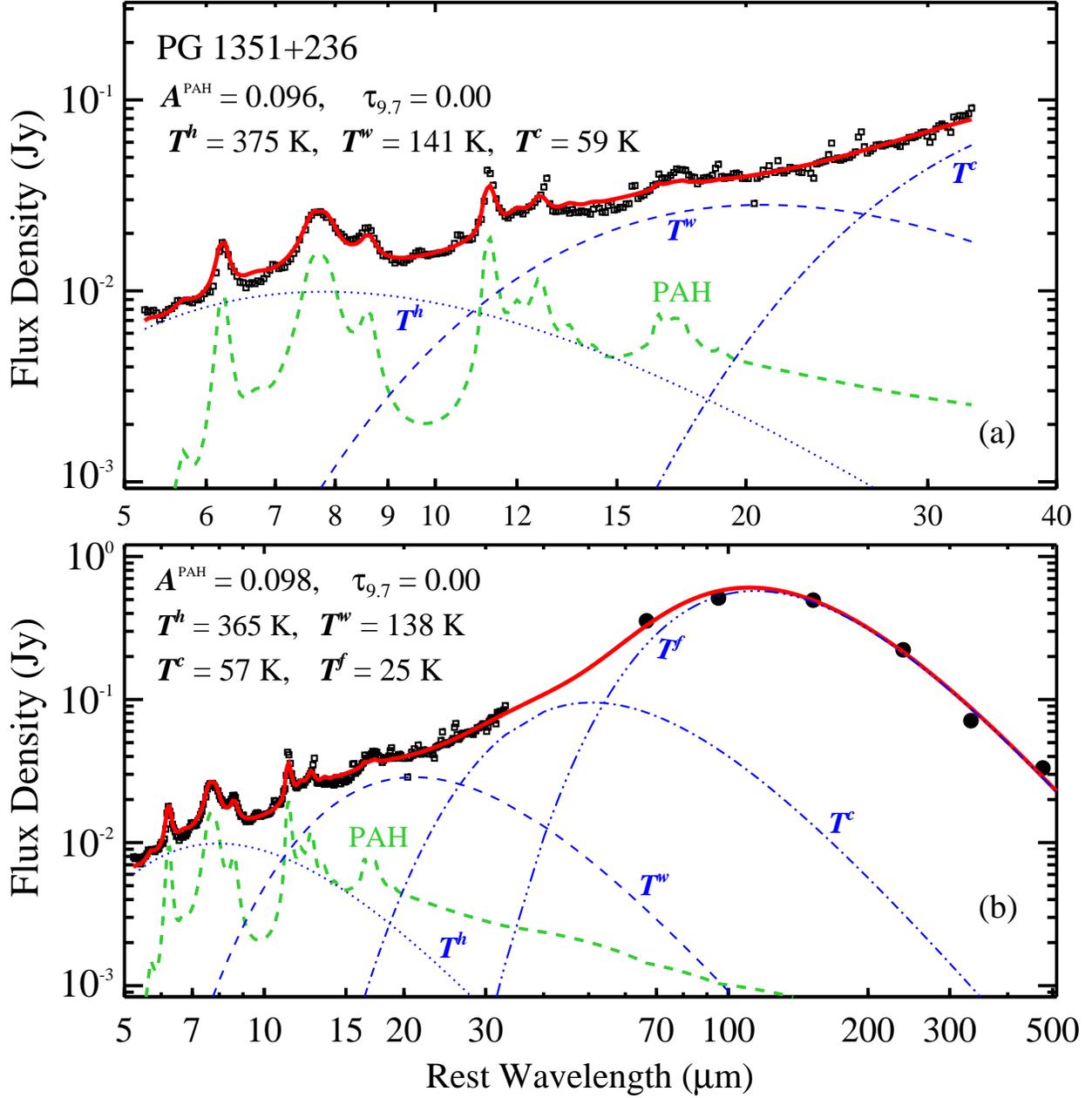}}
\caption{\footnotesize
({\it a}) Decomposition of the IRS spectrum of PG\,1351+236 (black
squares). The best-fit model (solid red line) comprises a theoretical PAH
template (green dashed line) and three modified blackbody components (blue
lines) for the hot (dotted), warm (dashed), and cold (dot-dashed) dust
continuum.  ({\it b}) Decomposition of the IRS spectrum of PG\,1351+236 
with the SED extended to the FIR using photometric measurements from
{\it Herschel}\ at 70, 100, 160, 250, 350, and 500$\mum$.  An additional
cold dust component with $T^f = 25$ K is required to fit the long-wavelength
data, but the best-fit parameters of the original fit to the IRS are not
significantly affected.
\label{fig:fir2}             }
\end{center}
\end{figure}

We choose the ULIRG IRAS F01364--1042 at $z = 0.048$ and the quasar 
PG 1351+236 at $z = 0.055$ for our tests (Table~\ref{tab:basic_unique}).
The FIR photometry, obtained using {\it Herschel}/PACS for 70, 100, and 160 
$\mum$ and {\it Herschel}/SPIRE for 250, 350, and 500$\mum$, are given in Chu 
\etal (2017) and Shangguan \etal (2018). A single modified blackbody with 
$\beta=2$ is adopted to account for the FIR SED.  The SEDs and their best fits 
are shown in Figure~\ref{fig:fir1} for IRAS F01364--1042 and in Figure 
\ref{fig:fir2} for PG 1351+236.  In each figure, the top panel (a) shows the SED 
from $\simali$5 to 40$\mu$m, while the bottom panel (b) shows the full IR SED from 
$\simali$5 to 500$\mu$m.  We see that a single, additional modified blackbody 
of temperature $T^f \approx 25-30$ K can well describe the FIR portion of the 
SED.  Nevertheless, this extra cold dust component has little noticeable 
effect on the temperature of the hot, warm, and cold dust components.  The 
derived extinction and the scale factor for the PAH template are also 
essentially unchanged.  This test verifies that our standard fitting 
technique, which is based exclusively on the IRS spectrum, not only gives 
robust measurements of PAH but also places reasonable constraints on the 
overall dust continuum emission in the MIR.

\subsubsection{ The Effects of Photon Density and PAH Parameters on 
                             the PAH Theoretical Template  \label{sec:discussion_U}}

\begin{figure} 
\begin{center}
\resizebox{0.7\vsize}{!}
%{\includegraphics{plot_u4_case.eps}}
{\includegraphics{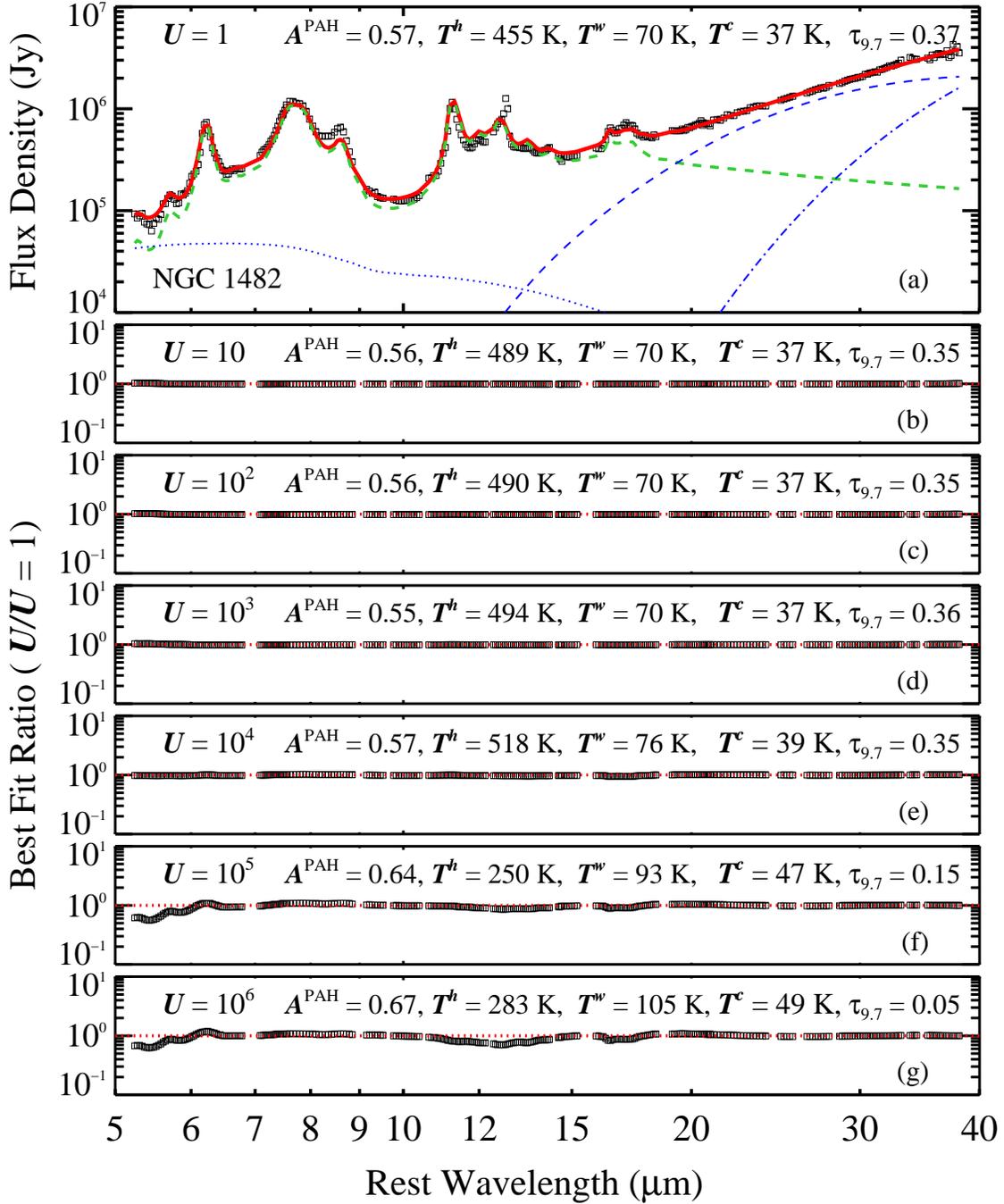}}
\caption{\footnotesize
({\it a}) Decomposition of the IRS spectrum of NGC\,1482 (black squares). The 
best-fit model (solid red line) comprises a theoretical PAH template 
calculated with $U\,=\,1$ and dust grains with sizes $a< 20\,\Angstrom$
(green dashed line) and three modified blackbody components (blue lines) for 
the hot (dotted), warm (dashed), and cold (dot-dashed) dust continuum.  
The bottom panels show the ratio of the best-fit model relative to the model 
with $U = 1$ for ({\it b}) $U = 10$, ({\it c}) $U = 10^2$, ({\it d}) 
$U = 10^3$, ({\it e}) $U = 10^4$, ({\it f}) $U = 10^5$, and ({\it g}) 
$U = 10^6$.  The derived model parameters are relatively insensitive
to $U$ within the range explored.
\label{fig:u4}             }
\end{center}
\end{figure}

All of the fits thus far are based on a theoretical PAH template calculated 
assuming a photon density of $U=1$.  In Section~\ref{sec:pah}, we showed
that, for grain sizes $a \simlt 20\Angstrom$ or $a \simlt 50\Angstrom$, the PAH 
spectrum for $\lambda$ \lax\ 15 $\mu$m is quite uniform for a wide range of 
photon densities from $U=1$ to $10^{6}$ (Figure 2). Beyond 15$\mum$,
the emission mainly originates from the continuum associated with PAHs.  
The power drops in high-$U$ environments because for $U \simgt 10^{4}$ the 
timescale for photon absorption becomes less than the cooling timescale for 
some grains due to the high photon density.  These grains will not cool completely 
and hence gradually will attain a temperature close to the equilibrium 
temperature. 
Nevertheless, the deviation of the PAH spectrum beyond 
15$\mum$ will not seriously affect the recovery of the observed PAH emission. 
This is demonstrated in Figure~\ref{fig:u4}, where we fit NGC\,1482 using 
theoretical templates of different $U$ at $a \simlt 20\Angstrom$. The uppermost 
panel shows the standard, default fit with $U=1$, and the subsequent panels 
display the best-fit ratio between higher values of $U$ and $U=1$. It can 
be seen that the scale factor for the PAH template varies by less than 20\% 
across six orders of magnitude in $U$.  As for the derived dust temperatures and 
extinction, they remain practically constant up till $U=10^{4}$; for
$U\simgt10^{5}$, the temperature and extinction vary markedly because of changes 
in the template spectra longer than 15$\mum$ under conditions of extremely 
high photon density.  We therefore suggest that, under most conditions, the 
$U = 1$ template is sufficient to measure accurate PAH strengths.

In the case of grains with a full size distribution, PAHs dominate the spectrum
at wavelengths \lax\ $15\mum$ for $U\simlt10^3$.  When $U$ increases to 
sufficiently large values (\eg $\simgt10^4$), some large grains 
($a \simgt 50\Angstrom$) will gain high temperatures and peak 
at short wavelengths, causing changes of spectral shape,
and the MIR emission will no longer be purely 
emitted from PAHs.  Repeating our analysis for all galaxies using the PAH 
template for $U=10^{4}$ and grains with a full size distribution, we find that 
the fits are considerably worse. PAH emission is generally underestimated,
and the 10$\mum$ silicate feature is not well reproduced.  

Besides photon density $U$, the characteristics of the theoretical PAH template
are determined by the combined effects of chemical composition, grain size 
distribution, gas temperature, electron density, and mass fraction locked in 
ultra-small grains.  The physical parameters for the theoretical PAH spectra 
used in the current paper are constrained by observations of the Milky Way and 
the Magellanic Clouds.  Although, in principle, these parameters should vary in 
different galaxies, currently we have very limited information on them for 
external galaxies because of the difficulty in determining the extinction 
curve and chemical depletion in distant galaxies. 
In spite of these 
complications, our study highlights the interesting fact that the PAH spectrum 
is surprisingly homogeneous across an extraordinarily wide range of galactic 
environments (\S~\ref{sec:discussion_caveat}).

\subsubsection{The Uniformity of the PAH Spectrum and 
     the Peculiar Emission Spectra in Elliptical Galaxies
     \label{sec:discussion_caveat}}

\begin{figure} 
\begin{center}
\resizebox{0.6\vsize}{!}
{\includegraphics{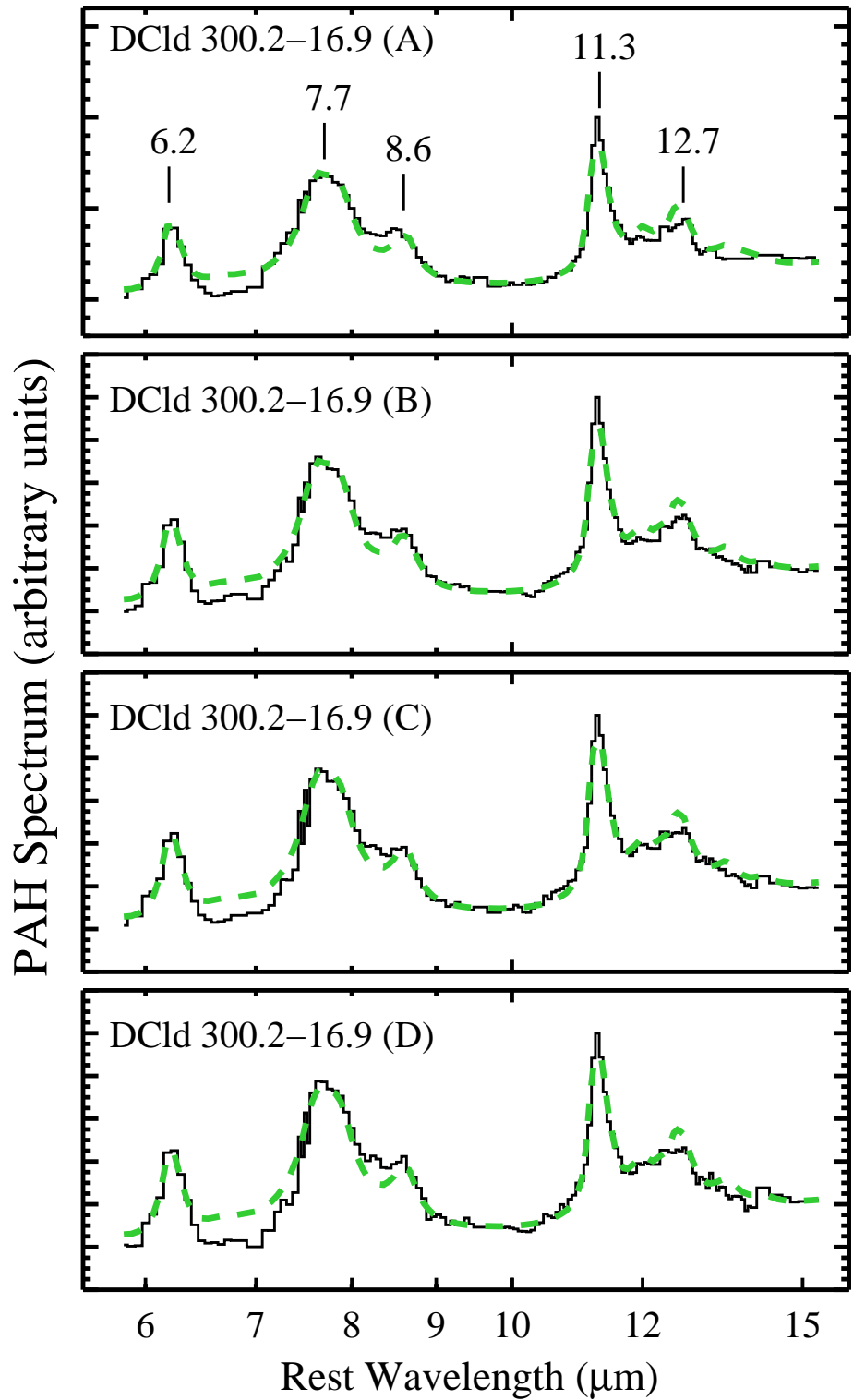}}
\caption{\footnotesize
Comparison of the observed PAH spectra for the HLCs (black histograms), 
obtained from our decomposition, with the best-fit theoretical template (green 
dashed line).  The five most prominent PAH features are labeled.
\label{fig:pah_spec_hgl} }
\end{center}
\end{figure}

\begin{figure} 
\begin{center}
\resizebox{0.8\vsize}{!}
{\includegraphics{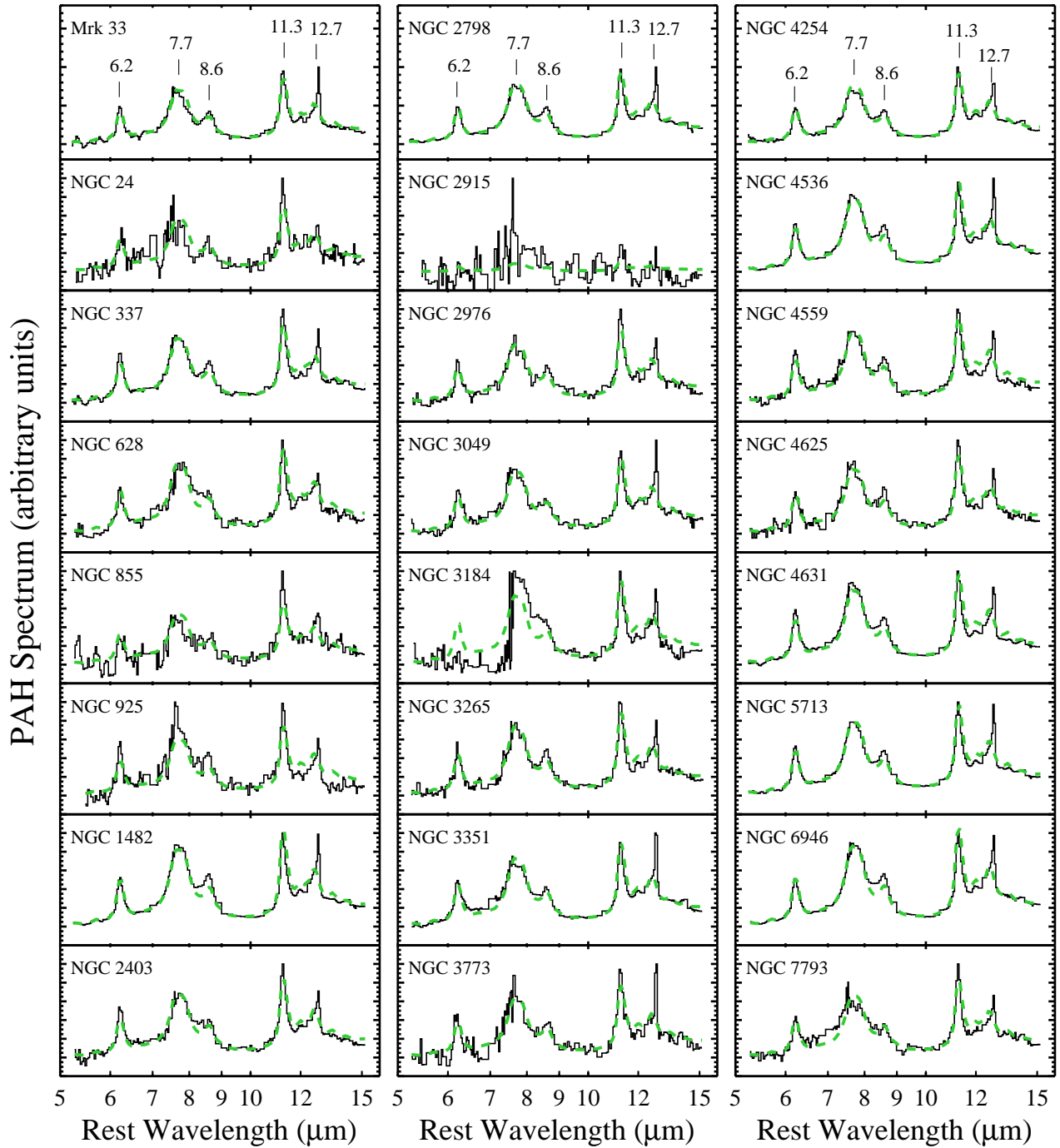}}
\caption{\footnotesize
Comparison of the observed PAH spectra for the SINGS H~{\scriptsize{II}} galaxies 
(black histograms), obtained from our decomposition, with the best-fit theoretical 
template (green dashed line).  The five most prominent PAH features are 
labeled. The PAH spectra are remarkably invariant.
\label{fig:pah_spec_hii} }
\end{center}
\end{figure}

\begin{figure} 
\begin{center}
\resizebox{0.8\vsize}{!}
{\includegraphics{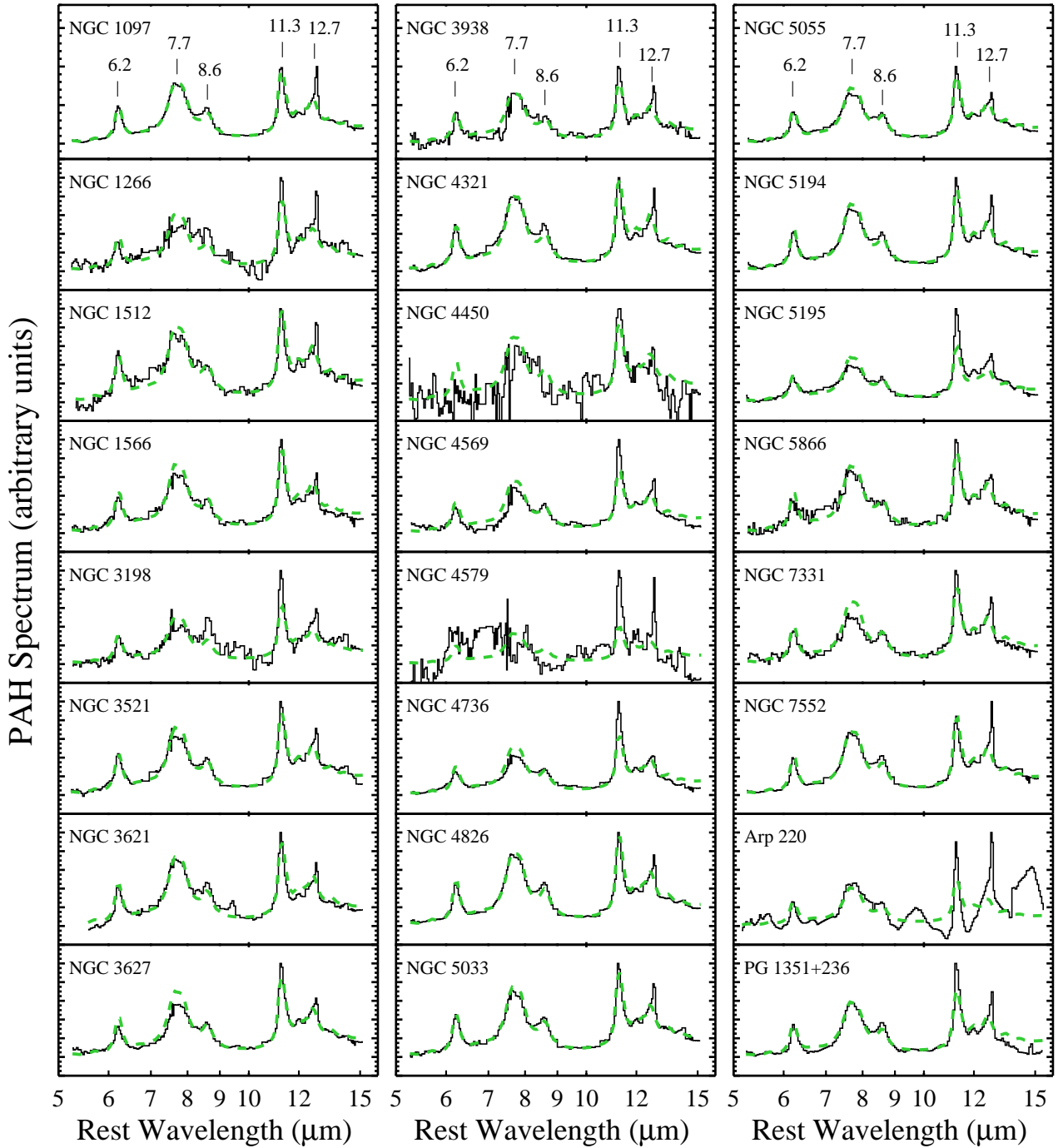}}
\caption{\footnotesize
Comparison of the observed PAH spectra for the SINGS AGNs, Arp\,220, and 
PG\,1351+236 (black histograms), obtained from our decomposition, with the 
best-fit theoretical template (green dashed line).  The five most prominent 
PAH features are labeled. The PAH spectra are remarkably invariant.
\label{fig:pah_spec_agn} }
\end{center}
\end{figure}

For each object with clear PAH detection, we derive two PAH spectra, one from 
the best-fit theoretical PAH template and the other directly from the residuals
between the observed total spectrum and best-fitting continuum model (after 
removal of isolated ionic emission lines).  The two versions of the PAH spectra
are compared directly in Figure~\ref{fig:pah_spec_hgl} for HLCs, in
Figure~\ref{fig:pah_spec_hii} for H~{\footnotesize{II}} nuclei, and in 
Figure~\ref{fig:pah_spec_agn}
for AGNs.  It is remarkable that a {\it single}\ theoretical PAH template can 
successfully match the observed PAH spectrum over such a diverse range of 
astrophysical environments.  
This implies that, to first order, the PAH spectrum is 
essentially invariant, demonstrating that, 
although the PAH spectral shape
(and, as a result, the PAH band ratio) does show some variations 
among different systems, it remains fairly constant in a wide variety 
of environments.
The most notable exception is for the feature at 8.6$\mum$, for 
which the model generally underestimates the PAH flux by $\simali$30\%.  
In several AGNs, the PAH 7.7$\mum$ band is overpredicted 
in the best-fit model, while the PAH 11.3$\mum$ band 
is underestimated. This phenomenon is quite obvious in 
NGC\,3627, 4569, 4736, 5195, 7331 and PG\,1351+236. 
For the sample of AGNs investigated here, the mean and 
standard deviation of the ratio of the 7.7$\mum$ band
to the 11.3$\mum$ band is 3.11$\pm$0.03 
from the best-fit template and 3.02$\pm$0.80 
from the observed spectrum.  
By contrast, for the six AGNs having an apparent deficit 
in the 7.7$\mum$ band, 
7.7\,$\mu$m/11.3\,$\mu$m\,$=$\,2.40$\pm$0.35,  which
is consistent with the ratio of 
2.15$\pm$0.52 derived from PAHFIT (\S~\ref{sec:compare}).

\begin{figure*}[ht] 
\begin{center}
\resizebox{0.8\hsize}{!}
{\includegraphics{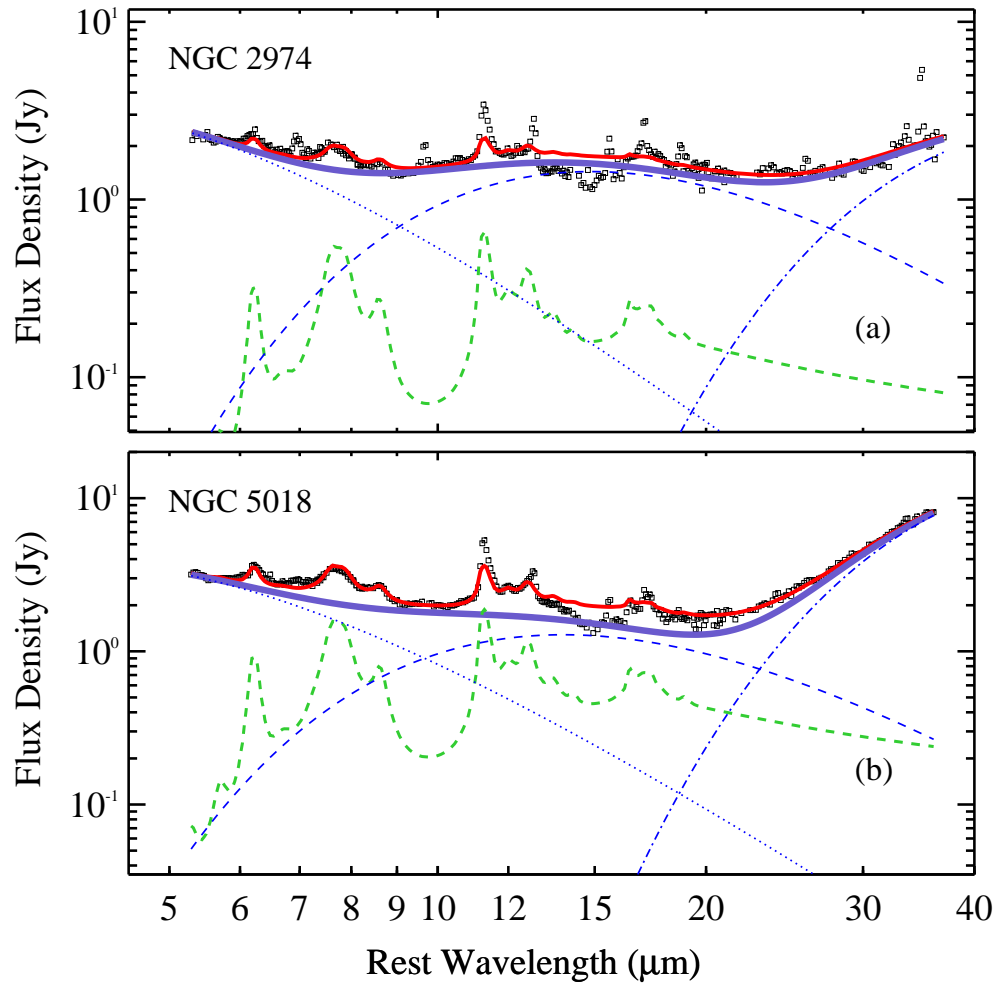}}
%\resizebox{0.8\hsize}{!}
%{\includegraphics{ngc5018_mbb3_kvt.eps}}
\caption{\footnotesize 
           Decomposition of the {\it Spitzer}/IRS spectra 
               of the elliptical galaxies ({\it a})
               NGC\,2974 and ({\it b}) NGC\,5018,
               which show peculiar PAH emission 
               (Kaneda et al.\ 2008).  
               For each source, we show 
               the observed spectrum (black squares) 
               and the best-fit model (solid red line),
              which comprises a theoretical PAH template 
              (green dashed line) and three modified 
              blackbody components (blue lines) 
              for the hot (dotted), warm (dashed), and 
              cold (dot-dashed) dust continuum.  
              Also shown is the overall continuum (purple line)
              obtained by summing the three modified 
              blackbody components.
\label{fig:elliptical_irs_fit} }      
\end{center}
\end{figure*}

We note that appreciable variations in the PAH spectral profiles 
have been detected, often in extreme environments
(\eg Diamond-Stanic \& Rieke 2010, Sales \etal 2010).
In some elliptical galaxies, the PAH 7.7$\mum$ 
band is significantly suppressed relative to 
the PAH 11.3$\mum$ band
(e.g., see Kaneda \etal 2005, 2008;  Rampazzo \etal 2013).
This is also true in the centers of active galaxies
(e.g., see Smith et al.\ 2007).
We have also applied our technique to
NGC\,2974 and NGC\,5018,
two elliptical galaxies exhibiting 
unusual PAH spectra (Kaneda et al.\ 2008).
As shown in Figure~\ref{fig:elliptical_irs_fit},
the current technique can reasonably well characterize 
the overall continuum emission 
underlying the PAH features.
However, the PAH emission features
are not closely reproduced 
by the PAH template spectrum:
both galaxies show a strong enhancement 
at the PAH 11.3$\mum$ band 
in comparison with the template.
Our technique also fails to reproduce 
the dip around 15$\mum$. 
This is not unexpected, since the theoretical template 
has a fixed PAH spectral shape.
The peculiar 11.3/7.7 band ratios seen 
in these elliptical galaxies indicate that 
the PAH sizes may be affected 
in the hot plasmas of ellipticals;
smaller PAHs may be preferentially destroyed 
while larger ones survive in the hostile environment. 
As a matter of fact, these sources will serve as 
a backbone to remove the contamination of old stars 
to star formation rates estimated from PAH emission
(Y.~Xie et al.\ 2018, in preparation). 

We note that, with a fixed PAH spectral shape,
the current technique by itself does not 
automatically yield PAH band ratios 
for sources whose PAH spectra differ 
considerably from the theoretical template.
Nevertheless, variations of the PAH band ratios 
among different astronomical environments can
be investigated from the residual spectrum once 
the underlying MIR continuum and extinction 
are self-consistently described and subtracted
(see, e.g., Figures~\ref{fig:pah_spec_hgl}$-$\ref{fig:pah_spec_agn}). 
The total luminosity of PAHs can be determined straightforwardly.  Our 
technique also allows us to place upper limits on the strength of PAH emission
in galaxies that are highly obscured and/or hosting luminous AGNs, for which 
PAH features are often difficult to detect.

%%%%%%

\subsection{Comparison with Other Techniques\label{sec:compare}}

\begin{figure} 
\begin{center}
\resizebox{0.6\vsize}{!}
{\includegraphics{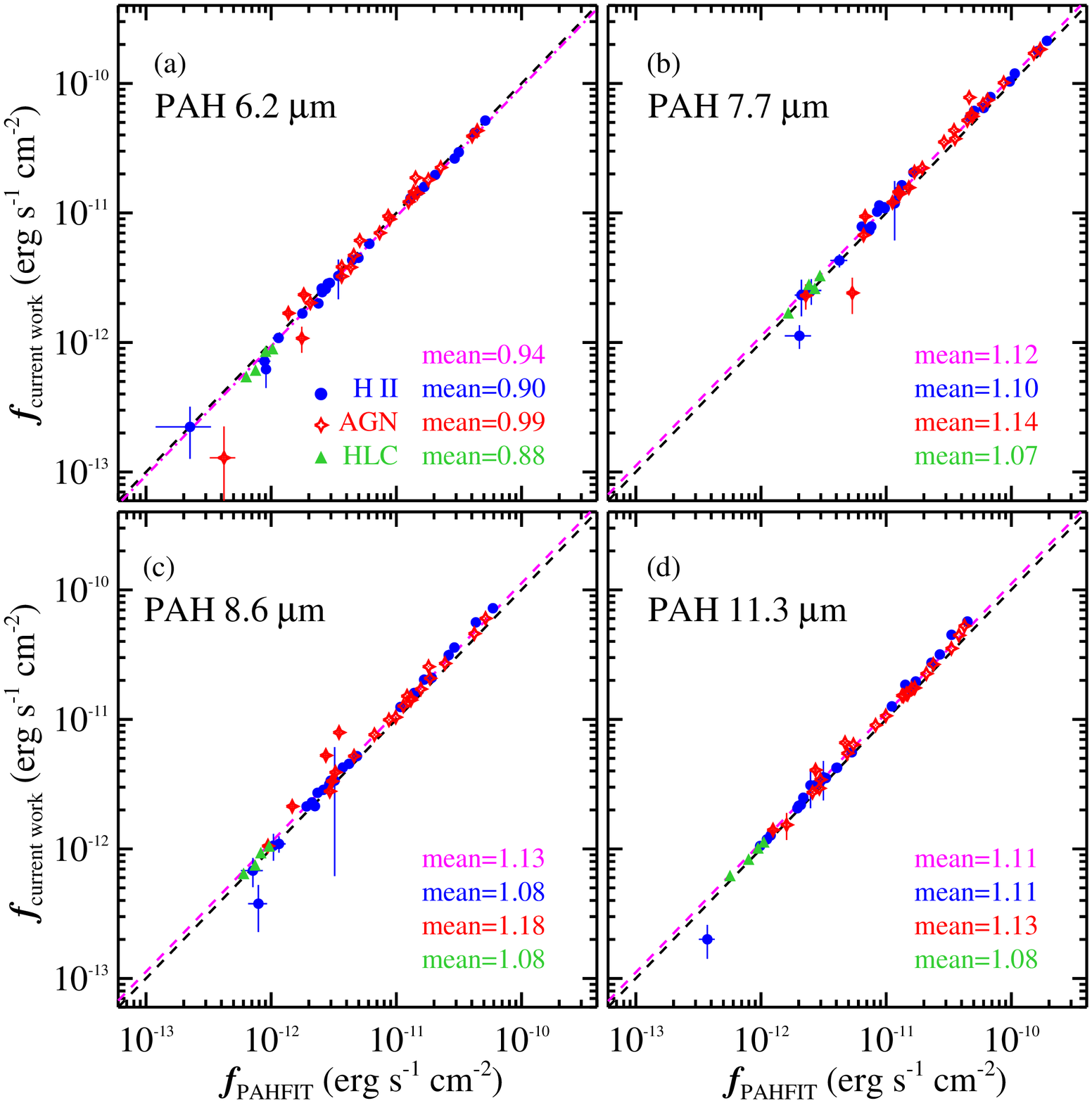}}
\caption{\footnotesize
Comparison of the PAH fluxes derived from 
the current template-fitting method with 
those from PAHFIT for the features 
at ({\it a}) 6.2$\mum$, ({\it b}) 7.7$\mum$, 
({\it c}) 8.6$\mum$, and ({\it d}) 11.3$\mum$.  
Our PAH fluxes were 
derived directly from the observed spectra, after subtraction 
of the best-fit model continuum.  In each panel, the black dashed line 
indicates the one-to-one relation, and the the magenta dashed line represents
the mean value of all SINGS H~{\scriptsize{II}} galaxies, AGNs, and HLCs  
calculated from the ratio of fluxes from our measurements and those derived from PAHFIT. 
Separate mean ratios for the HLCs, H~{\scriptsize{II}} galaxies, and AGNs are also displayed. 
\label{fig:s_pahfit} }
\end{center}
\end{figure}

\begin{figure} 
\begin{center}
\resizebox{0.6\vsize}{!}
{\includegraphics{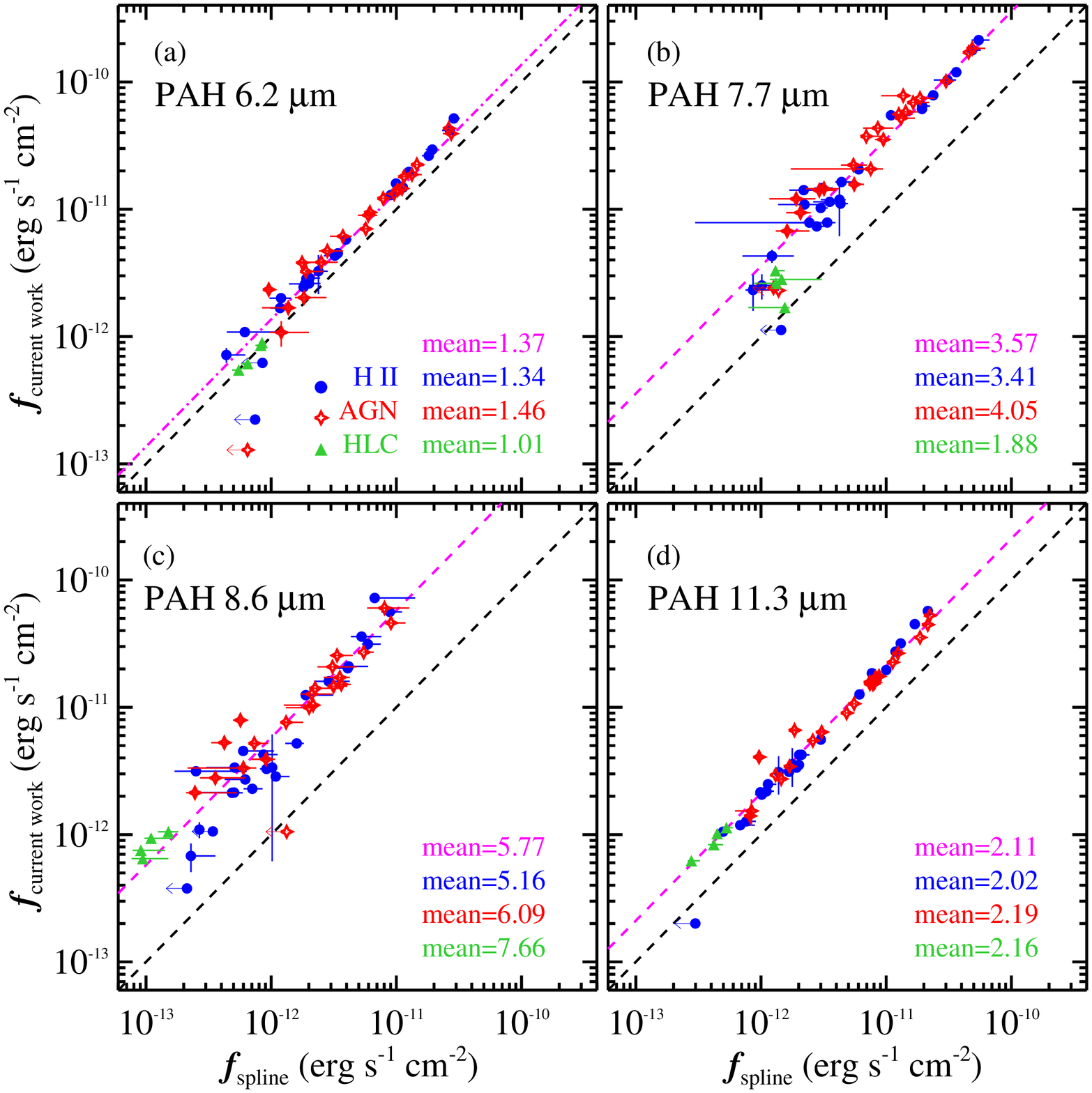}}
\caption{\footnotesize
Comparison of the PAH fluxes derived from 
the current template-fitting method with 
those from the spline fit method, 
for the features at ({\it a}) 6.2$\mum$, 
({\it b}) 7.7$\mum$, ({\it c}) 8.6$\mum$, 
and ({\it d}) 11.3$\mum$.  Our 
PAH fluxes were derived directly from the observed spectra 
after subtraction of the best-fit model continuum.  In each panel, the black 
dashed line indicates the one-to-one relation, and the the magenta dashed line 
represents the mean value of all SINGS H~{\scriptsize{II}} galaxies, AGNs, and HLCs  
calculated from the ratio of fluxes from our measurements and those derived 
from the spline method.  Separate mean ratios for the HLCs, H~{\scriptsize{II}} galaxies, and AGNs are also displayed.
\label{fig:s_spline} }
\end{center}
\end{figure}

\begin{figure}
\begin{center}
\resizebox{0.6\vsize}{!}
{\includegraphics{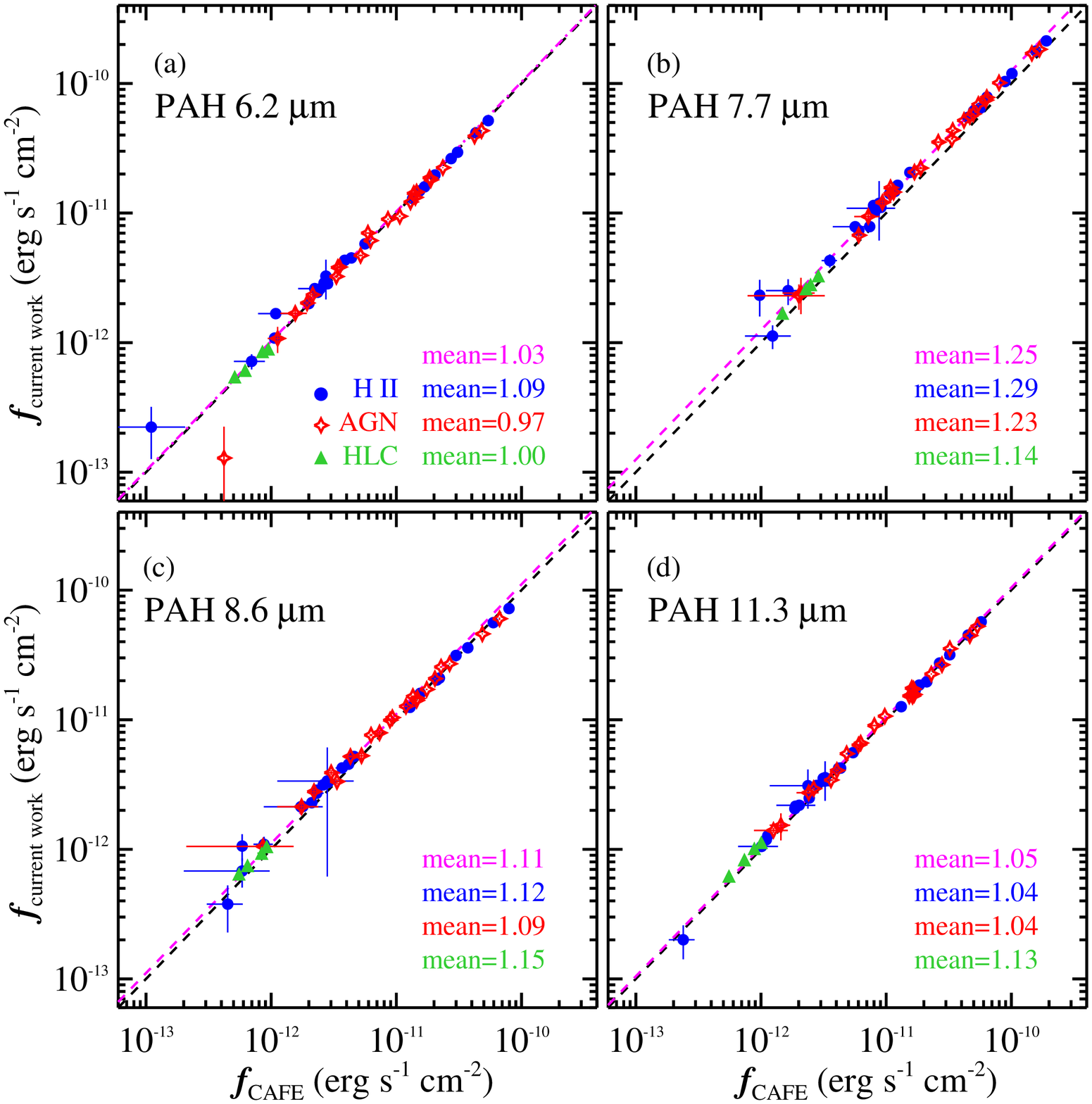}}
\caption{\footnotesize
Comparison of the PAH fluxes derived from the current template-fitting method
with those from CAFE for the features
at ({\it a}) 6.2$\mum$, ({\it b}) 7.7$\mum$,
({\it c}) 8.6$\mum$, and ({\it d}) 11.3$\mum$.  Our PAH fluxes were
derived directly from the observed spectra after subtraction
of the best-fit continuum model.  In each panel, the black dashed line
indicates the one-to-one relation, and the the magenta dashed line represents
the mean value of all SINGS H~{\scriptsize{II}} galaxies, AGNs, and HLCs, calculated from
the ratio of fluxes from our measurements and those derived from CAFE.
Separate mean ratios for the HLCs, H~{\scriptsize{II}} galaxies, and AGNs are also displayed.
\label{fig:s_cafe} }
\end{center}
\end{figure}

\begin{figure}
\begin{center}
\resizebox{0.66\vsize}{!}
{\includegraphics{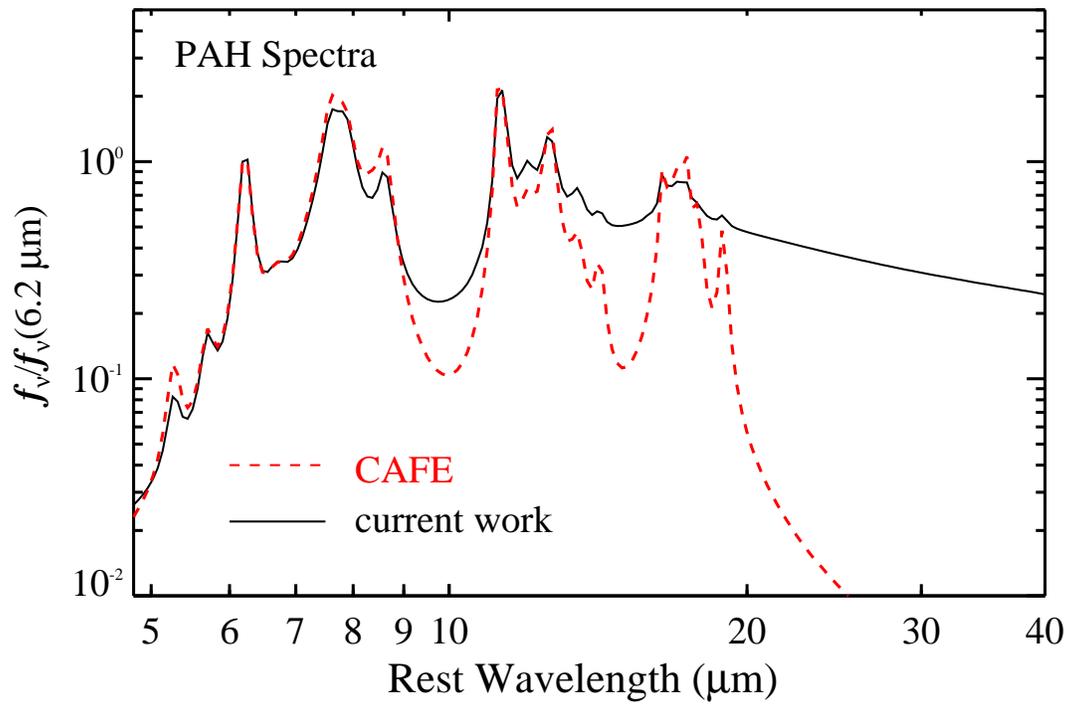}}
\caption{\footnotesize
Comparison of the PAH template spectrum adopted
in the current work (solid black line) with that
used in CAFE (red dashed line; Marshall \etal 2007). }
\label{fig:cafe_dl07}
\end{center}
\end{figure}

We compare the fluxes of the individual PAH features
at 6.2, 7.7, 8.6, and 11.3$\mum$ measured with our technique 
(Table~\ref{tab:spec_flux}) with those obtained from PAHFIT 
(Table~\ref{tab:pahfit_flux}; 
Figure~\ref{fig:s_pahfit}),
spline fit (Table~\ref{tab:spline_flux}; Figure~\ref{fig:s_spline}), and
CAFE (Table~\ref{tab:cafe_flux}; Figure~\ref{fig:s_cafe})\footnote{
A number of studies (\eg Rieke \etal 2009;  da Cunha \etal 2010; 
Wu \etal 2010; Brown \etal 2014) have used starburst templates to 
explore the multiwavelength SEDs of galaxies.  However, 
most of these templates were derived directly from observed spectra, 
which contain both dust continuum and line emission and therefore 
do not directly yield fluxes for the PAH features.  Here we limit our discussion 
to the three methods (PAHFIT, spline fit, and CAFE) that are closest 
in methodology to ours. 
}.  
We duplicated the PAH template of CAFE based on
the parameters tabulated in Table~1
of Marshall \etal (2007). 
%
%\textcolor{red}{
The fluxes for individual PAH bands 
were derived directly from the observed spectra 
by fitting them with Drude profiles  
with FWHM parameters taken from Draine \& Li (2007), 
after subtraction of the best-fit continuum model,
while the total PAH flux was derived from integrating the observed,
continuum-subtracted PAH spectra from 5 to 20$\mum$ 
where PAH emission domininates the radiation.
In general, our new technique yields roughly 
the same level of PAH strength as PAHFIT 
for the individual bands
at 6.2, 7.7, 8.6, and 11.3$\mum$,
suggesting that the two methods derive a
similar level of continuum emission. 
By contrast, spline fit systematically and 
significantly underestimates the PAH strength 
by factors of $\simali$1.5--2 
for the 6.2 and 11.3$\mum$ features 
and by as much as factors of $\simali$4--6 
for the 7.7 and 8.6$\mum$ features. 
The large discrepancy arises from the fact that spline fit 
does not properly account for the flux in the wings of the lines.  
As for the comparison with CAFE, the PAH fluxes of the 6.2, 8.6, and 
11.3$\mum$ bands are consistent with ours within $\simali$10\%, but those of 
the 7.7$\mum$ band are $\sim$20\% lower.  
%\textcolor{red}{
As shown in Figure~\ref{fig:cafe_dl07}, 
the PAH template used in CAFE matches our template 
fairly well at main PAH features at 6.2, 7.7, 8.6 and 11.3$\mum$, 
except for the deep drop at $\simali$10$\mum$.  
At $\lambda>20\mum$, the CAFE template  declines 
precipitously relative to our theoretical template. 
Both discrepancies can be accounted for with the  
continuum emission associated with the PAH molecules 
(see Li \& Draine 2001).
%}

\begin{figure} 
\begin{center}
\resizebox{0.6\vsize}{!}
{\includegraphics{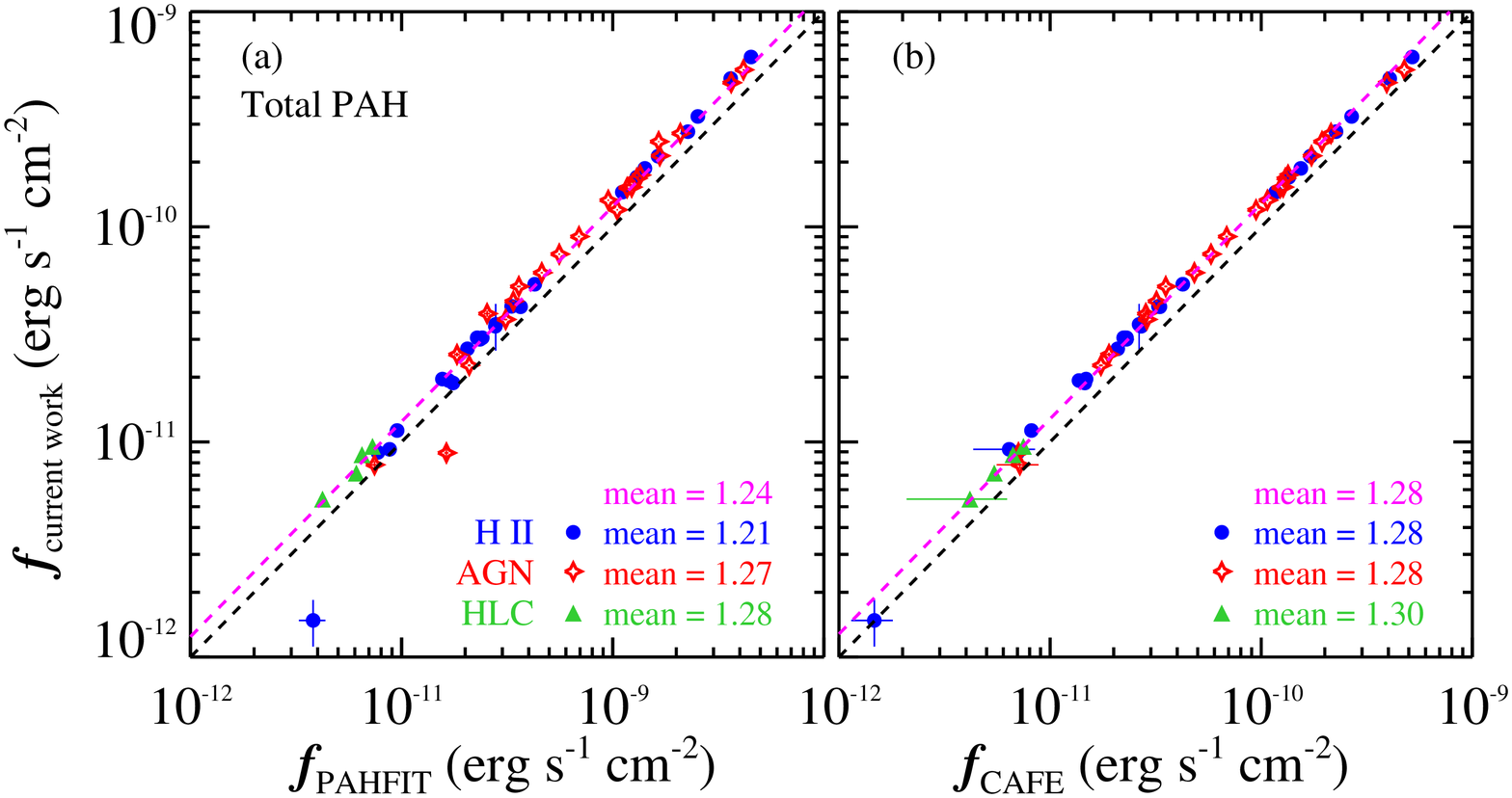}}
\caption{\footnotesize
Comparison of the total PAH flux derived from the current template-fitting method with 
	those from (a) PAHFIT and (b) CAFE, considering PAH emission between 5 and 20$\mum$. 
	The black dashed line indicates the one-to-one relation, and the magenta dashed line 
	represents the mean value of all SINGS H~{\scriptsize{II}} galaxies, AGNs, and HLCs  
	calculated from the ratio of fluxes from our measurements and those of the other methods.
	Separate mean ratios for the HLCs, H~{\scriptsize{II}} galaxies, and AGNs are also 
	given.}
\label{fig:s_cafe_pahfit_total}
\end{center}
\end{figure}

\begin{figure} 
\begin{center}
\resizebox{0.59\vsize}{!}
{\includegraphics{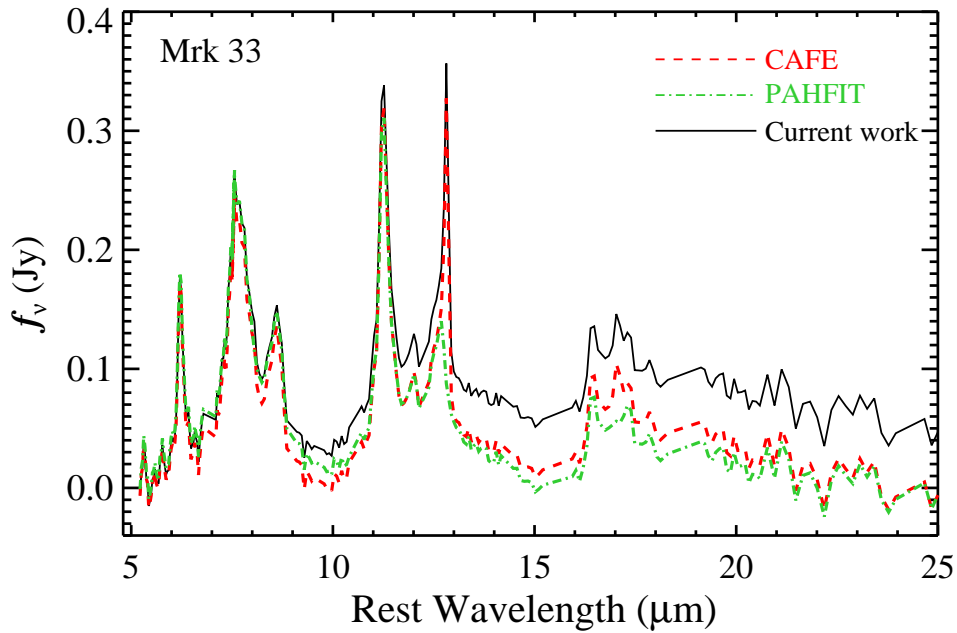}}
\caption{\footnotesize
Comparison of the ``residual'' PAH spectra
of Mrk\,33, a star-forming galaxy,
obtained by subtracting 
(from its {\it Spitzer}/IRS spectrum)
the best-fit continuum
derived, respectively, from our technique (black solid line),
PAHFIT (green dot-dashed line), and
CAFE (red dashed line). 
}
\label{fig:Ours_CAFE_PAHFIT} 
\end{center}
\end{figure}

Figure~\ref{fig:s_cafe_pahfit_total} examines the total flux
from all PAH bands as measured using PAHFIT and CAFE.
Our technique recovers $\simali$30\% more total PAH flux 
than either of these two methods; 
the difference varies from $\simali$20\% 
to $\simali$40\% as the integration interval 
changes from 5--15 to 5--30$\mum$. 
%
%\textcolor{red}{
This primarily reflects our treatment of
the long-wavelength continuum emission of PAHs.
To account for the laboratory-measured
long-wavelength skeletal vibrations of PAHs 
(e.g., see Moutou et al.\ 1996),
Draine \& Li (2007) incorporated 
a ``continuum'' opacity
from $\simali$13$\mum$ longward, 
represented by a Drude profile peaking at
15$\mum$ with an FWHM of 12$\mum$
(see Table~1 in Draine \& Li 2007). 
This is clearly demonstrated 
in Figure~\ref{fig:Ours_CAFE_PAHFIT},
where we compare the ``residual'' PAH spectra
of the star-forming galaxy Mrk\,33,
obtained by subtracting 
(from the observed {\it Spitzer}/IRS spectrum)
the best-fit continuum
derived, respectively, from 
our technique, PAHFIT, and CAFE.
It is apparent that the ``residual'' PAH spectra
at $\lambda\simlt13\mum$ 
are essentially the same 
for all three methods.
This explains why the fluxes
for individual PAH bands 
derived from all three methods 
agree with each other fairly well.
However, the PAH continuum emission
at $\lambda\simgt13\mum$ from our
method considerably exceeds that
of PAHFIT and CAFE. 
With the PAH continuum emission included, 
our technique requires a smaller amount 
of thermal continuum emission and thus 
a larger amount of PAH emission. 
%\footnote{\textcolor{red}{
%We note that the [Ne II] 12.81$\mum$, 
%$\rm H_{2}$ $S$ (4) 8.03$\mum$ and 
%$\rm H_{2}$ $S$ (1) 17.03$\mum$ are blending 
%in the observed PAH spectra, the total flux contribution of these three lines 
%are $\leq 2\%$ in median and  therefore will not affect our main conclusion.}}.
%}
The extinction, as discussed below,
only plays a minor role, since most of 
our sample galaxies are only mildly extinguished. 

We suggest that the total PAH flux
(both features and continuum) 
emitted in the $\simali$5--20$\mum$
wavelength range 
is a better star formation rate indicator 
compared to the flux of an individual PAH feature 
(\eg the 7.7$\mum$ feature), since the latter is 
sensitive to PAHs of a particular size 
and a particular charging status 
(\eg neutral or ionized). 
%\textcolor{red}{
Also, physically speaking, 
the PAH continuum is excited by stellar photons 
in the same way as the PAH features.
The absorption of stellar UV photons
is balanced by the emission of both PAH features 
and PAH continuum.
Therefore, a proper counting of the absorbed UV
starlight (and thus the star formation rate) should also include 
the PAH continuum emission.
%}
%
For a given source, the template method proposed here 
determines the dust continuum emission 
and therefore readily allows us to 
derive the total PAH flux 
(both PAH features and PAH continuum)
by simply integrating the ``residual'' spectrum 
(\eg continuum-subtracted observational spectrum) 
over $\simali$5--20$\mum$. 
%
%\textcolor{red}{
To allow for a fair comparison 
with other techniques (\eg PAHFIT), 
a scaling factor of $\sim$1.3 is needed
because our template technique recovers $\simali$30\% 
more total PAH flux than PAHFIT.
%}

\begin{figure} 
\begin{center}
\resizebox{0.59\vsize}{!}
%{\includegraphics{plot_tau_v1_size1.eps}}
{\includegraphics{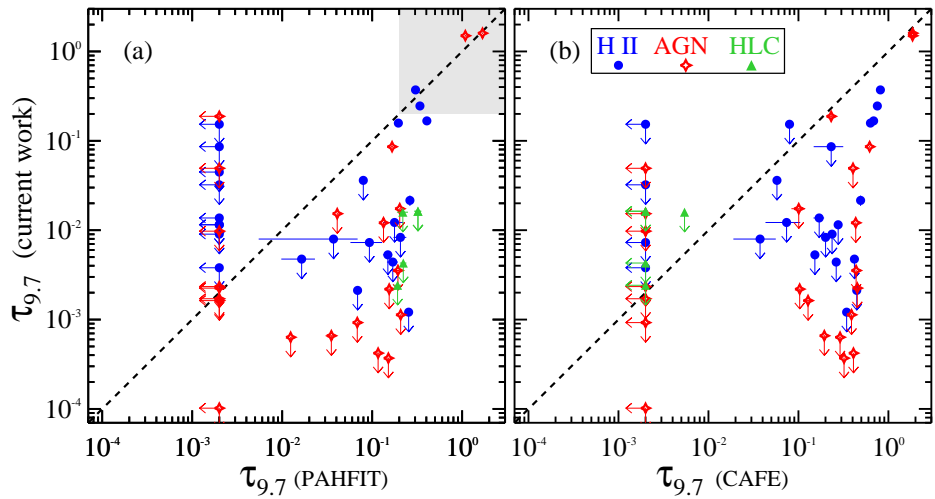}}
\caption{\footnotesize
Comparison of the 9.7$\mum$ optical depth derived from our 
template-fitting method with those from (a) PAHFIT and (b) CAFE.  
Objects that have extinction values equal to zero from PAHFIT or CAFE 
are assigned an artificial value of $2 \times 10^{-3}$. 
In panel (a), the two methods give comparable extinction values 
when $\tau_{9.7} > 0.2$ (shaded region).} 
\label{fig:extinction} 
\end{center}
\end{figure}

%
Lastly, we compare the extinction 
derived from our technique 
with that from PAHFIT 
(Table~\ref{tab:extinction}; Figure~\ref{fig:extinction}a).  
For SINGS galaxies that 
have significant silicate absorption ($\tausil\simgt0.2$; \eg NGC\,1266, 1482, 
3198, and 6946), the two techniques obtain comparable results.  On the other 
hand, for galaxies for which PAHFIT derives small ($\tausil\simlt0.2$) but 
nonzero extinction, our technique obtains systematically lower
values, most of them are upper limits
(e.g., the average extinction for HLCs from PAHFIT is 
$\simali$0.2, whereas our technique only derives 
an upper limit for all of them).   
%\textcolor{red}{
Similar trends appear when our extinctions
are compared to those derived using the CAFE template 
(Figure~\ref{fig:extinction}b).%}
This is consistent with the optical analysis of the SINGS galaxies by Dale \etal 
(2006), who derived $A_V \approx 1\magni$.  
A. Mishra \etal (2018, in preparation) also report a 
small optical extinction for HLCs, with $A_V\simlt1\magni$.  
Assuming $A_{V}/\tausil = 18$ for the diffuse ISM (Roche \& Aitken 1984), 
SINGS galaxies and HLCs have a typical MIR extinction of 
$\langle \tausil \rangle \approx0.05$.

\section{Summary \label{sec:summary}}

We present a new technique to decompose PAH emission from MIR spectra by 
making use of a theoretical PAH template in conjunction with three continuum 
emission components and an extinction term.  Our future goals are to 
understand the physical processes responsible for the production of the 
pervasive PAH spectrum and to use the PAH strength as a quantitative measure of 
the star formation rate in different extragalactic environments.  The main 
results of this paper are as follows.

$\bullet$
We can robustly extract the PAH spectrum in a wide range of galactic 
environments, from the simplest Galactic HLCs to nearby 
star-forming galaxies, low-luminosity active galaxies, luminous quasars, and 
highly obscured IR-luminous galaxies.  When PAH is undetected, an upper 
limit can be placed on its strength.

$\bullet$
We can simultaneously describe the associated continuum reasonably well and 
derive the MIR extinction.  Galaxies that have strong emission or absorption 
features at 9.7 and 18$\mum$ appear to have diverse silicate dust 
properties (grain size and chemical composition) and extinction curves.

$\bullet$
To achieve a robust decomposition of the PAH spectrum from the MIR SED, 
FIR observations are not needed .

$\bullet$
The PAH spectrum is surprisingly invariant with galactic environment,
suggesting a uniform set of conditions responsible for its excitation.

$\bullet$
The PAH fluxes of individual bands derived from the current method 
are consistent with those obtained using PAHFIT and CAFE,
whereas the total PAH fluxes derived from our method 
are as much as $\simali$30\% higher. 
The primary cause for the difference in total PAH flux is 
the long-wavelength continuum emission of PAHs 
associated with their skeletal vibrations, which is
incorporated in the theoretical PAH template 
used in our work. 
A scaling factor of $\simali$1.3 is needed when one compares 
the total PAH flux derived from our template method 
with that from PAHFIT. 
The PAH fluxes obtained from the spline fit method 
are lower than ours because it underestimates the 
wings of the PAH features.

$\bullet$
This technique is motivated to decompose PAH emission and 
derive upper limits in complicated systems, which will set 
a uniform baseline for calibrating PAH emission to 
star formation rate. In parallel, PAHFIT, CAFE, and  
other techniques have their own advantages in dealing with 
different systems by allowing the band ratios to 
change inherently.

\acknowledgments
We thank the referee, J. D. Smith, for stimulating comments 
and suggestions that helped to improve
the quality and presentation of our paper, as well as for 
 his careful explanation of the
IRS spectra of SINGS galaxies.
H.~Kaneda kindly shared
the IRS spectra of NGC\,2974 and NGC\,5018, and 
J.~Ingalls sent us the IRS spectra of the four Galactic high-latitude clouds.
B.~T.~Draine and L.~H.~Jiang provided helpful discussions and suggestions. 
This work was supported by the National Key R\&D Program of China 
(2016YFA0400702) and the National Science Foundation of China (11473002, 
11721303).  
Y.~X. is supported by China Postdoctoral Science Foundation Grant 2016 M591007. 
The Cornell Atlas of \spitzerirs Sources (CASSIS) is a product of the Infrared 
Science Center at Cornell University, supported by NASA and JPL.

\clearpage

%%%% Figure 16 %%%%

%%%% Figure 17 %%%%

%%%% Figure 18 %%%%

%%%% Figure 19 %%%%

%%%% Figure 20 %%%%

%%%% Figure 21 %%%%

%%%% Figure 22 %%%%

%%%% Figure 23 %%%%

\clearpage

%%% Table 1: Basic parameters for HGLs%%%
\begin{deluxetable}{lccccc}
\tabletypesize\tiny
\tablecolumns{6}
\tablecaption{\label{tab:basic_hgl}
      Basic Parameters for Galactic High-latitude Clouds
              }
\tablewidth{0pt}
\tablehead{
\colhead{Source} & \colhead{R.A. (J2000)} & \colhead{Decl. (J2000)} & 
	\colhead{Type} & \colhead{Distance} & \colhead{Extraction Aperture} \\
\colhead{}  & \colhead{(hh mm ss.s)} & \colhead{(dd mm ss)} & 
	\colhead{}  & \colhead{(pc)}  & \colhead{(arcsec $\times$ arcsec)} \\
	\colhead{(1)} & \colhead{(2)} & \colhead{(3)} & \colhead{(4)} & \colhead{(5)} & \colhead{(6)}
}
\startdata
DCld~300.2-16.9 (A) & 11 54 23.1 & $-$79 31 42 & HLC & 70$\pm$15   &  $ 142.8 \times 10.7$ \\
DCld~300.2-16.9 (B) & 11 52 08.3 & $-$79 09 33 & HLC & 70$\pm$15   &  $ 142.8 \times 10.7$ \\
DCld~300.2-16.9 (C) & 11 48 24.4 & $-$79 18 00 & HLC & 70$\pm$15   &  $ 142.8 \times 10.7$ \\
DCld~300.2-16.9 (D) & 11 55 33.8 & $-$79 20 54 & HLC & 70$\pm$15   &  $ 142.8 \times 10.7$ \\
\enddata
\end{deluxetable}
%%% Table: Basic parameters for HGLs%%%

%%% Table 2: Basic parameters for SINGS%%%
\begin{deluxetable}{lrrccc}
\tabletypesize\tiny
\tablecolumns{6}
\tablecaption{\label{tab:basic_sings}
 	 Basic Parameters for SINGS Sample
              }
\tablewidth{0pt}
\tablehead{
\colhead{Source} & \colhead{R.A. (J2000)} & \colhead{Decl. (J2000)} & 
	\colhead{Type} & \colhead{Distance}  &  \colhead{Extraction Aperture}\\
\colhead{}  & \colhead{(hh mm ss.s)} & \colhead{(dd mm ss.s)} & 
	\colhead{}  & \colhead{(Mpc)}  &  \colhead{(arcsec $\times$ arcsec)} \\
	\colhead{(1)} & \colhead{(2)} & \colhead{(3)} & \colhead{(4)} & \colhead{(5)}  & \colhead{(6)}
}
\startdata
Mrk~~33 ................... & 10 32 31.82 & +54 24 02.5  &  H II &  22.9   &  $52 \times 30$  \\   
NGC~24 ...................  & 00 09 56.37 & $-$24 57 51.2 &  H II &   7.3  &  $52 \times 30$  \\
NGC~337 .................   & 00 59 50.20 &$-$07 34 45.8 &  H II &  22.4   &  $52 \times 30$  \\
NGC~628 .................   & 01 36 41.60 & +15 47 00.0  &  H II &   7.3   &  $52 \times 30$  \\
NGC~855 .................   & 02 14 03.70 & +27 52 38.4  &  H II &   9.6   &  $52 \times 30$  \\
NGC~925 .................   & 02 27 17.25 & +33 34 41.6  &  H II &   9.1   &  $52 \times 30$  \\
NGC~1482 ...............   & 03 54 38.88 & $-$20 30 07.1  &  H II &  23.2  &  $52 \times 30$  \\
NGC~2403 ...............   & 07 36 49.95 & +65 36 03.5   &  H II &   3.2   &  $52 \times 30$  \\
\enddata
\end{deluxetable}
%%% Table: Basic parameters for SINGS%%%

%%% Table 3: Basic parameters for Unique Galaxies%%%
\begin{deluxetable}{lcccc}
\tabletypesize\tiny
\tablecolumns{5}
\tablecaption{\label{tab:basic_unique}
              Basic Parameters for Quasars and Obscured Galaxies
              }
\tablewidth{0pt}
\tablehead{
\colhead{Source} & \colhead{R.A. (J2000)} & \colhead{Decl. (J2000)} & 
\colhead{Type} & \colhead{Distance} \\
\colhead{}  & \colhead{(hh mm ss.s)} & \colhead{(dd mm ss.s)} & 
\colhead{}  & \colhead{(Mpc)}  \\
\colhead{(1)} & \colhead{(2)} & \colhead{(3)} & \colhead{(4)} & \colhead{(5)}
}
\startdata
Arp~220           & 15 34 57.23 & +23 30 11.3   & AGN+starburst &  78.6  \\
PG~1244+026       & 12 46 35.25 & +02 22 08.8   & Quasar        &  213.1 \\
PG~1351+236       & 13 54 06.43 & +23 25 49.1   & Quasar        &  245.4 \\
PG~1617+175       & 16 20 11.29 & +17 24 27.7   & Quasar         &  529.6 \\
IRAS~F01364--1042 & 01 38 52.79 & $-$10 27 12.1  & AGN+starburst &  213.1 \\
IRAS~F08572+3915  & 09 00 25.35 &  +39 03 54.0  & AGN+starburst &  261.2 \\
\enddata
\end{deluxetable}
%%% Table: Basic parameters for Unique Galaxies%%%

%%% Table 4: PAH Flux Measured from PAH Residuals (our method) %%%
\begin{landscape}
\begin{deluxetable}{rcccccccc}
\tabletypesize\tiny
\tablecolumns{9}
\tablecaption{\label{tab:spec_flux}
              PAH Fluxes Measured from Current Method
              }
\tablewidth{0pt}
\tablehead{
\colhead{Source} & \colhead{PAH (6.2$\mum$)} & \colhead{Err (6.2$\mum$)}  & 
                                \colhead{PAH (7.7$\mum$)} & \colhead{Err (7.7$\mum$)}  &
                                \colhead{PAH (8.6$\mum$)} & \colhead{Err (8.6$\mum$)}  &
                                \colhead{PAH (11.3$\mum$)} & \colhead{Err (11.3$\mum$)}  \\ 
\colhead{}  & \colhead{$\rm (erg~s^{-1} cm^{-2})$}  & \colhead{$\rm(erg~s^{-1} cm^{-2})$} & 
                     \colhead{$\rm (erg~s^{-1} cm^{-2})$}  & \colhead{$\rm (erg~s^{-1} cm^{-2})$} &
                     \colhead{$\rm (erg~s^{-1} cm^{-2})$}  & \colhead{$\rm (erg~s^{-1} cm^{-2})$} &
                     \colhead{$\rm (erg~s^{-1} cm^{-2})$}  & \colhead{$\rm (erg~s^{-1} cm^{-2})$} \\
\colhead{(1)} & \colhead{(2)} & \colhead{(3)} & \colhead{(4)} & \colhead{(5)} & 
\colhead{(6)} & \colhead{(7)} & \colhead{(8)} & \colhead{(9)}
}
\startdata
% obj, pah62_yx_spec, pah62_yx_spec_err, pah77_yx_spec, pah77_yx_spec_err, pah86_yx_spec, pah86_yx_spec_err, pah112_yx_spec, pah112_yx_spec_err
 Mrk~~33...................      &    3.26E-12 &  1.11E-12  &  1.19E-11 &  5.74E-12  &  3.37E-12 &  2.75E-12  &  3.57E-12 &  1.20E-12 \\
 NGC~24...................       &    7.17E-13 &  9.56E-14  &  2.52E-12 &  5.66E-13  &  6.80E-13 &  1.72E-13  &  1.05E-12 &  6.88E-14 \\
 NGC~337.................        &    5.78E-12 &  8.24E-14  &  2.06E-11 &  5.61E-13  &  5.20E-12 &  1.19E-13  &  5.57E-12 &  4.88E-14 \\
 NGC~628.................        &    2.61E-12 &  7.57E-14  &  1.09E-11 &  4.79E-13  &  3.36E-12 &  1.21E-13  &  3.10E-12 &  1.04E-12 \\
 NGC~855.................        &    6.21E-13 &  1.76E-13  &  2.32E-12 &  7.30E-13  &  1.06E-12 &  2.48E-13  &  1.19E-12 &  7.75E-14 \\
 NGC~925.................        &    2.00E-12 &  8.63E-14  &  7.33E-12 &  5.51E-13  &  2.14E-12 &  1.40E-13  &  2.06E-12 &  5.77E-14 \\
 NGC~1482 ..............         &    4.16E-11 &  1.34E-13  &  1.78E-10 &  7.70E-13  &  5.61E-11 &  2.80E-13  &  4.50E-11 &  1.83E-13 \\
 NGC~2403 ..............         &    4.50E-12 &  7.90E-14  &  1.41E-11 &  4.68E-13  &  4.54E-12 &  1.42E-13  &  4.23E-12 &  5.72E-14 \\
\enddata
\end{deluxetable}
\end{landscape}
%%% Table 4: PAH Flux Measured from PAH Residuals %%%

%%% Table 5: PAH Fluxes Measured from PAHFIT %%%
\begin{landscape}
\begin{deluxetable}{rcccccccc}
\tabletypesize\tiny
\tablecolumns{9}
\tablecaption{\label{tab:pahfit_flux}
              PAH Fluxes Measured from PAHFIT  
              }
\tablewidth{0pt}
\tablehead{
\colhead{Source} & \colhead{PAH (6.2$\mum$)} & \colhead{Err (6.2$\mum$)}  & 
                                \colhead{PAH (7.7$\mum$)} & \colhead{Err (7.7$\mum$)}  &
                                \colhead{PAH (8.6$\mum$)} & \colhead{Err (8.6$\mum$)}  &
                                \colhead{PAH (11.3$\mum$)} & \colhead{Err (11.3$\mum$)}  \\ 
\colhead{}  & \colhead{$\rm (erg~s^{-1} cm^{-2})$}  & \colhead{$\rm(erg~s^{-1} cm^{-2})$} & 
                     \colhead{$\rm (erg~s^{-1} cm^{-2})$}  & \colhead{$\rm(erg~s^{-1} cm^{-2})$} &
                     \colhead{$\rm (erg~s^{-1} cm^{-2})$}  & \colhead{$\rm (erg~s^{-1} cm^{-2})$} &
                     \colhead{$\rm (erg~s^{-1} cm^{-2})$}  & \colhead{$\rm (erg~s^{-1} cm^{-2})$} \\
\colhead{(1)} & \colhead{(2)} & \colhead{(3)} & \colhead{(4)} & \colhead{(5)} & 
\colhead{(6)} & \colhead{(7)} & \colhead{(8)} & \colhead{(9)}
}
\startdata
% obj, pah62_pahfit, pah62_pahfit_err, pah77_pahfit, pah77_pahfit_err, pah86_pahfit, pah86_pahfit_err, pah112_pahfit, pah112_pahfit_err
 Mrk~~33...................    &  3.44E-12 &  8.56E-14  &  1.17E-11 &  4.89E-13  &  3.21E-12 &  1.06E-13  &  3.15E-12 &  4.44E-14 \\
 NGC~24...................     &  8.85E-13 &  9.97E-14  &  2.52E-12 &  5.09E-13  &  7.15E-13 &  1.42E-13  &  9.83E-13 &  6.21E-14 \\
 NGC~337.................      &  6.08E-12 &  8.10E-14  &  1.65E-11 &  5.60E-13  &  4.84E-12 &  1.15E-13  &  5.31E-12 &  5.66E-14 \\
 NGC~628.................      &  2.53E-12 &  7.75E-14  &  9.75E-12 &  4.99E-13  &  2.99E-12 &  1.14E-13  &  2.48E-12 &  5.16E-14 \\
 NGC~855.................      &  9.09E-13 &  7.58E-14  &  2.10E-12 &  2.43E-13  &  1.04E-12 &  1.32E-13  &  1.11E-12 &  6.27E-14 \\
 NGC~925.................      &  2.38E-12 &  9.33E-14  &  7.34E-12 &  5.39E-13  &  2.24E-12 &  1.27E-13  &  1.95E-12 &  5.49E-14 \\
 NGC~1482...............       &  4.22E-11 &  1.27E-13  &  1.63E-10 &  5.23E-13  &  4.32E-11 &  1.49E-13  &  3.33E-11 &  9.29E-14 \\
 NGC~2403...............       &  4.98E-12 &  7.73E-14  &  1.26E-11 &  4.58E-13  &  4.18E-12 &  1.41E-13  &  4.06E-12 &  6.58E-14 \\
\enddata
\end{deluxetable}
\end{landscape}
%%% Table 5: PAH Fluxes Measured from PAHFIT %%%

%%% Table 6: PAH Fluxes Measured from SPLINE %%%
\begin{landscape}
\begin{deluxetable}{lrccrccrccrcc}
\tabletypesize\tiny
\tablecolumns{13}
\tablecaption{\label{tab:spline_flux}
              PAH Fluxes Measured from Spline Technique
              }
\tablewidth{0pt}
\tablehead{
\colhead{Source} & \colhead{PAH (6.2$\mum$)} & \colhead{lower} & \colhead{upper}  & 
                                \colhead{PAH (7.7$\mum$)} & \colhead{lower} & \colhead{upper}  & 
                                \colhead{PAH (8.6$\mum$)} & \colhead{lower} & \colhead{upper}  &  
                                \colhead{PAH (11.3$\mum$)} & \colhead{lower} & \colhead{upper}  \\                                
\colhead{}   & \colhead{$\rm (erg~s^{-1} cm^{-2})$}  & \colhead{limit} &  \colhead{limit} & 
                      \colhead{$\rm (erg~s^{-1} cm^{-2})$}  & \colhead{limit} &  \colhead{limit} & 
                      \colhead{$\rm (erg~s^{-1} cm^{-2})$}  & \colhead{limit} &  \colhead{limit} &  
                      \colhead{$\rm (erg~s^{-1} cm^{-2})$}  & \colhead{limit}  & \colhead{limit} \\                       
\colhead{(1)} & \colhead{(2)} & \colhead{(3)} & \colhead{(4)} & \colhead{(5)}  & \colhead{(6)} & \colhead{(7)} & 
\colhead{(8)} & \colhead{(9)} & \colhead{(10)} & \colhead{(11)} & \colhead{(12)} & \colhead{(13)}    }  
\startdata
%
% obj, pah62_spline, pah62_err_low, pah62_err_upp, pah77_spline, pah77_err_low, pah77_err_upp, pah86_spline, pah86_err_low, pah86_err_upp, pah112_spline, pah112_err_low, pah112_err_upp
 Mrk~~33 .................  &      2.38E-12  &  2.14E-12  &  2.83E-12  &      4.24E-12  &  3.00E-12  &  4.74E-12  &      1.02E-12  &  9.27E-13  &  1.09E-12  &      1.78E-12  &  1.72E-12  &  1.85E-12 \\
 NGC~24  .................  &      4.38E-13  &  4.37E-13  &  6.21E-13  &      1.02E-12  &  8.01E-13  &  1.35E-12  &      2.28E-13  &  2.12E-13  &  3.58E-13  &      4.97E-13  &  4.95E-13  &  5.61E-13 \\
 NGC~337  ...............   &      3.95E-12  &  3.66E-12  &  4.37E-12  &      6.02E-12  &  5.81E-12  &  7.70E-12  &      1.60E-12  &  1.30E-12  &  1.82E-12  &      3.00E-12  &  2.83E-12  &  3.11E-12 \\
 NGC~628  ...............   &      2.01E-12  &  1.73E-12  &  2.47E-12  &      2.22E-12  &  1.37E-12  &  3.10E-12  &      5.10E-13  &  3.04E-13  &  6.18E-13  &      1.38E-12  &  1.36E-12  &  1.45E-12 \\
 NGC~855  ...............   & $<$  8.52E-13  &     --     &     --     &      8.63E-13  &  7.53E-13  &  1.39E-12  & $<$  3.42E-13  &     --     &     --     &      6.82E-13  &  6.49E-13  &  8.92E-13 \\
 NGC~925  ...............   &      1.20E-12  &  9.70E-13  &  1.45E-12  &      2.78E-12  &  2.42E-12  &  3.16E-12  &      5.05E-13  &  2.72E-13  &  5.60E-13  &      1.01E-12  &  8.96E-13  &  1.11E-12 \\
 NGC~1482 ..............    &      2.68E-11  &  2.32E-11  &  2.80E-11  &      4.87E-11  &  4.74E-11  &  5.70E-11  &      8.86E-12  &  7.60E-12  &  1.11E-11  &      1.69E-11  &  1.62E-11  &  1.74E-11 \\
 NGC~2403 ..............    &      3.40E-12  &  2.69E-12  &  3.73E-12  &      2.19E-12  &  1.68E-12  &  3.90E-12  &      5.97E-13  &  5.63E-13  &  1.05E-12  &      2.11E-12  &  2.02E-12  &  2.29E-12 \\
\enddata
\end{deluxetable}
\end{landscape}
%%% Table 6: PAH Flux Measured from SPLINE %%%

%%% Table 7: PAH Flux Measured from CAFE %%%
\begin{landscape}
\begin{deluxetable}{rcccccccc}
\tabletypesize\tiny
\tablecolumns{9}
\tablecaption{\label{tab:cafe_flux}
%\textcolor{red}{
	      PAH Fluxes Measured Using Template in CAFE} 
%              }
\tablewidth{0pt}
\tablehead{
\colhead{Source} & \colhead{PAH (6.2$\mum$)} & \colhead{Err (6.2$\mum$)}  & 
                                \colhead{PAH (7.7$\mum$)} & \colhead{Err (7.7$\mum$)}  &
                                \colhead{PAH (8.6$\mum$)} & \colhead{Err (8.6$\mum$)}  &
                                \colhead{PAH (11.3$\mum$)} & \colhead{Err (11.3$\mum$)}  \\ 
\colhead{}  & \colhead{$\rm (erg~s^{-1} cm^{-2})$}  & \colhead{$\rm(erg~s^{-1} cm^{-2})$} & 
                     \colhead{$\rm (erg~s^{-1} cm^{-2})$}  & \colhead{$\rm (erg~s^{-1} cm^{-2})$} &
                     \colhead{$\rm (erg~s^{-1} cm^{-2})$}  & \colhead{$\rm (erg~s^{-1} cm^{-2})$} &
                     \colhead{$\rm (erg~s^{-1} cm^{-2})$}  & \colhead{$\rm (erg~s^{-1} cm^{-2})$} \\
\colhead{(1)} & \colhead{(2)} & \colhead{(3)} & \colhead{(4)} & \colhead{(5)} & 
\colhead{(6)} & \colhead{(7)} & \colhead{(8)} & \colhead{(9)}
}
\startdata
 Mrk~~33...................       &  2.74E-12 &  3.88E-12  &  8.79E-12 &  1.50E-11  &  2.80E-12 &  9.63E-12  &  3.25E-12 &  9.47E-12 \\
 NGC~24...................        &  6.98E-13 &  1.93E-13  &  1.65E-12 &  5.50E-13  &  5.85E-13 &  3.85E-13  &  1.01E-12 &  3.54E-13 \\
 NGC~337.................         &  5.61E-12 &  2.40E-13  &  1.55E-11 &  1.23E-12  &  4.60E-12 &  4.83E-13  &  5.43E-12 &  4.05E-13 \\
 NGC~628.................         &  2.23E-12 &  5.84E-13  &  8.31E-12 &  3.48E-12  &  2.84E-12 &  1.72E-12  &  2.37E-12 &  1.19E-12 \\
 NGC~855.................         &     --    &     --     &  9.75E-13 &  1.61E-12  &  5.86E-13 &  1.22E-12  &  1.10E-12 &  1.15E-12 \\
 NGC~925.................         &  2.00E-12 &  8.69E-14  &  6.04E-12 &  5.61E-13  &  1.79E-12 &  1.50E-13  &  1.86E-12 &  6.41E-14 \\
 NGC~1482 ..............          &  4.32E-11 &  1.08E-13  &  1.57E-10 &  6.32E-13  &  5.95E-11 &  2.37E-13  &  4.52E-11 &  1.36E-13 \\
 NGC~2403 ..............          &  4.37E-12 &  1.55E-13  &  1.07E-11 &  7.50E-13  &  4.15E-12 &  3.29E-13  &  4.32E-12 &  2.33E-13 \\
\enddata
\end{deluxetable}
\end{landscape}
%%% Table 7: PAH Flux Measured from CAFE %%%

%%% Table 8: Extinctions %%%
\begin{deluxetable}{lrccrcccc}
\tabletypesize\tiny
\tablecolumns{9}
\tablecaption{\label{tab:extinction}
%\textcolor{red}{
	Extinctions Measured from 
        the Current Method, PAHFIT, and CAFE 
              }
%              }
\tablewidth{0pt}
\tablehead{
\colhead{} & \multicolumn{2}{c} {This Work}  & &  
                      \multicolumn{2}{c} {PAHFIT}  & &  
                      \multicolumn{2}{c} {CAFE} \\
%\colhead{} & \colhead{} & \colhead{} & \colhead{} & \colhead{}  \\ 
\cline{2-3} 
\cline{5-6} 
\cline{8-9}
\colhead{Source} & \colhead{$\tau_{9.7}$}   &  \colhead{Err($\rm \tau_{9.7}$)}  & &
\colhead{$\tau_{9.7}$}   &  \colhead{Err($\rm \tau_{9.7}$)}  & &
\colhead{$\rm \tau_{9.7}$}  &  \colhead{Err($\rm \tau_{9.7}$)}   \\ 
\colhead{(1)} & \colhead{(2)} & \colhead{(3)} & & \colhead{(4)} & \colhead{(5)} & &
\colhead{(6)} & \colhead{(7)} 
}
\startdata
% obj, tau_yx, tau_yx_err , tau_pahfit, tau_pahfit_err, tau_cafe, tau_cafe_err
 Mrk~~33..................      &     $<$  1.53E-01 &     --    &  &     --    &     --      &  &        7.96E-02  &  2.38E-01 \\
 NGC~24..................       &     $<$  8.60E-02 &     --    &  &     --    &     --      &  &        2.31E-01  &  8.35E-02 \\
 NGC~337................        &     $<$  5.30E-03 &     --    &  &  1.49E-01 &  1.77E-02   &  &        1.52E-01  &  1.36E-02 \\
 NGC~628................        &     $<$  7.96E-03 &     --    &  &  3.72E-02 &  3.17E-02   &  &        3.75E-02  &  1.85E-02 \\
 NGC~855................        &     $<$  3.61E-02 &     --    &  &  7.97E-02 &  8.04E-02   &  &        5.77E-02  &  7.36E-02 \\
 NGC~925................        &     $<$  1.15E-02 &     --    &  &     --    &     --      &  &   $<$  5.67E-03  &     --    \\
 NGC~1482..............         &          3.71E-01 &  1.99E-03 &  &  3.03E-01 &  6.97E-03   &  &        8.15E-01  &  1.94E-03 \\
 NGC~2403..............         &     $<$  4.40E-03 &     --    &  &  1.70E-01 &  2.35E-02   &  &        2.63E-01  &  1.52E-02 \\
\enddata
\end{deluxetable}
%%% Table 8: Extinctions %%%


\clearpage

\begin{thebibliography}{}

\expandafter\ifx\csname natexlab\endcsname\relax\def\natexlab#1{#1}\fi

\bibitem[Acke et al.(2010)]{2010ApJ...718..558A} Acke, B., Bouwman, J., Juh{\'a}sz, A., et al.\ 2010, \apj, 718, 558 

\bibitem[Allamandola et al.(1985)]{1985ApJ...290L..25A} Allamandola, L.~J., Tielens, A.~G.~G.~M., \& Barker, J.~R.\ 1985, \apjl, 290, L25 

\bibitem[Allamandola et al.(1999)]{1999ApJ...511L.115A} Allamandola, L.~J., Hudgins, D.~M., \& Sandford, S.~A.\ 1999, \apjl, 511, L115

%\bibitem[Alonso-Herrero et al.(2016)]{2016MNRAS.455..563A} Alonso-Herrero, A., Esquej, P., Roche, P.~F., et al.\ 2016a, \mnras, 455, 563 

\bibitem[Alonso-Herrero et al.(2016)]{2016MNRAS.463.2405A} Alonso-Herrero, A., Poulton, R., Roche, P.~F., et al.\ 2016, \mnras, 463, 2405 

\bibitem[Alonso-Herrero et al.(2014)]{2014MNRAS.443.2766A} Alonso-Herrero, A., Ramos Almeida, C., Esquej, P., et al.\ 2014, \mnras, 443, 2766 

\bibitem[Andrews et al.(2015)]{2015ApJ...807...99A} Andrews, H., Boersma, C., Werner, M.~W., et al.\ 2015, \apj, 807, 99

\bibitem[Armus et al.(2007)]{2007ApJ...656..148A} Armus, L., Charmandaris, V., Bernard-Salas, J., et al.\ 2007, \apj, 656, 148

\bibitem[Bakes \& Tielens(1994)]{1994ApJ...427..822B} Bakes, E.~L.~O., \& Tielens, A.~G.~G.~M.\ 1994, \apj, 427, 822 

\bibitem[Barcos-Mu{\~n}oz et al.(2015)]{2015ApJ...799...10B} Barcos-Mu{\~n}oz, L., Leroy, A.~K., Evans, A.~S., et al.\ 2015, \apj, 799, 10 

\bibitem[Batejat et al.(2012)]{2012A&A...542L..24B} Batejat, F., Conway, J.~E., Rushton, A., et al.\ 2012, \aap, 542, L24 

\bibitem[Bauschlicher et al.(2010)]{2010ApJS..189..341B} Bauschlicher, C.~W., Jr., Boersma, C., Ricca, A., et al.\ 2010, \apjs, 189, 341-351 

\bibitem[Blitz et al.(1984)]{1984NASCP2345..231B} Blitz, L., Magnani, L., \& Mundy, L.\ 1984, NASA Conference Publication, 2345, 231

\bibitem[Boersma et al.(2013)]{2013ApJ...769..117B} Boersma, C., Bregman, J.~D., \& Allamandola, L.~J.\ 2013, \apj, 769, 117

\bibitem[Boersma et al.(2014)]{2014ApJ...795..110B} Boersma, C., Bregman, J., \& Allamandola, L.~J.\ 2014, \apj, 795, 110 

\bibitem[Brandl et al.(2006)]{2006ApJ...653.1129B} Brandl, B.~R., Bernard-Salas, J., Spoon, H.~W.~W., et al.\ 2006, \apj, 653, 1129 

\bibitem[Brown et al.(2014)]{2014ApJS..212...18B} Brown, M.~J.~I., Moustakas, J., Smith, J.-D.~T., et al.\ 2014, \apjs, 212, 18 

\bibitem[Cami(2011)]{2011EAS....46..117C} Cami, J.\ 2011, EAS Publications Series, 46, 117 

\bibitem[Chiar \& Tielens(2006)]{2006ApJ...637..774C} Chiar, J.~E., \& Tielens, A.~G.~G.~M.\ 2006, \apj, 637, 774

\bibitem[Chu et al.(2017)]{2017ApJS..229...25C} Chu, J.~K., Sanders, D.~B., Larson, K.~L., et al.\ 2017, \apjs, 229, 25 

\bibitem[da Cunha et al.(2010)]{2010A&A...523A..78D} da Cunha, E., Charmandaris, V., D{\'{\i}}az-Santos, T., et al.\ 2010, \aap, 523, A78 

\bibitem[Dale et al.(2012)]{2012ApJ...745...95D} Dale, D.~A., Aniano, G., Engelbracht, C.~W., et al.\ 2012, \apj, 745, 95 

\bibitem[Dale et al.(2006)]{2006ApJ...646..161D} Dale, D.~A., Smith, J.~D.~T., Armus, L., et al.\ 2006, \apj, 646, 161 
%\textcolor{red}{
\bibitem[Diamond-Stanic \& Rieke(2010)]{2010ApJ...724..140D} Diamond-Stanic, A.~M., \& Rieke, G.~H.\ 2010, \apj, 724, 140 %}

\bibitem[Draine(2003)]{2003ARA&A..41..241D} Draine, B.~T.\ 2003, \araa, 41, 241 

\bibitem[Draine \& Lee(1984)]{1984ApJ...285...89D} Draine, B.~T., \& Lee, H.~M.\ 1984, \apj, 285, 89 

\bibitem[Draine \& Li(2007)]{2007ApJ...657..810D} Draine, B.~T., \& Li, A.\ 2007, \apj, 657, 810 

\bibitem[Draine \& Li(2001)]{2001ApJ...551..807D} Draine, B.~T., \& Li, A.\ 2001, \apj, 551, 807 

\bibitem[Geers et al.(2006)]{2006A&A...459..545G} Geers, V.~C., Augereau, J.-C., Pontoppidan, K.~M., et al.\ 2006, \aap, 459, 545 

\bibitem[Gillett et al.(1975)]{1975ApJ...200..609G} Gillett, F.~C., Forrest, W.~J., Merrill, K.~M., Soifer, B.~T., \& Capps, R.~W.\ 1975, \apj, 200, 609 

\bibitem[Genzel et al.(1998)]{1998ApJ...498..579G} Genzel, R., Lutz, D., Sturm, E., et al.\ 1998, \apj, 498, 579 

\bibitem[Hao et al.(2009)]{2009ApJ...704.1159H} Hao, L., Wu, Y., Charmandaris, V., et al.\ 2009, \apj, 704, 1159 

\bibitem[Heckman(1980)]{1980A&A....87..152H} Heckman, T.~M.\ 1980, \aap, 87, 152 

%\bibitem[Helou et al.(2004)]{2004ApJS..154..253H} Helou, G., Roussel, H., Appleton, P., et al.\ 2004, \apjs, 154, 253

\bibitem[Hony et al.(2001)]{2001A&A...370.1030H} Hony, S., Van Kerckhoven, C., Peeters, E., et al.\ 2001, \aap, 370, 1030 

\bibitem[Houck et al.(2004)]{2004SPIE.5487...62H} Houck, J.~R., Roellig, T.~L., Van Cleve, J., et al.\ 2004, \procspie, 5487, 62 

\bibitem[Huang et al.(2009)]{2009ApJ...700..183H} Huang, J.-S., Faber, S.~M., Daddi, E., et al.\ 2009, \apj, 700, 183 

\bibitem[Hudgins \& Allamandola(2004)] HHudgins, D. M., \& Allamandola, L. J. 2004, in ASP Conf. Ser. 309, Astrophysics of Dust, ed. A.N. Witt, G.C. Clayton, \& B.T. Draine (San Francisco, CA: ASP), 665

\bibitem[Hudgins \& Allamandola(2005)] HHudgins, D. M., \& Allamandola, L. J. 2005, in IAU Symp. 231, Astrochemistry: Recent Successes and Current Challenges, ed D.C. Lis, G.A. Blake,  \& E. Herbst (Cambridge: Cambridge Univ. Press), 443

%\bibitem[Hudgins \& Allamandola(2004)]{2004ASPC..309..665H} Hudgins, D.~M., \& Allamandola, L.~J.\ 2004, Astrophysics of Dust, 309, 665

%\bibitem[Hudgins \& Allamandola(2005)]{2005IAUS..231..443H} Hudgins, D.~M., \& Allamandola, L.~J.\ 2005, Astrochemistry: Recent Successes and Current Challenges, 231, 443

%\bibitem[Hudgins et al.(2005)]{2005ApJ...632..316H} Hudgins, D.~M., Bauschlicher, C.~W., Jr., \& Allamandola, L.~J.\ 2005, \apj, 632, 316

\bibitem[Ingalls et al.(2011)]{2011ApJ...743..174I} Ingalls, J.~G., Bania, T.~M., Boulanger, F., et al.\ 2011, \apj, 743, 174 

\bibitem[Iwasawa et al.(2011)]{2011A&A...529A.106I} Iwasawa, K., Sanders, D.~B., Teng, S.~H., et al.\ 2011, \aap, 529, A106 

\bibitem[Iwasawa et al.(2001)]{2001MNRAS.326..894I} Iwasawa, K., Matt, G., Guainazzi, M., \& Fabian, A.~C.\ 2001, \mnras, 326, 894 

\bibitem[Kaneda et al.(2005)]{2005ApJ...632L..83K} Kaneda, H., Onaka, T., \& Sakon, I.\ 2005, \apjl, 632, L83

\bibitem[Kaneda et al.(2008)]{2008ApJ...684..270K} Kaneda, H., Onaka, T., Sakon, I., et al.\ 2008, \apj, 684, 270

\bibitem[Kemper et al.(2004)]{2004ApJ...609..826K} Kemper, F., Vriend, W.~J., \& Tielens, A.~G.~G.~M.\ 2004, \apj, 609, 826 

\bibitem[Kennicutt et al.(2003)]{2003PASP..115..928K} Kennicutt Jr., R.~C., Armus, L., Bendo, G., et al.\ 2003, \pasp, 115, 928 

\bibitem[Kennicutt \& Evans(2012)]{2012ARA&A..50..531K} Kennicutt Jr., R.~C., \& Evans, N.~J.\ 2012, \araa, 50, 531 

\bibitem[Leger \& Puget(1984)]{1984A&A...137L...5L} L\'eger, A., \& Puget, J.~L.\ 1984, \aap, 137, L5 

\bibitem[Lebouteiller et al.(2011)]{2011ApJS..196....8L} Lebouteiller, V., Barry, D.~J., Spoon, H.~W.~W., et al.\ 2011, \apjs, 196, 8 

\bibitem[Li(2004)] LLi, A. 2004, in ASP Conf. Ser. 309, Astrophysics of Dust, ed. A.N. Witt, G.C. Clayton, \& B.T. Draine (San Francisco, CA: ASP), 417

\bibitem[Li(2009)]{LiAG09}Li, A. 2009, in Small Bodies in Planetary Sciences (Lecture Notes in Physics Vol. 758), ed. I. Mann, A. Nakamura, \& T. Mukai (Springer), 167

\bibitem[Li \& Draine(2001)]{2001ApJ...554..778L} Li, A., \& Draine, B.~T.\ 2001, \apj, 554, 778 

\bibitem[Li \& Mann(2012)]LLi, A., \& Mann, I. 2012, in Astrophys. Space Sci. Library, Vol. 385, Nanodust in the Solar System: Discoveries and Interpretations (Berlin: Springer-Verlag), 5

\bibitem[Low et al.(1984)]{1984ApJ...278L..19L} Low, F.~J., Young, E., Beintema, D.~A., et al.\ 1984, \apjl, 278, L19 

%\bibitem[Lu et al.(2003)]{2003ApJ...588..199L} Lu, N., Helou, G., Werner, M.~W., et al.\ 2003, \apj, 588, 199

\bibitem[Lutz et al.(1998)]{1998ApJ...505L.103L} Lutz, D., Spoon, H.~W.~W., Rigopoulou, D., Moorwood, A.~F.~M., \& Genzel, R.\ 1998, \apjl, 505, L103 

\bibitem[Magdis et al.(2011)]{2011A&A...534A..15M} Magdis, G.~E., Elbaz, D., Dickinson, M., et al.\ 2011, \aap, 534, A15   

\bibitem[Magnani et al.(1985)]{1985ApJ...295..402M} Magnani, L., Blitz, L., \& Mundy, L.\ 1985, \apj, 295, 402 

\bibitem[Markwardt(2009)]{2009ASPC..411..251M} Markwardt, C.~B.\ 2009, Astronomical Data Analysis Software and Systems XVIII, 411, 251

\bibitem[Marshall et al.(2007)]{2007ApJ...670..129M} Marshall, J.~A., Herter, T.~L., Armus, L., et al.\ 2007, \apj, 670, 129 

\bibitem[Mathis et al.(1983)]{1983A&A...128..212M} Mathis, J.~S., Mezger, P.~G., \& Panagia, N.\ 1983, \aap, 128, 212 
%\bibitem[Mattioda et al.(2005)]{2005ApJ...629.1188M} Mattioda, A.~L., Hudgins, D.~M., \& Allamandola, L.~J.\ 2005, \apj, 629, 1188
%\textcolor{red}{
\bibitem[Moutou et al.(1996)]{1996A&A...310..297M} Moutou, C., L\'eger, A., \& D'Hendecourt, L.\ 1996, \aap, 310, 297% }

\bibitem[Norris(1988)]{1988MNRAS.230..345N} Norris, R.~P.\ 1988, \mnras, 230, 345 

\bibitem[Onaka et al.(1996)]{1996PASJ...48L..59O} Onaka, T., Yamamura, I., Tanabe, T., Roellig, T.~L., \& Yuen, L.\ 1996, \pasj, 48, L59

\bibitem[Peeters et al.(2004)]{2004ApJ...613..986P} Peeters, E., Spoon, H.~W.~W., \& Tielens, A.~G.~G.~M.\ 2004, \apj, 613, 986 

\bibitem[Rampazzo et al.(2013)]{2013MNRAS.432..374R} Rampazzo, R., Panuzzo, P., Vega, O., et al.\ 2013, \mnras, 432, 374

\bibitem[Rieke et al.(2009)]{2009ApJ...692..556R} Rieke, G.~H., Alonso-Herrero, A., Weiner, B.~J., et al.\ 2009, \apj, 692, 556

\bibitem[Roche \& Aitken(1984)]{1984MNRAS.208..481R} Roche, P.~F., \& Aitken, D.~K.\ 1984, \mnras, 208, 481

\bibitem[Rosenberg et al.(2011)]{2011A&A...532A.128R} Rosenberg, M.~J.~F., Bern{\'e}, O., Boersma, C., Allamandola, L.~J., \& Tielens, A.~G.~G.~M.\ 2011, \aap, 532, A128 
%\textcolor{red}{
\bibitem[Sales et al.(2010)]{2010ApJ...725..605S} Sales, D.~A., Pastoriza, M.~G., \& Riffel, R.\ 2010, \apj, 725, 605%}

\bibitem[Sargsyan et al.(2011)]{2011ApJ...730...19S} Sargsyan, L., Weedman, D., Lebouteiller, V., et al.\ 2011, \apj, 730, 19  

\bibitem[Seok \& Li(2015)]{2015ApJ...809...22S} Seok, J.~Y., \& Li, A.\ 2015, \apj, 809, 22


\bibitem[Schmidt \& Green(1983)]{1983ApJ...269..352S} Schmidt, M., \& Green, R.~F.\ 1983, \apj, 269, 352

\bibitem[Shangguan et al.(2018)]{2018ApJ...854..158S} Shangguan, J., Ho, L.~C., \& Xie, Y.\ 2018, \apj, 854, 158

\bibitem[Shannon et al.(2016)]{2016ApJ...824..111S} Shannon, M.~J., Stock, D.~J., \& Peeters, E.\ 2016, \apj, 824, 111 

\bibitem[Shi et al.(2014)]{2014ApJS..214...23S} Shi, Y., Rieke, G.~H., Ogle, P.~M., Su, K.~Y.~L., \& Balog, Z.\ 2014, \apjs, 214, 23 

\bibitem[Smith et al.(2007)]{2007ApJ...656..770S} Smith, J.~D.~T., Draine, B.~T., Dale, D.~A., et al.\ 2007, \apj, 656, 770 

\bibitem[Spoon et al.(2006)]{2006ApJ...638..759S} Spoon, H.~W.~W., Tielens, A.~G.~G.~M., Armus, L., et al.\ 2006, \apj, 638, 759 

\bibitem[Tanaka et al.(1996)]{1996PASJ...48L..53T} Tanaka, M., Matsumoto, T., Murakami, H., et al.\ 1996, \pasj, 48, L53

\bibitem[Tielens(2008)]{2008ARA&A..46..289T} Tielens, A.~G.~G.~M.\ 2008, \araa, 46, 289 

\bibitem[Veilleux et al.(2009)]{2009ApJS..182..628V} Veilleux, S., Rupke, D.~S.~N., Kim, D.-C., et al.\ 2009, \apjs, 182, 628

\bibitem[Vermeij et al.(2002)]{2002A&A...382.1042V} Vermeij, R., Peeters, E., Tielens, A.~G.~G.~M., \& van der Hulst, J.~M.\ 2002, \aap, 382, 1042 

\bibitem[Wang et al.(2015)]{2015ApJ...811...38W} Wang, S., Li, A., \& Jiang, B.~W.\ 2015, \apj, 811, 38 

\bibitem[Weingartner \& Draine(2001a)]{2001ApJ...548..296W} Weingartner, J.~C., \& Draine, B.~T.\ 2001a, \apj, 548, 296 

\bibitem[Weingartner \& Draine(2001b)]{2001ApJ...553..581W} Weingartner, J.~C., \& Draine, B.~T.\ 2001b, \apj, 553, 581 

\bibitem[Werner et al.(2004)]{2004ApJS..154....1W} Werner, M.~W., Roellig, T.~L., Low, F.~J., et al.\ 2004, \apjs, 154, 1 

\bibitem[Wu et al.(2010)]{2010ApJ...723..895W} Wu, Y., Helou, G., Armus, L., et al.\ 2010, \apj, 723, 895

%\bibitem[Xie et al.(2018)]{Xie2018} Xie, Y., Ho, L. C., Li, A., \& Shangguan, J. 2018, in preparation

\bibitem[Xie et al.(2017)]{2017ApJS..228....6X} Xie, Y., Li, A., \& Hao, L.\ 2017, \apjs, 228, 6 

\bibitem[Yang et al.(2017)]{2017NewAR..77....1Y} Yang, X.~J., Glaser, R., Li, A., \& Zhong, J.~X.\ 2017a, \nar, 77, 1

\bibitem[Yang et al.(2017)]{2017ApJ...837..171Y} Yang, X.~J., Li, A., Glaser, R., \& Zhong, J.~X.\ 2017b, \apj, 837, 171


\bibitem[Yuan et al.(2010)]{2010ApJ...709..884Y} Yuan, T.-T., Kewley, L.~J., \& Sanders, D.~B.\ 2010, \apj, 709, 884 

\end{thebibliography}
\end{document}